\documentclass[eqsecnum,showkeys,secnumarabic,showpacs, twocolumn]{revtex4}

\usepackage{amsfonts}
\usepackage{amsmath}
\usepackage{amssymb}
\usepackage{graphicx}

\usepackage{dcolumn}

\usepackage{bm}

\begin{document}

\preprint{...}

\title{Game Theory and Topological Phase Transition}

\author{Tieyan Si}\email{sity@itp.ac.cn}

\affiliation{Institute of Theoretical Physics, Chinese Academy of
Sciences, Beijing, 100080, China\\
Graduate School of the Chinese Academy of Sciences, Beijing,
100039, China}

\date{\today}

\begin{abstract}

Phase transition is a war game. It widely exists in different
kinds of complex system beyond physics. Where there is revolution,
there is phase transition. The renormalization group
transformation, which was proved to be a powerful tool to study
the critical phenomena, is actually a game process. The phase
boundary between the old phase and new phase is the outcome of
many rounds of negotiation between the old force and new force.
The order of phase transition is determined by the cutoff of
renormalization group transformation. This definition unified
Ehrenfest's definition of phase transition in thermodynamic
physics. If the strategy manifold has nontrivial topology, the
topological relation would put a constrain on the surviving
strategies, the transition occurred under this constrain may be
called a topological one. If the strategy manifold is open and
noncompact, phase transition is simply a game process, there is no
table for topology. An universal phase coexistence equation is
found, it sits at the Nash equilibrium point. Inspired by the
fractal space structure demonstrated by renormalization group
theory, a conjecture is proposed that the universal scaling law of
a general phase transition in a complex system comes from the
coexistence equation around Nash equilibrium point. Game theory
also provide us new understanding to pairing mechanism and
entanglement in many body physics.

\end{abstract}

 \maketitle

\tableofcontents

\section{Introduction}

When physicists encounter millions of interacting molecules or
atoms, they can not control the trajectory or momentum of each
individual particles, so they study the macroscopic states at
different temperature or other physical parameter. In most cases,
there would be some significant change of the macroscopic states
at certain value of parameters, they call it a phase transition.
Phase transition are common phenomena in all branches of physics.
People's interest in phase transition can trace back to thousands
of years ago. A recent example is the superfluid-Mott-insulator
phase transition occurred in a gas of ultracold atoms in optical
lattice\cite{greiner}.

Statistic mechanics were developed to provide a theoretical
description of phase transition. The occurrence of phase
transition is related to the singularity of statistical functions
in the thermodynamic limit(see Ref. \cite{stanley} for review).
However statistical mechanics is not powerful enough to predict
all different kind of phase transition. For classical Hamiltonian
systems, the hypothesis connecting phase transitions to the change
of configuration space topology was proposed recent years(see Ref.
\cite{kastner} and references therein for review). Topological
order also rise in quantum systems (see Ref. \cite{wen} for
review), such as the fractional quantum Hall system. Different
fractional quantum Hall effect states all have the same symmetry,
that is beyond the Landau¡¯s symmetry breaking theory. The
topological order in quantum system is related to degenerate
ground states, quasiparticle statistics, edge states, or momentum
topology\cite{volovik}, et al. These topological order theory told
us many interesting phenomena in some special models and special
systems. However it is still far from providing us an universal
explanation to phase transition occurred in all different branches
of physics, much less in other complex system beyond physics.

Generally people believed that it is the thermal fluctuation that
drives the transition from one phase to another in classical phase
transition. As temperature is lowered, the thermal fluctuations
are suppressed. The quantum fluctuation began to play a vital role
in quantum phase transition. Unfortunately this kind of argument
does not hold in many quantum system. More over, some physicist
believed that phase transition is due to the competition between
two competing orders in a physical system, but people can always
propose some anomalous examples which can not be explained by two
competing orders.

The occurrence mechanism of phase transitions is still not clear
in countless systems. Similar critical phenomena arose in a broad
physical and social systems. Many unrelated models covering
physics, chemistry, biology present similar scaling laws. Anyone
who saw this can not help asking why. Is there a 'theory of
everything' that can give us an universal explanation? Such a
theory sounds like superstring theory.

What I am trying to do in this paper is not to establish the
superstring theory of phase transition, but to present an
universal theoretical explanation to phase transition occurred in
different systems based on renormalization group theory and game
theory, and beyond the two, such as topology, quantum field
theory.

The first step is to break the envelop of physics and extend the
concept of phase transition from physical system to complex system
including chemistry system, biological system, social system,
economic system, et al. Phase transition is a sudden jumping from
one stable state to another. A two player game always has two
stable states, the winning state and losing state. Thus we can
define phase transition as a war game. If we check the war game
carefully, one would see it has all the the same phenomena as
phase transition occurred in physical system. War is a conflict
between millions of soldiers who are armed and well organized. The
butterfly effect is a basic character of war going on. The final
destiny of the fighter is determined by some minor accidental
event. When the two groups with equivalent force are fighting
against each other but keeping at a draw state, if any one of them
is reinforced a little, he will win the war in a few seconds. This
is a phase transition.

So we can take phase transition as a war game, the strategies of
the players extended the strategy base manifold. When we are
studying the state evolution of the game corresponding to
different strategy, the topology of the strategy base manifold
comes in. The phase transition is a war game between new phase and
old phase, each of them is governed by a kind of dominant
interaction(it may has many minor affiliated companions).  It will
be shown in the main tex of this paper, the surviving strategy of
the two phases carry opposite winding number, the sum of these
winding numbers is a topological number on the strategy base
manifold. After the winding number are annihilated by pairs, the
last one winding number around the last surviving strategy is
decided by the topological number, this also decided who will be
the winner between the old phase and new phase. It is in this
sense, we call it a topological phase transition. In fact, phase
transition is always related to the topology of the strategy base
manifold when the strategy base manifold is compact. In some
cases, the base manifold is open and noncompact , there is no need
to consider the effect of topological constrain. In fact, the
compact strategy manifold like a finite region confined by
fencing, it put some constrains on the way we choose strategy.
There is singular point we can never eliminated smoothly in
strategy space, that is the fundamental origin of topological
phase transition.

The paper is organized as follows:

In section 2, the most general conception of phase transition in
complex system is defined.

In section 3, we established the game theory of renormalization
group transformation, and find the general solution of
renormalization group transformation equations. The fundament
classification of phase transition through symmetry losing is
presented.

In section 4, topological current theory of phase transition is
established, this theory spontaneously produced an universal
equation of phase coexistence. A conjecture on the universal
scaling law is proposed base on a topological hypothesis in
fractal strategy space of game theory. Further more, we
established the evolution equation of phases and the quantum phase
coexistence equation.

In section 5, we developed the quantum statistics of many player
game, and proposed a conjecture to find the fixed point of a many
player game using quantum density matrix. More over, a new
quantity to measure the entanglement of quantum states in a game
is found. The single direction of renormalization group follows
the second thermodynamic law, the renormalization flows to
increase the entropy as well as the quantum entanglement. More
over, the coexistence state of multi-player game is discussed.

In section 6, we developed the game theory of phase transition in
classical many body system as well as quantum many body system. A
new holographic topological quantity to characterize momentum
space is proposed. We studied the quantum many body theory of war
game and gave a new pairing mechanism base on prisoner dilemma.

The last section is devoted to a brief summary and outlook.

\section{Phase transition and war game}

\subsection{Phases of complex system}

A complex system consists of many different elements that are
connected or related, it appears like a black box to us. One can
obtain the information within the complex system by its responses
to external perturbation. These responses and perturbations are
macroscopic variables. Different stable phases of complex system
are characterized and distinguished by these observables.

For most condensed matter physicists, a liter of water is a
complex system, since it is hardly possible to find exact
analytical solution for the motion of $10^{27}$ $H_{2}O$
molecules. We can characterize its different phases by observing
its chemical composition and physical properties, such as volume,
pressure, temperature, density, crystal structure, index of
refraction, chemical potential, and so forth. One of the main task
of physicist is to find the relations among these macroscopic
variables via presumably well known microscopic interactions
between particles.

A stable phase has self-restoring ability when it is disturbed
from equilibrium states. During a relative long life time, the
chemical makeup does not decompose, and the physical properties
keep a good manner. If we know the position and momentum of every
particles in a dynamic system at a given time, the evolution of an
integral system is exactly predictable. Unfortunately, there is
few integrable many body system. The position and momentum of
particles can not be exactly measured at the same time due to the
Heisenberg uncertainty principal. So we define the state vector of
a stable phase by the complete set of observables.

Let $\vec{x}$ be $n$ independent states of a complex system. For a
physical system consists of $N$ interacting particles, the
components of an arbitrary vector $\vec{x}_{i}$ are consists of
$3N$ position coordinates $q_{1},q_{1},...,q_{3N}$, $3N$ momentum
coordinates $p_{1},p_{1},...,p_{3N}$, and a vector of other
physical parameters $\vec{\gamma}$, such as temperature
$\gamma_{1}=T$, pressure $\gamma_{2}=P$, particle density
$\gamma_{3}=N$, volume $\gamma_{4}=V$, chemical potential
$\gamma_{5}=\mu$,$\cdots$, conductivity $\gamma_{j}=\sigma$,
susceptibility $\gamma_{k}=\chi$, and so on. Here the state vector
$\vec{x}=(\vec{q},\vec{p})$ is a much more general conception than
that of statistical physics. It indicates the information inside
the black box of any complex system.

The response of the complex system are induced by external input
vector $\vec{\gamma}$. For physicist, $\gamma_{i}$ represents
those familiar external applied magnetic field, electric field,
pressure, neutrino current, electric current, detecting laser
beams, heaters, and so forth. For chemist or biologist, the input
vector represents something like enzymes, chemical accelerator.

Not all the components of the state vector are observable under
perturbation. The output vector $\vec{y}$ represents those
observables that people can definitely detected in laboratory. The
output is strongly depend on what people want to study. For
example, there are circulatory systems, nervous systems and
digestive systems within human body. If we are studying the human
population, there is no need to take into account of these
subsystems; one only counts the people, the output vector covers
the number of people, distribution of people, etc. If the subject
is about flu's spread, it may be best to discuss the immune
subsystem.

For a general dynamic system
\begin{equation}
\frac{d\vec{x}_{in}}{dt}=F_{in}(\vec{x},\vec{\gamma}),\;\;\;\;\;\;\;\vec{O}_{ut}=F_{out}(\vec{x},\vec{\gamma}),
\end{equation}
the output vector is not always differentiable for all degrees of
differentiation on the whole range of the parameter space
$M(\vec{x},\vec{\gamma})$. There exist critical points
$(\vec{x}^{\ast},\vec{\gamma}^{\ast})$ at which the $C^{k}$ output
functions blow up. The whole input space is divided into separated
blocks by these critical points,
\begin{equation}
M^{0}(\vec{r})=\{({r}_{1}^{\ast},{r}_{2}^{\ast})\cup({r}_{2}^{\ast},{r}_{3}^{\ast})
\cdots\cup({r}_{n}^{\ast},{r}_{n}^{\ast})\},
\end{equation}
here we denote $({r}:=(\vec{x},\vec{\gamma}))$.  The stable phase
are defined in these discrete blocks. The output functions present
very good behavior within the blocks but diverge at the very
boundary. These boundaries are where the phase transition occurs.

This general mathematical definition of phases for complex system
applies for many different fields. The most familiar Ehrenfest's
definition\cite{ehrenfest} of phase transition in thermodynamics
is a good exemplar. The output vector is only a function---free
energy. The input vectors are thermodynamic quantities,
temperature $T$ and pressures $P$. For the zeroth order stable
phase the free energy of the two phases $F(T, P)$ is $C^{\infty}$
in the whole input space $(T, P)\in[-\infty,+\infty]$. For the
first order phase transition, the free energy $F(T, P)$ is
continuous in the region $T\in[T_{i},T_{f}],P\in[P_{i},P_{f}]$,
but the first order derivative is not continuous,
\begin{eqnarray}\label{00-1st-order}
&&\frac{{\partial}F_{A}}{{\partial}T}\neq\frac{{\partial}F_{B}}{{\partial}T},\;\;
\frac{{\partial}F_{A}}{{\partial}P}\neq\frac{{\partial}F_{B}}{{\partial}P}, \nonumber\\
&& T_{A}\in(T_{i},T_{c}),T_{B}\in(T_{c},T_{f}), \nonumber\\
&& P_{A}\in(P_{i},P_{c}),P_{B}\in(P_{c},P_{f}).
\end{eqnarray}
The first order stable phase blocks are divided into smaller
blocks by the second order phase transition.

The phase transition of complex system may be defined from the
divergence of the $C^{k}$ output functions. They could be any
observable functions. For mathematician, a stable phase is marked
by a $C^{k}$ function in the discrete blocks on the input vector
space. The critical point is the phase boundary between the
separated blocks.

\subsection{Phase transition}\label{phasetransition}
\begin{figure}
\begin{center}\label{saddle3}
\includegraphics[width=0.20\textwidth]{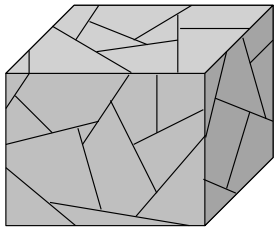}
\caption{ The whole phase diagram is split into discrete domains.
The stable phase exist in the inner region. The domains wall is
the phase coexistence region. In each domain, there is a dominant
interaction which is the king ruling over the other weaker
interaction. There is a war going on at the phase coexistence
boundary. When we tune the physical parameters, we are helping a
certain interaction to fight against the others, this certain
phase would grow stronger and stronger, it will finally unified
the whole phase diagram.}
\end{center}
\end{figure}

The stable output states of a dynamic complex system are confined
in different domains in the whole input state vector space. As the
input vector changes within a domain, the system is in a stable
state, it represents a kind of physical order. When the input
vector jumps from one domain to another, the system jumps from one
stable phase to another. The domain wall is the exact phase
boundary at which the old phase becomes unstable and decays, but
the new phase comes into being and finally leads to new stable
order. Phase transition occurs everywhere in nature. Every phase
transition indicates a revolution induced by the interaction
between the systems and their environment. the fittest states of
the old domain is replaced by the fittest states of the new
domain.

The lifeless nature world is much more intellectual than most
physicists thought. What physicist measures in experiment is
always the observable in equilibrium. Quantum theory tell us the
energy levels of molecules and atoms are discrete and quantized,
we can calculate the transitions probability between those levels
involving the absorption or emission of photons. But we don't know
how and why that occurs. In principal, the non-equilibrium process
of many particle system follows the same rules that govern the
living systems, such as ants society, honeybees group, traffic
system, etc.

If we focus on the behavior of particles before they reach
equilibrium, one would doubt that electrons are possibly
intelligent particles. One simple example is two resistors
connected in parallel in electric circuit. As all knew, the
current is proportional to the inverse of resistivity in subways.
In the beginning, the electron moves together in the main path.
When they reach the bifurcation point, they split into two
subways. Less current in the strong resistive way, and larger
current in the less resistive one. Like the cars in traffic, if
all the electrons choose the less resistive way, they block each
other until it becomes more resistive than the other subway. Then
some electrons withdraw and transferred to the other way. There is
no traffic jam in electric current network, because the cars are
everywhere under the control of a global electric field.

The non-equilibrium dynamic process shows up in the vicinity of
critical point, at which a small quantitative change of a
parameter would results in a qualitative change of the global
behavior of a complex system. Biological systems that adapt to
their environment thrive, those that fail to evolve fade away. The
environment are external input of the biological system. A stable
phase of a biological system only exist in a finite region of
environment parameters, so does a lifeless physical system. The
stable phase of a collection of $H_{2}O$ molecules is water
between 0$^{\circ}$C$\sim$100$^{\circ}$C, below 0$^{\circ}$C(273
K) is ice crystal, and it becomes gas above 100$^{\circ}$C.

Like any living creatures in nature, a lifeless physical system of
interacting $H_{2}O$ molecules evolutes following the rule---
"survival of the fittest". They took different collective
structures according to different pressure, temperature, impurity,
radiation, gravitation field, electric field, magnetic field,
density, etc. The stablest phase at certain range of the parameter
space survives as the fittest structure, the other phases in this
domain fade away. Out of this special domain, the molecules have
to reorganize their collective motion pattern in order to fit the
new environment. When the new phase is born, the old phase dies.

The phase transition most scientists observed in physics,
chemistry, biology, complex network, economics, $\cdots$ is a
revolution. Phase transition is a war between the force of old
phase and the force of new phase. The fundamental particles are
millions of armed soldiers. They are divided into different
large-scale armed groups who are fighting with one another. There
is always a critical region in which the fighters bet their bottom
dollar on one war, the winners take everything, the losers get
nothing. This critical point is the phase boundary. Sometimes, the
turning point is not so definite that it broadens into a finite
region, this indicates a crossover transition.

\section{Renormalization group theory and game theory}

\subsection{Game theory and Renormalization group}\label{renormal-game}

Wilson's renormalization group theory provides a fundamental
non-perturbative approach to quantum critical
phenomena\cite{wilson}. One basic character of quantum many body
system is that the microscopic particles are not distinguishable.
Army is the best social system for simulating collective behaviors
of quantum many particles in physics. Usually the soldiers are
identical particles in the eyes of a general, but are
distinguishable particles for a sergeant. The hierarchy structure
of army plays the same role as quantum numbers in physics. The
statistics of particles is scale dependent. One example is two
column of dipolar Bosonic atoms obeying anyonic statistics due to
the long range interaction. There are other biology system which
similar to human society, such as ant group, honeybee group, we
mainly take the army as a basic example to demonstrate
renormalizattion group theory.

In Kadanoff construction\cite{kadanoff}, a certain number of
neighboring particles are grouped into one cell which act as new
elementary particles of the renormalized Hamiltonian. At critical
point, the Hamiltonian is identical to the original Hamiltonian.
For an army, this coarse-graining procedure naturally take place.
An army has a hierarchical structure, the units of different size
include a collection of lower rank of subordination particles. 100
men are group into a company, each company acts like one particle
at higher rank marked by a captain. Every 10 companies form a
regiment, the regime particle may be named after by colonel. This
coarse-graining procedure take finite steps from brigade to
division, to corps, and finally to an army.

In fact, the coarse-graining procedure is a simplification of army
at war. In the microscopic level, it is the soldiers who are
fighting with each other. Since they do not hate each other
personally, they behave as indistinguishable identical particles,
and fight as a whole. Renormalization theory simplifies the war
between millions of soldiers to a war between thousands of
companies, to a war between hundreds of regiments, finally to the
war between two army. It is the war between two generals, it is
also the war between hundred of colonels as well as captains.

The mean field theory view the war as a fight between two full
generals dressed up by millions of soldiers. This is correct in
most cases, but it is not always accurate, for the general's
strategy is carried out by hundreds of colonels and captains
instead of the elementary soldiers. At the critical point, the
correlation between the members of an army extends to its maximal
value. As the butterfly effect says, the battle may be lost due to
a nail which fail to fix the shoes of the horse. For want of the
horse, a rider is lost. The lost of one rider may directly leads
to the loss of a battle.

The renormalization group starts from the most fundamental
particles of the army: soldiers. The first order renormalization
is to reduce the interaction between millions of soldiers to
hundred of captains which are dressed up by soldiers. The second
order renormalization procedure is to identify effectively the
captains belonging to one regiment with one colonel. This
particle-blocking process may continue, and finally end up with
full generals.

This explanation of renormalization group theory from war between
armies is not merely a parable. There is a rigorous mathematical
correspondence between the game theory of war and renormalization
group theory. Let's take the two-dimensional Ising model on
triangle lattice as one example. The soldiers are spins
$\sigma_{i}$, the battle field is the triangular lattice, the
Hamiltonian of this game reads
\begin{equation}\label{IsingHami}
H=\gamma_{1}\sum^{N}_{<ij>}\sigma_{i}\sigma_{j}+\gamma_{2}\sum^{N}_{i}\sigma_{i},
\;\;\;(\sigma_{i}=\pm1).
\end{equation}
$\gamma_{1}=J/k_{B}T$ denotes the interaction between the nearest
neighbor spins. $\gamma_{2}=\mu{B}/k_{B}T$ is the effective
external applied magnetic field. This model may be treated as
$N$-players game, the $N$ spins are $N$ players, each of them has
two strategy $\pm1$. The Hamiltonian is the payoff function. The
players take different strategy to minimize the energy function.
Another different modelling of this Ising model by game theory is
to take it as a two-player game. We christen the two players
${\gamma}_{1}$ and ${\gamma}_{2}$. ${\gamma}_{1}$'s task is to
choose a strategy $|{\gamma}_{1}\rangle$ in its strategy space
$\{{\gamma}_{1}\}$ to control the interaction between neighboring
soldiers, so that they act following his orders. ${\gamma}_{1}$
governs the soldiers by either ferromagnetic interaction or
anti-ferromagnetic interaction. ${\gamma}_{2}$ commands the spin
soldiers to keep strictly in the direction of external magnetic
field. ${\gamma}_{1}$ and ${\gamma}_{2}$ choose different
strategies to win this game.

A decision rule for ${\gamma}_{1}$ is a operator
$\hat{f}_{{\gamma}_{1}}:\{{\gamma}_{2}\}\Rightarrow\{{\gamma}_{1}\}$,
it associates each strategy
$|{\gamma}_{2}\rangle{\in}\{{\gamma}_{2}\}$ of ${\gamma}_{2}$ with
the strategies
$|{\gamma}_{1}\rangle{\in}\hat{f}_{{\gamma}_{1}}|{\gamma}_{2}\rangle$,
which may be played by ${\gamma}_{1}$ when he knows that
${\gamma}_{2}$ is playing $|{\gamma}_{2}\rangle$. Similarly, the
decision rule for $|{\gamma}_{2}\rangle$ is a map
$\hat{f}_{|{\gamma}_{2}\rangle}$ from $\{{\gamma}_{1}\}$ to
$\{{\gamma}_{2}\}$. When a pair of strategies
$(|\bar{\gamma}_{1}\rangle,|\bar{\gamma}_{2}\rangle)$ satisfies
\begin{equation}\label{a=b=a}
|\bar{\gamma}_{1}\rangle{\in}\hat{f}_{{\gamma}_{1}}|\bar{\gamma}_{2}\rangle,\;\;\;|\bar{\gamma}_{2}\rangle{\in}\hat{f}_{{\gamma}_{2}}|\bar{\gamma}_{1}\rangle,
\end{equation}
they form a consistent pair of strategies. The set of consistent
pair may be empty or very large or it may reduce to a small number
of bi-strategies. The problem of finding consistent strategy pairs
is so-called fixed-point problem. We may construct a consistent
map of the strategy pair,
\begin{equation}\label{ab=ba}
\forall(|{\gamma}_{1}\rangle,|{\gamma}_{2}\rangle)\in{\{{\gamma}_{1}\}}\times{\{{{\gamma}_{2}}\}},\;\;\;\hat{f}(|{\gamma}_{1}\rangle,|{\gamma}_{2}\rangle):=\hat{f}_{{\gamma}_{1}}\times{\hat{f}_{{\gamma}_{2}}},
\end{equation}
such that
\begin{eqnarray}\label{fab=fba}
&&\hat{f}_{AB}|{\gamma}_{1}\rangle=\hat{f}_{{\gamma}_{1}}{\hat{f}_{{\gamma}_{2}}}|{\gamma}_{1}\rangle=|{\gamma}'_{1}\rangle\in{\{{\gamma}_{1}\}},\;\;\;\\
&&\hat{f}_{BA}|{\gamma}_{2}\rangle=\hat{f}_{{\gamma}_{2}}{\hat{f}_{{\gamma}_{1}}}|{\gamma}_{2}\rangle=|{\gamma}'_{2}\rangle\in{\{{\gamma}_{2}\}}.
\end{eqnarray}
Then a consistent pair is explicitly written in the form,
\begin{equation}\label{matrix-ab}
\left(%
\begin{array}{c}
  |\bar{\gamma}_{1}\rangle\\
  |\bar{\gamma}_{2}\rangle\\
\end{array}%
\right)=\left(%
\begin{array}{cc}
  0 & \hat{f}_{{\gamma}_{1}} \\
  \hat{f}_{{\gamma}_{2}} & 0 \\
\end{array}%
\right)\left(%
\begin{array}{c}
  |\bar{\gamma}_{1}\rangle\\
  |\bar{\gamma}_{2}\rangle\\
\end{array}%
\right).
\end{equation}
This decision rule matrix of the two players is just the
renormalization group transformation matrix
\begin{equation}\label{game-renormal-map}
\hat{U}=\left(%
\begin{array}{cc}
  0 & \hat{f}_{{\gamma}_{1}} \\
  \hat{f}_{{\gamma}_{2}} & 0 \\
\end{array}%
\right)
\end{equation}
which can be deduced from the decimation procedure of
renormalization group transformation. The players may play many
steps to reach an agrement. Each time we group the spins in a sum
by Kadanoff blocks, the original degree of freedom is decimated
into the fewer degree of freedom. Every block is now a new giant
elementary particle whose spin is determined by the majority rule.
The Kadanoff-blocking process is actually a game process. After
the first step of Kadanoff-blocking, the two player accomplished
the first round of game, and their possible strategy space is
reduced to a smaller one due to the information they get from the
first round game.

The rescaled Hamiltonian is of the same form as the original one,
\begin{equation}\label{IsingHami'}
H'=\gamma'_{1}\sum^{N'/2}_{<ij>}\sigma'_{i}\sigma'_{j}+\gamma'_{2}\sum^{N'}_{i}\sigma'_{i}.
\end{equation}
The new parameters represent the new strategy of the two players
for the second round of the game, they follow the renormalization
group transformation,
\begin{equation}\label{k'=k}
\gamma'_{1}=\gamma'_{1}(\gamma_{1},\gamma_{2}),\;\;\;\gamma'_{2}=\gamma'_{2}(\gamma_{1},\gamma_{2}).
\end{equation}
We denote the strategy space of the two player as
$K=\{|\gamma_{1}\rangle,|\gamma_{2}\rangle\}$, the strategy vector
of the two players is
\begin{equation}\label{psi2player}
|\gamma\rangle= \left(
\begin{array}{c}
  |\gamma_{1}\rangle\\
 |\gamma_{2}\rangle\\
\end{array}
\right).
\end{equation}
The game operator is a map from the strategy space to itself,
\begin{equation}\label{psi-u-psi}
\hat{U}:\;|\gamma\rangle\in{K}\rightarrow|\gamma'\rangle\in{K}.
\end{equation}
This game operator is equivalent to the renormalization group
transformation. Applying Brouwer's fixed point theorem, we know if
the the strategy space is convex compact subsets of finite
dimensional vector space, there is at least one pair of consistent
strategy which satisfies
\begin{equation}\label{barpsi=psi}
|\bar{\gamma}\rangle=\hat{U}|\bar{\gamma}\rangle.
\end{equation}
$|\bar{\gamma}\rangle$ is the brouwer fixed point, or Nash
equilibrium point. This fixed is the saddle point of physical
observables on the manifold expanded by $(\gamma_{1},\gamma_{2})$.

This game operator for the two dimensional Ising model may be
derived from the recursion relation the semigroup transformation,
\begin{eqnarray}\label{k'=rk}
&&\vec{\gamma}'=\hat{{U}}_{L}\vec{\gamma},\\
&&\vec{\gamma}'=(\gamma'_{1},\gamma'_{2}),\;\vec{\gamma}=(\gamma_{1},\gamma_{2}).
\end{eqnarray}
here $L$ is rescaling factor of Kadanoff-blocking. We calculate a
statistical physical observable, such as free energy
\begin{equation}
F=-\frac{1}{\beta}ln\;Z,\;\;\;Z=Tr\;e^{-\beta{H(\gamma_i(t))}},
\end{equation}
using the rescaled Hamiltonian
$H'(\{\sigma'_{i}\},\gamma'_{1},\gamma'_{2},N')$ and the
$H'(\{\sigma_{i}\},\gamma_{1},\gamma_{2},N)$. Since the
Hamiltonian is of the same form as that before the scale
transformation, so does the free energy. Comparing the coefficient
function of the spin-spin coupling term of the effective
Hamiltonian, we get the transformation function $U_{L}$. In the
vicinity of the non-trivial fixed point $\vec{\gamma}^{\ast}$,
\begin{equation}\label{k=Uk}
\vec{\gamma}^{\ast}=\hat{{U}}_{L}\vec{\gamma}^{\ast},
\end{equation}
we perform Taylor expansion around
$\vec{\gamma}^{\ast}=(K^{\ast}_{1},\gamma^{\ast}_{2})$, and make a
truncation to the first order(the simplest case). The
renormalization group transformation is identical with
coordination transformation,
\begin{equation}\label{jacobi-renormal}
\left(%
\begin{array}{c}
 \gamma'_{1}-\gamma^{\ast}_{1} \\
  \gamma'_{2}-\gamma^{\ast}_{2} \\
\end{array}%
\right)=
\left(%
\begin{array}{cc}
  (\frac{\partial{\gamma'_{1}}}{\partial{\gamma_{1}}}) & (\frac{\partial{\gamma'_{1}}}{\partial{\gamma_{2}}}) \\
  (\frac{\partial{\gamma'_{2}}}{\partial{\gamma_{1}}}) & (\frac{\partial{\gamma'_{2}}}{\partial{\gamma_{2}}}) \\
\end{array}%
\right)_{\ast}
\left(%
\begin{array}{c}
  \gamma_{1}-\gamma^{\ast}_{1} \\
  \gamma_{2}-\gamma^{\ast}_{2} \\
\end{array}%
\right).
\end{equation}
We denote ${\delta{{\gamma}'}}=(\gamma'_{1}-\gamma^{\ast}_{1},
\gamma'_{2}-\gamma^{\ast}_{2})^{T}$ and
${\delta{{\gamma}}}=(\gamma_{1}-\gamma^{\ast}_{1},
\gamma_{2}-\gamma^{\ast}_{2})^{T}$, the element of the group is
\begin{equation}\label{re-jacobi-renormal}
{U}_{L}=
\left(%
\begin{array}{cc}
  (\frac{\partial{\gamma'_{1}}}{\partial{\gamma_{1}}}) & (\frac{\partial{\gamma'_{1}}}{\partial{\gamma_{2}}}) \\
  (\frac{\partial{\gamma'_{2}}}{\partial{\gamma_{1}}}) & (\frac{\partial{\gamma'_{2}}}{\partial{\gamma_{2}}}) \\
\end{array}%
\right)_{\ast}.
\end{equation}
This is the first order approximation of the game operator, the
exact game operator
${\delta{{\gamma}'}}=\hat{{U}}{\delta{{\gamma}}}$ may be obtained
by including the higher order approximations.

If we take this Ising model as a war game, the full generals of
the two army are $\gamma_{1}$ and $\gamma_{2}$. The spins confined
in the lattice sites are soldiers. The two generals take better
and better strategies to play through scale transformations. The
player delivered his message to his opponent through scale
transformation matrix. Then they adjust their physical parameters
$\gamma_{1}$ and $\gamma_{2}$ in the next round of game. This game
process is represented by a series of game operator, $I,
\hat{{U}}_{L}, (\hat{{U}}_{L})^{2},(\hat{{U}}_{L})^{3},\cdots.$
The game operator actually defines a flow from high energy to low
energy by the change of scale.

From physicist's point of view, high energy means hight momentum.
While the momentum is characterized by the fourier transformation
of lattice spacing on triangular lattice. If we divide the lattice
space smaller and smaller, its dominant momentum representation
grows higher and higher, and finally leads to divergency in the
continuum limit. In the war game, the full general may roughly
divide his army into tens of corps, and he is only in charge of
the tens of major generals, this is the low energy case for a
physicist. The full general may continue to dived his army into
hundred of regiments, and further into thousands of companies. If
he is powerful enough like god, he can directly take charge of the
millions of soldiers, this is the high energy part of a war game.

Effective low-energy theories can always be reached by integrating
the high-energy degrees of freedom. This is not merely a
conception in physics, it occurs every day everywhere in economics
. If we went to free market, the bargaining procedure between
sellers and buyers is actually a dynamical demonstration of
renormalization group transformation. In the beginning, the seller
would present a very high price to maximize his profit, the buyer
would try to lower it down in order to decrease his damage. This
is what happens: Buyer:"How much?",
Seller:"1000 dollars.",  \\
Buyer:"Too expensive, if you sell it at 500 dollars, I would buy
it.", \\
Seller:"No way, how about 500+$\frac{1}{2}500$
dollars",\\
Buyer:"500+$\frac{1}{2}500$-$\frac{1}{4}500$!",\\
Seller:"500+$\frac{1}{2}500$-$\frac{1}{4}500$+$\frac{1}{8}500$!",\\
Buyer:"500+$\frac{1}{2}500$-$\frac{1}{4}500$+$\frac{1}{8}500$-$\frac{1}{16}500$!",\\
$\cdots$ $\;\;\;$ $\cdots$ \\
Buyer:"500 (1+$\lim_{N\rightarrow\infty}\sum_{n=1}^{N}{(-1)^{n+1}}{2^{-n}}$)!",\\
Seller:"Done!".\\
The fixed point of this bargain is done when $N\rightarrow\infty$.
Usually this bargain does not go so far, people always cut it off
after two or three round of bargains. Here we made the assumption
that both the buyer and seller are rational. In reality, there are
various cases. If both the buyer and seller are adamant, they may
trapped in a $p$-period circular bargain, i.e., B: "1000 dollars",
S:"500 dollars", B: "1000 dollars", S: "500 dollars" ,$\cdots$, B:
"1000 dollars", S:"500 dollars", $\cdots$. In a more complex case,
the buyer bargains following a complex mapping rule, his $p$th
offer is
${\gamma}^{a}_{p}=({\gamma}^{b}_{p-1}+{{\gamma}^{b}}^{3}_{p-1}+{\gamma}^{b}_{p-2}+\cdots)$,
where ${\gamma}^{b}_{p-1}$ is the seller's $(p-1)$th offer, the
game may end up with chaos, bifurcation, all kinds of nonlinear
phenomena come in. In fact, these nonlinear interacting phenomena
indeed occur during a phase transition.

The renormalization group transformation is just the bargain game
rule. The bargainers in physical system are different
interactions. Practical physical system usually has good
negotiators. The bargain series converged into a fixed point. The
unit money on which the buyer and seller are bargaining becomes
smaller and smaller during the dynamic process of renormalization
group transformation. In physics, this means more and more high
energy effects are integrate into the effective low energy theory.

Attractive fixed points describe stable phases within the
renormalization group formalism. The critical point of phase
transition corresponds to saddle point at which the physical
parameters in the game reach equilibrium, namely the Nash
equilibrium. At the Nash equilibrium point, if any one of the
player take a wrong strategy, he would fail and flows to the
attractive fixed point, then he is confined in a stable phase. The
Nash equilibrium point is where all different phases coexist.

There is a theorem in game theory concerning on the existence of
the saddle fixed point. Applying the Brouwer's fixed-point
theorem, Aubin obtains a corollary\cite{aubin}: Suppose that the
behaviors of  ${\gamma}_{1}$ and ${\gamma}_{2}$ are described by
one-to-one continuous decision rules and that the strategy set
 $\{{\gamma}_{1}\}$ and $\{{\gamma}_{2}\}$ are convex compact subsets of
finite-dimensional vector space, then there is at least one
consistent pair, i.e., there exist at least one Nash equilibrium.
This theorem may help us to see whether phase transition exist or
not for a given game.
\\
\\
\noindent{\emph{{The application of Renormalization group to game
theory------an example: Cournot duopoly}}}
\\
\\
Usually it is very hard to find the Nash equilibrium of
multi-player games. Renormalization group theory provide us new
tools to find the Nash equilibrium solution for a multi-player
game. We first transfer the multi-player to an effective quantum
or classical many body system, and find its Hamiltonian, we can
apply various well-developed numerical renormalization group
calculation method to find the critical point of phase transition,
then we find the Nash equilibrium solution.

We consider Cournot duopoly game(see Appendix H). The simplest
case is there are only two players Alice and Bob, they are
manufacturers of the same kind of product. In the beginning, Alice
produces $\gamma^{a}_{0}$, and Bob produces $\gamma^{b}_{0}$. The
two payers competes with each other, each of them changes their
productions according to their opponent's production. One chooses
proper strategy to minimize his own cost function. Alice's
canonical decision rule is $\gamma^{a}_{1}=f_{A}(\gamma^{b}_{0}),
\;\;\gamma^{b}_{1}=f_{B}(\gamma^{a}_{0})$, we express this game
into matrix form
\begin{equation}\label{CAB1-matrix}
\left(%
\begin{array}{c}
  \bar{\gamma}^{a}_{1}\\
  \bar{\gamma}^{b}_{1}\\
\end{array}%
\right)=\left(%
\begin{array}{cc}
 0 & \hat{f}_{A} \\
 \hat{f}_{B} & 0 \\
\end{array}%
\right)\left(%
\begin{array}{c}
  \bar{\gamma}^{a}_{0}\\
  \bar{\gamma}^{b}_{0}\\
\end{array}%
\right).
\end{equation}
The second round game follows
$\gamma^{a}_{2}=f_{A}(\gamma^{b}_{1})=f_{A}(f_{B}(\gamma^{a}_{0})),\;\;
\gamma^{b}_{2}=f_{B}(\gamma^{a}_{1})=f_{B}(f_{A}(\gamma^{b}_{0})),$
i.e.,
\begin{equation}\label{CAB-n-u}
\left(%
\begin{array}{c}
  \bar{\gamma}^{a}_{2}\\
  \bar{\gamma}^{b}_{2}\\
\end{array}%
\right)=\left(%
\begin{array}{cc}
  \hat{f}_{A}{\hat{f}_{B}} & 0 \\
  0 & \hat{f}_{B}{\hat{f}_{A}} \\
\end{array}%
\right)\left(%
\begin{array}{c}
  \bar{\gamma}^{a}_{0}\\
  \bar{\gamma}^{b}_{0}\\
\end{array}%
\right).
\end{equation}
This equation has a more clear expression
$\vec{\gamma}_{2}=\hat{U}^{2}\vec{\gamma}_{0}$ after we introduced
the game operator $\hat{U}$ and strategy vector $\vec{\gamma}$,
\begin{equation}\label{gameU}
\hat{U}=\left(%
\begin{array}{cc}
  0 & \hat{f}_{A}  \\
 \hat{f}_{B} & 0 \\
\end{array}%
\right),\;\;\;
\vec{\gamma}_{n}=\left(%
\begin{array}{c}
  \bar{\gamma}^{a}_{n}\\
  \bar{\gamma}^{b}_{n}\\
\end{array}%
\right).
\end{equation}
Suppose the players paly alternatively, Alice is in the even
periods and Bob in the odd periods, when Bob produces
${\gamma}^{b}_{2n-1}$ in the period $2n-1$, Alice produces
${\gamma}^{a}_{2n}=f_{A}({\gamma}^{b}_{2n-1})$ in the period $2n$,
Alice then changes her production rate and produces
${\gamma}^{b}_{2n+1}=f_{B}({\gamma}^{a}_{2n})$, and so on. The
sequences of ${\gamma}^{a}_{2n}$ and ${\gamma}^{b}_{2n-1}$
satisfies the recursion relation,
\begin{equation}\label{gamerecursion}
2z_{n+1}+z_{n}=u,
\end{equation}
This sequences converge to $\frac{u}{3}$. $\frac{u}{3}$ is the
fixed point of $\hat{U}^{n+1},({n\rightarrow\infty})$,
\begin{equation}\label{Cfixed}
\lim_{n\rightarrow\infty}\vec{\gamma}_{n}=\lim_{n\rightarrow\infty}\hat{U}^{n+1}\vec{\gamma}_{0}={\gamma}^{\ast}.
\end{equation}
Now we see this game operator is the same as bargain game in the
sense of the renormalization group transformation.

\subsection{Solutions of Renormalization group transformation equation in game
theory}\label{exp(s)}

The game in reality always experience a dynamic process. For
example, the bargain on price between the seller and buyer still
exist in primary market. The advent of supermarket drive this kind
of observable negotiation process to the backstage market
investigation. The operator of the supermarket adjust its next
day's merchandize distribution according to last day's sell. A
good customer also compares today's price with previous price to
buy what he needs at a lower price. This is a mutual interacting
process which may be expressed by a nonlinear game operator,
\begin{equation}\label{CAB1-matrix}
\left(%
\begin{array}{c}
  \bar{\gamma}^{a}_{n}\\
  \bar{\gamma}^{b}_{n}\\
\end{array}%
\right)=\left(%
\begin{array}{cc}
 0 & \hat{f}_{A} \\
 \hat{f}_{B} & 0 \\
\end{array}%
\right)\left(%
\begin{array}{c}
  \bar{\gamma}^{a}_{n-1}\\
  \bar{\gamma}^{b}_{n-1}\\
\end{array}%
\right),
\end{equation}
where ${\gamma}^{a}$ is the seller, ${\gamma}^{b}$ is the buyer.
This game has an equivalent representation by a pair of difference
equation,
\begin{equation}
{\gamma}^{a}_{n}=f_{A}({\gamma}^{b}_{n-1}),\;\;\;\;\;\;
{\gamma}^{b}_{n}=f_{B}({\gamma}^{a}_{n-1}).
\end{equation}
In fact, the buyer not only consider the seller's price, but also
consider money amount he could pay. On the other hand, the seller
must reflet the amount and quality of the product or service he
could offer. So a complete difference equations of this game is
\begin{equation}\label{difference2}
{\gamma}^{a}_{n}=f_{A}({\gamma}^{a}_{n-1},{\gamma}^{b}_{n-1}),\;\;\;\;\;\;
{\gamma}^{b}_{n}=f_{B}({\gamma}^{a}_{n-1},{\gamma}^{b}_{n-1}).
\end{equation}
In fact, the players of a game mainly concerns about profit
difference between last round and next round, the most efficient
way is to take the profit of last round as a datum mark. So we can
always find a term like ${\gamma}^{a}_{n-1}$ on the right hand of
 Eq. (\ref{difference2}). Thus the game difference
equation is transformed into differential equations,
\begin{eqnarray}\label{differential2}
&&{\gamma'}^{a}_{n}=\lim_{h\rightarrow0}\frac{{\gamma}^{a}_{n}-{\gamma}^{a}_{n-1}}{h}=F_{A}({\gamma}^{a}_{n-1},{\gamma}^{b}_{n-1}),\nonumber\\
&&{\gamma'}^{b}_{n}=\lim_{h\rightarrow0}\frac{{\gamma}^{b}_{n}-{\gamma}^{b}_{n-1}}{h}=F_{B}({\gamma}^{a}_{n-1},{\gamma}^{b}_{n-1}),
\end{eqnarray}
$h$ is the step size of the tuning parameter. This equation may be
viewed as renormalization group transformation equation. The two
functions $F_{A/B}$ form the game operator which map one state in
the strategy space into another. Its trajectory depicts the
renormalization flow. The strategy space of two player game is two
dimensional parameter space. The corresponding game operator is a
general $2\times2$ matrix. This two dimensional matrix is an
element of renormalization group. For some special case, if the
game operator is an element of $spin(3)$, we can expand it in
terms of traceless Pauli matrix.

The game operator of a $N$-player game is a traceless $N\times{N}$
matrix,
\begin{equation}
\hat{U}=\left(%
\begin{array}{cccc}
  \hat{f}_{11} &  \hat{f}_{12} & \cdots &  \hat{f}_{1N} \\
   \hat{f}_{21} & \hat{f}_{22} & \cdots &  \hat{f}_{2N} \\
  \vdots & \vdots & \ddots & \vdots \\
   \hat{f}_{N1} &  \hat{f}_{N2} & \cdots & \hat{f}_{NN} \\
\end{array}%
\right).
\end{equation}
Following the same procedure as two player game, we can get an
equivalent $N$ dimensional difference equations. We summarize the
output field and input field into one state vector
$\gamma=(\vec{O}_{ut},\vec{r}_{in})$, a general system is governed
by the differential equations,
\begin{equation}
\partial_{s}{\gamma_{i}}=\hat{L}(\gamma,\partial_{\gamma})\gamma_{i}, \;\;\;\;\;i=1,2,...,n,
\end{equation}
where $s$ is the tuning parameter, it could be any input field. In
fact, the output and input are not absolutely distinguished, that
depends on what we have known, what we still do not know. We
usually take those we can manipulate as input, and those we will
detect as output. When physicist study phase transition of a
system, they usually tune some external parameter, such as
temperature, magnetic field, et al. They investigate how the
system response according to different physical parameters so that
we may control it for practical purpose. For example,
superconductor physicist study how the conductivity behaves when
people raise the temperature as high as room temperature, they do
not care much about the time. If the parameter $s$ is taken as
time, it is just the conventional dynamic system, a physicist's
approach to its solution is the algebra dynamic
arithmetic\cite{sjwang}.

The game operator $\hat{L}(\gamma,\partial_{\gamma})$ is the
infinitesimal generator of the translational transformation with
respect to parameter $s$. For a given initial strategy
$\gamma^{0}$, the evolution of the system along $s$ follows
\begin{equation}
{\gamma}(s)=e^{\hat{L}(\gamma,\partial_{\gamma})s}\gamma^{0}=\sum_{n=0}^{\infty}\frac{s^{n}}{n!}\hat{L}^{n}(\gamma,\partial_{\gamma})\gamma^{0}.
\end{equation}
This equation enveloped the whole process of renormalization group
transformations. Take the the bargain game as an example, $n=0$ is
the first round of bargain, the buyer and seller put their initial
cards on the table. $n=1$ is the second round of bargain, after
they have settled the main business down, they began the
bargaining on minor business now, the amplitude of this round is
$1$. The amplitude of the third round of bargain is $1/2!$. As the
renormalization group transformation goes on, the amplitude of the
$n$th round of bargain decays following $1/n!$. When they made an
agrement on all the problems from the dominant ones to the
ignorable ones, peace arrived.

The ending point of this game is fixed point which is given by the
exact evolution state vector $\gamma(s)$. The unstable fixed point
corresponds to the Nash equilibrium point, any one of the players
takes one more aggressive step to increase his own profit would
result in the breakdown of the deal. The Nash equilibrium point is
where the phase transition occurs. The stable fixed point
indicates stable phase.

\subsection{Symmetry losing as a classification of phase transition}

We gave a very general definition of phase transition in section
(\ref{phasetransition}), phase transition is a game between old
phase and new phase. Whenever the winner becomes loser or vices
versa, phase transition occurs. Transition is always accompanied
by the transfer of award from loser to winner. The award is
quantized, so we observe sudden change of output across the phase
boundary.

Each time we make a renormalization group transformation, the
players accomplished one round of game. If someone lose, someone
must win. The temporary phase boundary invades from the winner
into the loser. The amplitude of boundary's change in the $n$th
round of game is proportional to $1/n!$. If $n=1$, that is the
$1$th order phase transition, the amplitude of sudden change is
most significant, the phase boundary is roughly fixed except some
unsettled regions. Then we need the second round of combat to
negotiate the main part of the unsettled region. As this kind of
negotiation goes on to higher order, the unsettled boundary
becomes smaller and smaller, its amplitude decays following
$1/n!$. When all the unsettled boundary are fixed, we reached a
fixed point. If this fixed point is stable, we are in a stable
phase. If this is an unstable fixed point, we are at a critical
state, to be or not to be, it lies in the hand of this point. Any
minor deviation from this point would decide who is winner, who is
loser.

In the region of stable phase, the state vector $\gamma$ of the
game is $C^{p}$ continuous function, i.e., the derivative of the
state vector $\frac{d^{p}\gamma}{ds^{p}}$ is continuous up to the
$p$th order. This differentiability of the $p$th order derivative
beaks at some singular points, at which we would observe a $p$th
order phase transition.

We usually take the $p$th order non-differentiability of output
vector to measure phase transition. These output vectors
$\vec{O}_{ut}(\vec{u})$ are the subset of the state vector,
usually they are the cost function of the game. The control
vectors $|\vec{u}\rangle$ are game players. For a quantum many
body system, the output vectors $\vec{O}_{ut}(\vec{u})$ could be
any statistical observables or any external response, such as free
energy ${O}_{1}=F$, ground state energy ${O}_{2}=E_{g}$, thermal
potential ${O}_{3}=\Omega$, susceptibility ${O}_{4}=\chi$,
specific heat ${O}_{5}=C_{H}$, correlation length ${O}_{6}=\xi$,
compressibility ${O}_{7}=\kappa_{T}$, $\cdots$. The players of the
game is control vector $|\vec{u}\rangle$, its component includes
all input variables, for example, temperature ${u}_{1}=T$,
pressure ${u}_{2}=P$, effective external magnetic
${u}_{3}=\gamma_{2}=\mu{B}/k_{B}T$, spin-spin interaction
${u}_{4}=\gamma_{1}=J/k_{B}T$, electric field ${u}_{5}=\emph{E}$,
chemical potential ${u}_{5}=\mu$,$\cdots$.

The Nash equilibrium of this $N$-player game is to find the
eigenvector of the operator $\hat{U}^{n+1}$ in strategy space.
This is tantamount to solve the equation
\begin{equation}
(\hat{U}^{n+1}-I)|\psi\rangle=0.
\end{equation}
This equation indicates an infinitesimal transformation around the
identity. A finite transformation is constructed by the repeated
application of this infinitesimal transformation. For a realistic
game, it is always impossible to get an exact matrix of
$\lim_{n\rightarrow\infty}\hat{U}^{n+1}$ up to the infinite order.
The output function encounter divergence on the input base
manifold. These singular points are where the transition from
loser to winner occurs. Out of these singular regions, the output
function has very good behavior.

Renormalization group is a semigroup, because its elements have no
inverse. The ordinary group is a subset of the Renormalization
group. So they only provide us some basic understanding to phase
transition in certain special cases.

In critical phenomena, the correlation length between particles
goes to infinity at the phase transition point\cite{stanley},
there is an obvious conformal transformation symmetry. This is a
kind of symmetry when we take the physical particles as players of
a game. But a practical physical system always contain billions of
particles, this is not a good choice.

A more practical way to study quantum many body system as a game
is taking the different interactions as players. The interaction
parameters form the strategy vector of the game. As we know, the
strategy vector at fixed point is invariant under the operation of
game operator. The holonomy group at the Nash equilibrium point is
a special subset of the game operator space. We may check the
discontinuity of the output field under holonomy group to check
the phase transition point. The strategy space is expanded by the
strategies of all players, the holonomy group transformation is
actually the transformation on the strategy manifold.

Lie group is more familiar to most physicists, it is also a subset
of renormalization group. When the game has a continuum of
players, the strategy space construct a manifold. The payoff
functions is a cross section of the fibre bundle established on
this base manifold. We can define the order of phase transition
from the loss of Lie group symmetry.

We take the two dimensional Ising model as an example to
demonstrate the explicit procedure. The output vector could be
chosen to have only one component---the free energy, and the input
vectors are $\gamma_{1}$ and $\gamma_{2}$, they are
correspondingly the coupling interaction and magnetic field.
$\gamma_{1}$ and $\gamma_{2}$ choose the best strategy at each
step to increase his own welfare and decrease his loss. They first
choose the initial strategy pair of arbitrary value. Then they
behaves following the decision matrix in the next round of game.
Repeating $n$ rounds of this game, they reach the fixed point. At
the Nash equilibrium point, both the two parameters has no further
steps to increase his welfare any more.

There are only two independent parameters under the constrain of
the equation of state, any finite transformation around the fixed
point could be reached by repeating infinitesimal of $SO(2)$ whose
generator is
\begin{eqnarray}\label{L}
\hat{L}={\gamma_{2}}\frac{\partial}{\partial{\gamma_{1}}}-{\gamma_{1}}\frac{\partial}{\partial{\gamma_{2}}}.
\end{eqnarray}
The group element of $SO(2)$ is expressed by the exponential map,
$Lie: \hat{I}{\Rightarrow}Exp(\hat{I}),$
\begin{eqnarray}\label{R-SO(2)}
&&U(\theta)=e^{\theta{\hat{L}}}
=\sum_{0}^{n}\frac{1}{n!}(\hat{L}\theta)^{n}\nonumber\\
&&=I+\hat{L}\theta+\frac{1}{2}\hat{L}^{2}\theta^{2}+\cdots.
\end{eqnarray}
We expand the group element up to the $p$th order, i.e, $U(\theta)
=\sum_{0}^{p}\frac{1}{p!}(\hat{L}\theta)^{p}$, the definition for
the $p$th order of phase transition reads,
\begin{eqnarray}\label{RFRA=RFRB}
&&\sum_{0}^{p-1}\frac{1}{(p-1)!}(\hat{L}\theta)^{p-1}\vec{O}_{ut}^{A}=\sum_{0}^{p-1}\frac{1}{(p-1)!}(\hat{L}\theta)^{p-1}\vec{O}_{ut}^{B},\nonumber\\
&&\sum_{0}^{p}\frac{1}{p!}(\hat{L}\theta)^{p}\vec{O}_{ut}^{A}{\neq}\sum_{0}^{p}\frac{1}{p!}(\hat{L}\theta)^{p}\vec{O}_{ut}^{B}.
\end{eqnarray}
As $p\rightarrow\infty$, it reaches the exact $SO(2)$ group
element $U(\theta)$. When we choose the output vector as free
energy $F$, and the input vector $\gamma_{1}$ as temperature $T$
and $\gamma_{2}$ as pressure $P$, this definition of phase
transition has unified all orders of Ehrenfest's definition into
one equation. For example, for $n=0$, $U^{(0)}=I$, it yields
$F^{A}=F^{B}$. For $n=1$, $U=I+\hat{L}\theta$, then
$(I+\hat{L}\theta)F^{A}=(I+\hat{L}\theta)F^{B},$ we derived
\begin{eqnarray}
&&{\Rightarrow}\;\;\;\;F^{A}=F^{B},\;\;\;\nonumber\\
&&\frac{\partial{F}^{B}}{\partial{T}}-\frac{\partial{F}^{A}}{\partial{T}}=0,\;\;
\frac{\partial{F}^{B}}{\partial{P}}-\frac{\partial{F}^{A}}{\partial{P}}=0.
\end{eqnarray}
Therefore the essence of Ehrenfest's definition for different
order of phase transition actually depend on how many order of the
symmetry of free energy is preserved during the phase transition.
We denote $U^{(p)}=\sum_{0}^{p}\frac{1}{p!}(\hat{L}\theta)^{p}, $
Eq, (\ref{RFRA=RFRB}) may be reduced to
\begin{eqnarray}\label{Rp=/Rp}
U^{(<p)}\vec{O}_{ut}^{A}=U^{(<p)}\vec{O}_{ut}^{B},\;\;\;U^{(p)}\vec{O}_{ut}^{A}{\neq}U^{(p)}\vec{O}_{ut}^{B},
\end{eqnarray}
So when $p\rightarrow\infty$, it reaches the exact $SO(2)$
transformation. That means the output vector is differentialable
to infinite order. Since there is no discontinuity, we are not
able to observe it from external responses.

Eq. (\ref{Rp=/Rp}) has provided us qualitative understanding to
how the symmetry loss induced a phase transition. It hold for the
more general case that the output vector is a cross section of
fibre bundle on a manifold expanded by many parameters. For
example, if $\vec{O}_{ut}$ is a vector field of ($\gamma_{1}$,
$\gamma_{2}$,$\gamma_{3}$), the simplest choice is to introduce
the generators of $SO(3)$ which are the three angular momentum
operator,
\begin{eqnarray}\label{GG-L123}
&&L_{1}=-\gamma_{3}\frac{\partial}{\partial{\gamma_{2}}}+\gamma_{2}\frac{\partial}{\partial{\gamma_{3}}},\nonumber\\
&&L_{2}=-\gamma_{1}\frac{\partial}{\partial{\gamma_{3}}}+\gamma_{3}\frac{\partial}{\partial{\gamma_{1}}},\nonumber\\
&&L_{3}=-\gamma_{2}\frac{\partial}{\partial{\gamma_{1}}}+\gamma_{1}\frac{\partial}{\partial{\gamma_{2}}},
\end{eqnarray}
The order of phase transition is characterized by the group
element of $SO(3)$, $U=e^{i\theta\vec{n}\cdot{\vec{L}}}$. We
expand $U$ to the $p$th order, and investigate the equation
\begin{eqnarray}\label{333-Coexis-RFRA=RFRB}
&&\phi=\frac{\delta^{p}[U(\theta)(\vec{O}_{ut}^{A}-\vec{O}_{ut}^{B})]}{\delta\theta^{p}}|_{\theta=0}=\frac{\delta^{p}[U(\theta){\delta}\vec{O}_{ut}]}{\delta\theta^{p}}{|}_{\theta=0}.\nonumber\\
\end{eqnarray}
If the field $\phi\neq0$ for the $p$th order transformation, but
vanishes for all the transformation under the $p$th order, we
define the phase transition as the $p$th order.

The game operator $\hat{L}$ in renormalization group elements
$U=e^{i\theta\vec{n}\cdot{\vec{L}}}$ could take arbitrary
sophisticate formulas. $\hat{L}$ can be expressed as polynomial
operator expanded by $\gamma$ and $\partial_{\gamma}$,
\begin{equation}
\hat{L}\propto\sum\;\gamma^{p}_{i}\frac{\partial^{m}}{\partial\gamma^{m}_{i}}
...\gamma^{l}_{j}\frac{\partial^{n}}{\partial\gamma^{n}_{j}}.
\end{equation}
The specific form of $\hat{L}$ relies on specific systems. No
matter how complex it is, we can always check the expansion of the
exact solution
${\gamma}(s)=e^{\hat{L}(\gamma,\partial_{\gamma})s}\gamma^{0}$ to
fond out the singular points which separate the whole space into
discrete regions.

It must be pointed out here, the symmetry loss here is different
from the conventional spontaneous symmetry breaking in physics.
Take the familiar $\psi^{4}$ model for example, there is a
spontaneous symmetry breaking from $SU(2)$ to $U(1)$ for vacuum
state, it is the lagrangian of $\psi^{4}$ model
$\mathcal{L}=\frac{1}{2}(\partial_{\mu}\psi)^{2}-\frac{1}{2}m^{2}\psi^{2}-\frac{\lambda}{4}\psi^{4}$
that has $SU(2)$ symmetry. But when the $\psi^{4}$ model is
studied using the game theory of renormalization group
transformation developed in this paper, we do no care about the
$\psi$-field at all. We just take the mass $m$ and coupling
constants $\lambda$ as two players, and take the physical
observables calculated from the partition function as output
vector. It is on the manifold expanded by the mass and coupling
constant we introduce the SO(2) transformation around the critical
point.

\section{The topological theory of universal phase transition}

In physics, the renormalization group transformation is going on
under the constrain that the Hamiltonian of the system must has
the same form after transformation as it is before transformation
at the critical point,
$H(\gamma'_{1},\gamma'_{2})=H(\gamma_{1},\gamma_{2})$. So the
partition function and free energy function also maintain their
form during the transformation. Following the spirit of special
relativity, Einstein's principle of general covariance states that
all coordinate systems are equivalent for the formulation of the
general laws of nature. Mathematically, this suggests that the
laws of physics should be tensor equations. In this sense, the
laws that governs the motion of everything in universe should not
depend on coordination, no matter it is in physical world or
social world.

Phase transition is perhaps the most common phenomena in nature as
well as in human society. The basic law of phase transition does
not depend on its base manifold, on which we established the
equation of the states for a certain system, i.e., physical
system, chemistry system, biology system, or social system. What
we face is a black box, the only source of information about the
inside of the black box is the output vector which responses when
we alternate the input.

When we confine our issue in physical system, the output vectors
$\vec{O}_{ut}(\vec{u})$ are macroscopic observables which may be
decided in the frame work of physical science or directly measured
by conducting experiment, such as free energy ${O}_{1}=F$, ground
state energy ${O}_{2}=E_{g}$, thermal potential ${O}_{3}=\Omega$,
susceptibility ${O}_{4}=\chi$, specific heat ${O}_{5}=C_{H}$,
correlation length ${O}_{6}=\xi$, compressibility
${O}_{7}=\kappa_{T}$, $\cdots$. The input vector are also
macroscopic observables, such as temperature $R_{1}=T$, pressure
$R_{2}=P$, particle density $R_{3}=N$, volume $R_{4}=V$, chemical
potential $R_{5}=\mu$,$\cdots$, conductivity $R_{j}=\sigma$,
susceptibility $R_{k}=\chi$, and so on. The input vector and
output vector are relative, they are no fixed once for ever, it
depends on our subject. For instance, if we study how the volume
of a gas changes when we change the pressure, the volume is the
output vector, and the pressure is the input. If we intend to
study the inverse relation between volume and pressure, the input
and the output exchange their roles.

In this section, we choose the base manifold of output vectors.
Our interest focus on the intrinsic geometric and topological
quantities, for they do not rely on the local coordination system.

\subsection{Topological current theory of phase transition}

The input vectors are players of a game, the output vectors are
the cost function or payoff function. The inputs take different
values to maximize their profits, the order of the strategy they
played during the game is of crucial importance. About 2300 years
ago in Chinese history, general Sun play horse racing with the
King, both of the two players have three horses, a weak one, a
regular one and a strong one. A match consists of three rounds,
each of the three horses must take part in at least one round. The
King's horse in each class is more powerful than that of Sun's.
But Sun finally won, he used his regular horse race against the
King's strong horse, he certainly lost this round. But his regular
beat the King's weak, and his strong beat the King's regular. If
Sun change the order of any two of his horses, he would lose the
game. Different order of strategy lead to totally different
results.

A physical process always depends on many external parameters
which entangled with one another. These external input are game
players, their different values represent different strategy. If
we try to liquify a realistic gas confined in a rubber container
by applying high pressure and cooling. There are two ways, one way
is first to press it, and then to cool it down, the other way is
first cooling it and then pressing it. These two ways do not
always lead to the exact same structure except some special cases.
More over, the speed of the cooling has inevitable effect to the
final result. A medium carbon steel is transformed to austenite at
about 1,550 degrees. If it is allowed to cool slowly back to room
temperature, it has the ferritic structure. If it is rapidly
cooled, the austenite quenched to another shape which has
high-strength structure.

So a general output vector field is established on curved
manifold, for physicist, the output vectors are physical
observables on curved parameter space. The strategies are not
commutable in the curved output space. The physical parameters are
commutable only if the base manifold is homeomorphic to flat
Euclidean space. Its physical representation is a description of
adiabatic physical transition process which is reversible in
classical thermodynamics.

To study the intrinsic geometric properties of phase transition,
we choose the most fundamental manifold of output field
${O}_{ut}(\vec{\gamma})$. The flow of vector field of
renormalization group transformation point out where the Nash
equilibrium point is. If the vector field is continuous everywhere
on the whole base manifold, one observe nothing. If only there
appears a discontinuity, then we make sure that there is something
happening. The fundamental vector field to detect the
discontinuity is
\begin{eqnarray}\label{fei}
\vec{\phi}={\frac{\delta^{p}[U(\theta){O}_{ut}(\vec{\gamma})]}{\delta\theta^{p}}}\bigr|_{\theta=0}.
\end{eqnarray}
We first take two component of this vector field to study the
$p$th order phase transition. They are the $p$th order tangent
vector field on ${O}_{ut}(\vec{\gamma})$,
\begin{eqnarray}\label{phi-12}
&&\phi_{1}=\partial^{p-1-a}_{\gamma_{1}}\partial^{p-1+a}_{\gamma_{2}}{O}_{ut}(\vec{\gamma}),\;\;(a\neq{b})\nonumber\\
&&\phi_{2}=\partial^{p-1-b}_{\gamma_{1}}\partial^{p-1+b}_{\gamma_{2}}{O}_{ut}(\vec{\gamma}),
\end{eqnarray}
where $p$ is arbitrary number, $\gamma_{i}$ are input vectors.
Here we first take the simplest case that has only two input field
as an example to present the basic relation between topology and
phase transition.

In the neighborhood of the critical point, the manifold is
approximately holomorphic to flat Euclidean space, the translation
operator $\partial_{\gamma}$ is good enough to describe the
transportation of output field. But topology concerns about the
global geometry of the manifold, we have to walk out of the
vicinity of the saddle point, then we need to introduce the
covariant derivative,
\begin{equation}\label{A-TP}
D_{\gamma_{1}}=\partial_{\gamma_{1}}+iA_{\gamma_{1}},\;
\;\;\;D_{\gamma_{2}}=\partial_{\gamma_{2}}+iA_{\gamma_{2}}.
\end{equation}
$A_{\gamma_{1}}$ and $A_{\gamma_{2}}$ is the gauge potential which
connects the vector field between different domains. The covariant
derivative of the tangent vector field on ${O}_{ut}(\vec{\gamma})$
is
\begin{equation}\label{partial=cova}
{D_{\gamma_{i}}}{\phi_{i}}=\partial_{\gamma_{i}}{\phi_{i}}+iA_{i}{\phi_{i}}.
\end{equation}
The commutator of the two covariant derivative produce the
Gaussian curvature on the two dimensional Riemannian manifold,
$[D_{i},D_{j}]=-i\Omega_{ij}$ with where
$\Omega_{ij}=\partial_{ij}=\partial_{i}A_{j}-\partial_{j}A_{i}$.
The Gaussian curvature may decompose into the product of two
principal curvatures, $\kappa_{1}(\gamma^{0})$ and
$\kappa_{2}(\gamma^{0})$. When $\kappa_{1}>0$ and $\kappa_{2}>0$,
it is a elliptic surface, when $\kappa_{1}>0$ and $\kappa_{2}=0$,
it is parabolic, for $\kappa_{1}>0$ and $\kappa_{2}<0$, it is
hyperbolic. If the Gaussian curvature is zero, it means the
manifold is flat.

As it is well known, Euler characteristic number on the compact
two-dimensional surface is defined by Gaussian curvature $G$,
i.e., $\chi(M)=\frac{1}{2\pi}\int_{M}G\sqrt{g}d^{2}x$. In the
differential geometry, the Euler characteristic is just a special
case of the Gauss-Bonnet-Chern theorem which defines a topological
invariant on the $2n$ dimensional surface. In terms of the
Riemannian curvature tensor, the Gaussian curvature $\Omega$ is
written as
\begin{equation}\label{gauss}
\Omega=-\frac{1}{4}\frac{\epsilon^{\mu\nu}}{\sqrt{g}}\frac{\epsilon^{\lambda\sigma}}{\sqrt{g}}R_{\mu\nu\lambda\sigma}.
\end{equation}
As all know, the Riemannian curvature tensor correspond to the
gauge field tensor in gauge field theory\cite{nash}. When we set
up the vielbein $e_{\mu}^{a}$ on the surface, the Riemannian
curvature tensor can be expressed in terms of the gauge field
tensor as
$R_{\mu\nu\lambda\sigma}=-e_{\lambda}^{a}e_{\sigma}^{b}F_{\mu\nu}^{ab}$.
We introduce the $SO(2)$ spin connection for the tangent vector
Eq. (\ref{phi-12}),
\begin{equation}\label{tangent-p}
\vec{\phi}=\phi_{1}T_{1}+\phi_{2}T_{2}.
\end{equation}
Then we can define an Gauss mapping $\vec{n}$ with
$n^{a}={\phi^{a}}/{||\phi||},\;\;n^{a}n^{a}=1,$ where
$||\phi||=\sqrt{\phi^{a}\phi^{a}}\;(a=1,2)$. Considering the
symmetry of the unit vector field $\vec{n}$, we introduce the
$SO(2)$ spin connection $\omega_{\mu}^{ab}$, the covariant
derivative of $\vec{n}$ is defined as
$D_{\mu}n^{a}=\partial_{\mu}n^{a}-\omega_{\mu}^{ab}n^{b}$.
Noticing $SO(2)$ is homeomorphic to $U(1)$. There is a one to one
correspondence between $SO(2)$ spin connection and U(1)
connection. Both of them have only one independent component. As
shown by Duan et al\cite{DuanSLAC}, one consider parallel
transportation from $D_{i}\phi=\partial_{i}\phi-iA_{i}\phi=0$ and
its conjugate equation
$D_{i}\phi^{\dag}=\partial_{i}\phi^{\dag}+iA_{i}\phi^{\dag}=0$, it
is easy to obtain the U(1) gauge potential
$A_{i}={\epsilon_{ab}n^{a}\partial_{i}n^{b}}$.

It was proved that the Gaussian curvature $G$ can be expressed as
\begin{equation}\label{kb}
\Omega={-\frac{1}{2}\frac{\epsilon^{\mu\nu}}{\sqrt{g}}F_{\mu\nu}},\quad\quad
F_{\mu\nu}={(\partial_{\mu}A_{\nu}-\partial_{\nu}A_{\mu})},
\end{equation}
here $F_{\mu\nu}$ is $U(1)$ gauge field tensor. Substitute  the
U(1) gauge potential $A_{i}={\epsilon_{ab}n^{a}\partial_{i}n^{b}}$
into the Gaussian curvature, one may express the Gaussian
curvature into a topological current
\begin{equation}\label{J}
\Omega=\sum_{i,j,a,b=1}^{2}\epsilon^{ij}\epsilon_{ab}\frac{{\partial}n^{a}}{\partial{\gamma_{i}}}\frac{{\partial}n^{b}}{\partial{\gamma_{j}}},
\end{equation}
This topological current appeared in a lot of condensed matter
systems. In differential geometry, the integral of the gauge field
2-form is the first Chern number $C_{1}=\int_{M}{\Omega}$, it is
the Euler characteristic number on a compact Riemannian manifold.

From the unit vector field $\vec{n}$, one sees that the zero
points of $\vec{\phi}$ are the singular points of the unit vector
field $\vec{n}$ which describes a 1-sphere in the $\vec{\phi}$
vector space. In light of Duan's $\phi-$mapping topological
current theory\cite{DuanSLAC}, using
$\partial_{i}\frac{\phi^{a}}{\parallel\phi\parallel}=\frac{\partial_{i}\phi^{a}}{\parallel\phi\parallel}+\phi^{a}\partial_{i}\frac{1}{\parallel\phi\parallel}$
and the Green function relation in $\phi-$space :
$\partial_{a}\partial_{a}\ln||\phi|| =2\pi\delta^{2}(\vec{\phi}),
(\partial_{a}={\frac{\partial}{\partial\phi^{a}}})$, one can prove
that
\begin{equation}
\Omega=\delta^2(\vec{\phi})D(\frac{\phi}{\gamma})=\delta^2(\vec{\phi})\{\phi^{1},\phi^{2}\},\label{jidelta}
\end{equation}
where
$D(\phi/q)=\frac{1}{2}\sum_{i,j,a,b=1}^{2}{\epsilon}^{jk}{\epsilon_{ab}\partial}_j{\phi}^a{\partial_k\phi}^b$
is the Jacobian vector. In the extra-two dimensional space, this
Jacobian vector is just the Poisson bracket of $\phi^{1}$ and
$\phi^{2}$,
$\{\phi^{1},\phi^{2}\}=\sum_{i}(\frac{\partial{\phi^{1}}}{\partial{\gamma_{i}}}\frac{\partial{\phi^{2}}}{\partial{\gamma_{j}}}-
\frac{\partial{\phi^{1}}}{\partial{\gamma_{i}}}\frac{\partial{\phi^{2}}}{\partial{\gamma_{j}}})$.
As we defined above, the vector field $\vec{\phi}$ is the tangent
vector field on the output manifold. This tangent vector field may
be viewed as a projection of a source vector field, $\vec{\Psi}$
which satisfy $\vec{\Psi}\cos\theta=\vec{\phi},$ $\theta$ is angle
between the vector $\Psi$ and the tangent plane at point $p$,
Obviously when $\phi=0$, the source vector field points vertically
up. If we draw the configuration of the vector field around point
$p$, one would see a Skyrmion configuration. There is a sharp peak
around the point at which $\phi=0$. These sharp peaks bears a
topological origin.

Eq. (\ref{jidelta}) provides us an important conclusion
immediately: $\Omega=0,\;iff\;\vec{\phi}\neq 0;\;\Omega\neq
0,\;iff\;\vec{\phi}=0.$ In other words, the solutions of the
equations
\begin{eqnarray}\label{delF12=0}
&&\phi_{1}=\partial^{p-1-a}_{\gamma_{1}}\partial^{p-1+a}_{\gamma_{2}}{O}_{ut}(\vec{\gamma})=0,\;\;(a\neq{b})\nonumber\\
&&\phi_{2}=\partial^{p-1-b}_{\gamma_{1}}\partial^{p-1+b}_{\gamma_{2}}{O}_{ut}(\vec{\gamma})=0,
\end{eqnarray}
decides whether the phase transition exist or not. If Eq.
(\ref{delF12=0}) has solutions, it means that there are some
points on the domain wall at which the tangent vector field
$\vec{\phi}$ is continuous. According to our general definition
about phase transition, a phase transition is a revolution, it
happens when the system can not survival without changing itself
to fit the new environment. If there is any strategy that the
system could take to survive, it will not take any risk to face
revolution. Therefore as long as there exist solutions for
$\vec{\phi}=0$, we will not observe any sudden change, this state
is marked by a non-zero Euler number. On the contrary, if Eq.
(\ref{delF12=0}) has no solution over all strategy space, this
means the tangent vector field across the domain wall encounter a
barrier, it has to jump over the barrier with finite hight to
access another domain. In the game theory, this indicates the
system can not find any strategy to help itself move from one
domain to another domain, a reform or revolution is required. This
indicates a phase transition, this state of system is marked by a
zero Euler number.

Now we see Eq. (\ref{jidelta}) actually describes topological
configuration of vector flow around the surviving strategy. Each
surviving strategy present a peak on the tangent vector plane.
More peaks means longer life time for the old phase, if peaks
become less and less, that means the old phase is dying. The
system becomes more and more unstable, and began to collapse. When
there is no peak left, the system has to start a revolution up,
the system is totally unstable now, chaos effect come into action.

The number of solutions of equation $\phi=0$ counts the number of
surviving strategy for the old phase. The implicit function theory
shows, under the regular condition\cite{goursat}
$D(\phi/\gamma)\neq 0$, we can solve the equations $\phi=0$ and
derive $n$ isolated solutions, which is denoted as
$\vec{z}_{k}=(\gamma^{1}_{k},\gamma^{2}_{k}),\;(k=1, 2,\ldots n)$
At the critical point $z_{k}$, the Jacobian $D(\frac{\phi}{\gamma})$ can be
expressed by Hessian matrix ${M}_{\delta{{O}_{ut}(\vec{\gamma})}}$
of $\delta{{O}_{ut}(\vec{\gamma})}$, i.e.,
$D(\frac{\phi}{\gamma})|_{z_{k}}=det{M}_{\delta{{O}_{ut}(\vec{\gamma})}}(z_{k})$. According to
the $\delta-$function theory\cite{Schouten}, one can
expand $\delta(\vec{\phi})$ at these solutions,
$\delta^{2}(\vec{\phi})=\sum_{k=1}^{l}\beta_{k}\frac{\delta^{2}(\vec{\gamma}-\vec{z}_{k})}
{\mid{D(\frac{\phi}{\gamma})}\mid_{z_{k}}}.$ Then the topological
current becomes
\begin{equation}\label{J=sum}
\Omega=\sum_{k=1}^{l}
\beta_{k}\frac{detM_{\delta{{O}_{ut}(\vec{\gamma})}}(z_{k})}{\mid{detM_{\delta{{O}_{ut}(\vec{\gamma})}}
(z_{k})}\mid}\delta(\gamma^{2}-\gamma^{2}_{k})\delta(\gamma^{1}-\gamma^{1}_{k}),
\end{equation}
here $\beta_{k}$ is the Hopf index and the Brouwer degree
$\eta_{k}=$sign$D(\phi/q)_{z_{k}}=$$\pm1$, i.e., sign$detM_{\delta{F}}(z_{k})=\pm1$.
.Applying \emph{Morse} theory, we can obtain the topological charge
of the transition points from Eq. (\ref{J}).

Following Duan's $\phi-$mapping method, it is easy to
prove\cite{DuanSLAC} that $\beta_{k}\eta_{k}=W_{k}$ is
the winding number around the $k$th critical point.
The winding number measure how many times the vector
flow surround the isolated surviving strategy. When
the output manifold extend a compact oriented
Riemannian manifold in strategy space, the total winding number
\begin{equation}\label{ch}
Ch=\int{\Omega}d^{2}q=\sum_{k=1}^{l}{W_{k}}
\end{equation}
is just the Euler number. Euler number is a topological number,
it strongly relies on the topology of the base manifold.
Especially when the output manifold is a compact orientable Riemannian
manifold, such as sphere, torus, or a disc with boundary, and so
on, the total topological charge of the transition points is the
Euler number. The Euler number of a 2-sphere is $Ch=2$.
According to Eq. (\ref{ch}), we see if there are two peaks of
the vector field distributed on output manifold, each point is
assigned with a winding number $W=1$ to make sure
the Euler characteristic number of the 2-sphere. If there is
only one peak, the winding number must be $2$.

As we know, each peak represents a surviving strategy, more
strategies provide more surviving opportunities for the old phase.
But there is a topological constrain from the base manifold which
requires that the sum of winding number around each strategy must
be equal to the Euler number. If we require the winding number
must be positive, we see if there are four strategies, each of
them must carries half winding number $W_{i}=1/2,{(i=1,2,3,4)}$.
The optimal distribution of the four strategies should make them
separated as far as possible. For if they are at a crowd, it would
be dangerous, their enemy does not need to spend much energy to
block them all, then the old phase dies. The four critical points
of the same sign repel each other to reach the minimal of the
system's total energy, so they will be separated as far as
possible. When the equilibrium is reached, the most likely
distribution is the four critical points are situated at the
vertices of a tetrahedron. As for the two strategies case, one
sits at the North Pole, the other sits at the South Pole. If there
is only one strategy with $W=2$, it is unstable and is apt to
split into two or four. If there is a strategy with $W=+3$, the
strategy with negative winding number would appear, but these
state are very unstable.

The output manifold may jump from a sphere to a torus, and to a torus
with many holes, the Euler number would jumps from $Ch=2$ to
$Ch=0$, then to $Ch=2(1-h)$ with $h$ as the number of holes of the
torus. The topological change of base manifold would either
kill the old phase or save its life, so topology plays a very important role
in phase transition.

\begin{figure}
\begin{center}\label{sphere}
\includegraphics[width=0.35\textwidth]{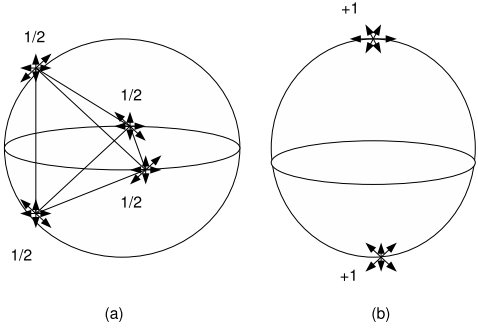}
\caption{(a) The distribution of four surviving strategy with topological charge $+1/2$ on a 2-sphere.
(b) Two surviving strategies with topological charge $W=+1$.}
\end{center}
\end{figure}

What we presented in the discussions above is the simplest case,
there is only one output field with two input vectors. For the
most general case, the output vector has $m$ components with $m$
input vectors. There exist a $m$ dimensional tangent vector field
for every component of the output vector,
\begin{equation}\label{MQdeltaF}
\phi^{i}=\partial^{p-1-a_{i1}}_{\gamma^{1}}\partial^{p-1-a_{i2}}_{\gamma^{2}}...\partial^{p-1-a_{im}}_{\gamma^{m}}{O}_{ut}(\vec{\gamma}),\nonumber\\
\end{equation}
where $\sum_{i}a_{i}=0.$ We can define a gauss map, i.e., an unit
vector field $\vec{n}$ with
$n^{a}={\phi^{a}}/{||\phi||},\;\;n^{a}n^{a}=1,$ where
$||\phi||=\sqrt{\phi^{a}\phi^{a}}\;(a=1,2,...,m)$. Following the
topological field theory\cite{DuanSLAC}, we can find a topological
current,
\begin{equation}\label{MJ}
\Omega=\sum_{i,j,...,k;
a,b,...,c=1}^{m}\epsilon^{ij...k}\epsilon_{ab...c}\frac{{\partial}n^{a}}{\partial{\gamma_{i}}}\frac{{\partial}n^{b}}{\partial{\gamma_{j}}}...\frac{{\partial}n^{c}}{\partial{\gamma_{k}}},
\end{equation}
where $(a,b,...,c=1,2,...,m)$, $(i,j,...,k=1,2,...,m)$,
$\epsilon_{ab...c}$ is antisymmetric tensor. On even dimensional
manifold, this topological current is exactly equivalent to the
Riemanian curvature tensor which directly leads to the Gaussian
curvature in two dimensions. Applying Laplacian Green function
relation, it can be proved that
$\Omega=\delta(\vec{\phi})D(\frac{\phi}{\gamma})$, where the
Jacobian $D(\frac{\phi}{\gamma})$ is defined as
\begin{equation}
D(\frac{\phi}{\gamma})=\sum_{i,j,...,k;
a,b,...,c=1}^{m}\epsilon^{ij...k}\epsilon_{ab...c}\frac{{\partial}\phi^{a}}{\partial{\gamma_{i}}}
\frac{{\partial}\phi^{b}}{\partial{\gamma_{j}}}...\frac{{\partial}\phi^{c}}{\partial{\gamma_{k}}}.\nonumber
\end{equation}
In $m=2n$ dimensional manifold, it was proved that the topological
charge of this current is the Chern number
$Ch=\int{\Omega}d^{2}q=\sum_{k=1}^{l}{W_{k}}$, which is the sum of
the winding number around the surviving strategies for
multi-player game.

\subsection{The universal equation of coexistence curve in phase
diagram}\label{coex-section}

A phase transition is a war, is a game, is a revolution. No matter
where it takes place, it becomes landmark in history of a system.
The renormalization group transformation theory told us a phase
transition point is the Nash equilibrium solution of a game. As
shown in the bargain game, the profit of seller is the loss of the
buyer and vice versa. When the seller takes proper strategies to
maximize his profit, the buyer is trying to minimize it by taking
strategies from a different space. So the seller is approaching to
the maximal point of the payoff function, in the meantime the
buyer is looking for its minimal point. A Nash equilibrium appears
at their intersection. The Nash equilibrium point is a saddle
point, the output field reaches its maximal point in $\gamma_{1}$
direction, but get a minimal value in $\gamma_{2}$ direction. The
derivative of the output field corresponds to $\gamma_{1}$ and
$\gamma_{2}$ must be of opposite sign. This leads to the
coexistence equation for different phases.

In previous sections, when solving the equation of tangent vector
field $\phi=0$ to find the surviving strategies, we applied a
regular condition $D(\phi/\gamma)\neq 0$, which comes from the
implicit function theorem\cite{goursat}. When the regular
condition is violated, i.e., $D(\phi/q)=0$, a definite solution of
equation $\phi=0$ is not available. Then the branch process of the
solutions function occurs. A mathematical demonstration of the
branch process could be found in Ref.\cite{DuanSLAC}. This
bifurcation may be understand from game theory. Under the regular
condition $D(\phi/\gamma)\neq 0$, if the solutions of $\phi=0$
exist, that means the old phase still has strategies to survive,
if there is no solutions, the old phase can not find any strategy
to make a living, it has to die. When the regular condition fails,
$D(\phi/\gamma)=0$, the old phase and new phase is at an
equilibrium war, if the old phase win, it find ways to survive, if
it loses, no surviving strategy exist, the old phase dies.
Therefore, it is at the very battlefield of $D(\phi/\gamma)=0$,
the two phases have equivalent power, nobody wins, nobody lose,
but they are fighting against each other. Several roads branched
out of this battlefield, to be or not to be, the old phase has to
make a choice when passing this critical region.

In two dimensional input space, the coexistence curve equation
$D(\phi/\gamma)=0$ is just the familiar Poisson bracket for the
tangent vector field
$\phi^{i}=\partial^{p-1-i_{1}}_{\gamma^{1}}\partial^{p-1-i_{2}}_{\gamma^{2}}
...\partial^{p-1-i_{m}}_{\gamma^{m}}{O}_{ut}(\vec{\gamma})$,
\begin{equation}\label{poisson}
\{\phi^{1},\phi^{2}\}=0.
\end{equation}
It is an unification of the special coexistence equations of
different order phase transition. As all know, in quantum
mechanics, if two operator  $\hat{A}$ and $\hat{B}$ commutate with
each other, i.e., $[\hat{A},\hat{B}]=0$, they share the same
eigenfunction. Eq. (\ref{poisson}) means that the two classical
field $\phi^{1}$ and $\phi^{2}$ are commutable at the phase
transition point.

When the game has three players, the output field is a function of
three parameters, each of them holds the life of an old phase. The
three phases intersects with one another at the coexistence points
which sit at the solutions of
\begin{equation}\label{poissongeneral3}
\{\phi^{i},\phi^{j},\phi^{k}\}=0,
\end{equation}
where $\{\phi^{i},\phi^{j},\phi^{k}\}$ is the generalized Poisson
bracket. Its quantum correspondence id the Jacobi identity
\begin{equation}
[A,[B,C]]+[C,[A,B]]+[B,[C,A]]=0.
\end{equation}
For a $n$-player game, we need to introduce a $n$-dimensional
renormalization group transformation on the output manifold. The
transformation operator expand the tangent vector space around the
identity on the manifold. We denote a vector operator as
$\vec{L}$, a group element is given by
$U=e^{i\theta\vec{n}\cdot{\vec{L}}}$. The basic tangent vector
field for phase transition is
\begin{eqnarray}\label{phi-=-O}
&&\vec{\phi}=(\phi_{\gamma_{1}},\phi_{\gamma_{2}},...,\phi_{\gamma_{n}})=\frac{\delta^{p}[U(\theta)\delta{\langle0|\hat{O}|0\rangle}]}{\delta\theta^{p}}|_{\theta=0}.\nonumber\\
\end{eqnarray}
The most general definition of generalized Poisson
bracket\cite{nambu} for $n$ component vector field is
\begin{equation}\label{poissongeneral=0}
\{\phi^{1},\phi^{2},...,\phi^{n}\}=\frac{\partial(\phi^{1},\phi^{2},...,\phi^{n})}
{\partial(\gamma_{1},\gamma_{2},...,\gamma_{n})}.
\end{equation}
The coexistence surface equation for n-player game is the Jacobian
field for n-component output field, it is equivalent to the
n-dimensional generalized Poisson bracket,
\begin{equation}\label{poissongeneral-expan}
\{\phi^{1},\phi^{2},...,\phi^{n}\}=\sum_{i,j,...,k}^{n}\epsilon^{ij...k}\frac{{\partial}\phi^{1}}{\partial{\gamma_{i}}}
\frac{{\partial}\phi^{2}}{\partial{\gamma_{j}}}...\frac{{\partial}\phi^{c}}{\partial{\gamma_{n}}}=0.
\end{equation}
This coexistence equation of $m$ vector field
$\{\phi_{i},(i=1,2,...,m)\}$ may be decomposed as a group equation
of two field equations $\{\{\phi^{i},\phi^{j}\}=0,i,j=1,2,...,m\}$
where $i$ and $j$ must runs over all component of the vector
field.

In order to verify the universal coexistence curve equation, we
take the two-phase coexistence equation $\{\phi^{1},\phi^{2}\}=0$
as an example, and apply it to thermodynamic physics. The output
field is the difference of free energy
$\hat{O}_{ut}(\gamma)=\delta{F}=F^{A}-F^{B}$, the input vector are
temperature $\gamma_{1}=T$ and pressure $\gamma_{2}=P$. It will be
shown, the universal coexist equation $\{\phi^{1},\phi^{2}\}=0$
unified all the coexistence equations in classical phase
transitions.

We first verify the second order phase transition. The order
parameter of the second order phase transition is
$\phi^{1}=\partial_{T}\delta{F}$ and
$\phi^{2}=\partial_{P}\delta{F}$, substituting them into the
Jacobian vector
\begin{eqnarray}\label{D(phi/x)}
\{\phi^{1},\phi^{2}\}=D(\phi/q)=\frac{\partial\phi^{1}}{\partial{T}}\frac{\partial\phi^{2}}{\partial{P}}
-\frac{\partial\phi^{1}}{\partial{P}}\frac{\partial\phi^{2}}{\partial{T}}=0,
\end{eqnarray}
and using the relations
\begin{eqnarray}\label{dtt=cp}
\partial_{T}\partial_{T}{\delta}F&=&\frac{{C_{p}^{A}}-{C_{p}^{B}}}{T},\;\;
\partial_{P}\partial_{P}{\delta}F=V(\kappa_{T}^{A}-\kappa_{T}^{B}),\nonumber\\
\partial_{P}\partial_{T}{\delta}F&=&V(\alpha^{B}-\alpha^{A}),
\end{eqnarray}
we arrive
\begin{equation}\label{D(phi/x)=cpcp-vv}
D(\phi/q)=\frac{V}{T}({C_{p}^{B}}-{C_{p}^{A}})(\kappa_{T}^{B}-\kappa_{T}^{A})-(V\alpha^{B}-V\alpha^{A})^{2}.
\end{equation}
Recalling the Ehrenfest equations
\begin{equation}\label{ehrenfest}
\frac{dP}{dT}=\frac{\alpha^{B}-\alpha^{A}}{\kappa^{B}-\kappa^{A}},\;\;\;\;
\frac{dP}{dT}=\frac{C_{p}^{B}-C_{p}^{A}}{TV(\alpha^{B}-\alpha^{A})},
\end{equation}
it is easy to verify that the equation above is in consistent with
the bifurcation condition Eq. (\ref{D(phi/x)=cpcp-vv}). So the
bifurcation equation $D(\phi/q)=0$ is an equivalent expression of
the coexistence curve equation. The solution of this equation is a
two dimensional coexistence surface, the Ehrenfest equations
actually indicates the normal vector of phase A and B is of
equivalent value with opposite direction, i.e.,
$|\vec{n}_{B}|=-|\vec{n}_{A}|$, that means they reach balance on
this surface.
\begin{figure}
\begin{center}\label{saddle}
\includegraphics[width=0.26\textwidth]{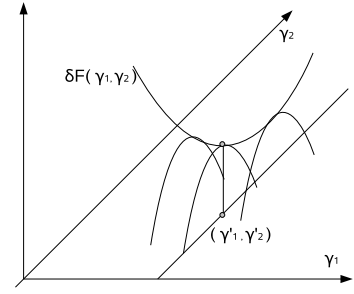}
\caption{The saddle surface of the free energy around the critical
point.}
\end{center}
\end{figure}

For the first order phase transition, we chose the vector order
parameter as $\phi=\partial^{0}\delta{F}$, here $'0'$ means no
derivative of the free energy. The generalized Jacobian vector of
the first order phase transition with $\phi=\partial^{0}\delta{F}$
is given by
\begin{equation}\label{D(phi/x)=TT-PP}
D(\frac{\phi}{q})=(\frac{\partial{F}^{B}}{\partial{T}}-\frac{\partial{F}^{A}}{\partial{T}})
+(\frac{\partial{F}^{B}}{\partial{P}}-\frac{\partial{F}^{A}}{\partial{P}})=0,
\end{equation}
in mind of the relation $\frac{\partial{F}}{\partial{T}}=-S$ and
$\frac{\partial{F}}{\partial{P}}=V$, and considering
$D(\frac{\phi}{q})=0$, we have
\begin{equation}\label{clapeyron}
\frac{dP}{dT}=\frac{(S^{B}-S^{A})}{(V^{B}-V^{A})}.
\end{equation}
This is the famous Clapeyron equation. The critical point is a
saddle point. So it is the maximal point of free energy difference
in $\gamma_{1}$ direction, and it is minimal point in $\gamma_{2}$
direction, their first order derivative must obey
\begin{equation}\label{delFsaddle11}
\frac{d\delta{F}}{d\gamma_{1}}\frac{d\delta{F}}{d\gamma_{2}}<0.
\end{equation}
The first order phase transition requires they must share the same
absolute value at the critical point, then
\begin{equation}\label{1+2=0}
\frac{d\delta{F}}{d\gamma_{1}}+\frac{d\delta{F}}{d\gamma_{2}}=0.
\end{equation}
In fact, the free energy difference between the two sides of the
coexistence acts as the phase potential, its first derivative is
force, the force of the two phases must be of the same value but
pointing in the opposite direction at the critical point.

The bifurcation equation $D(\phi/q)=0$ can be naturally
generalized to a higher-order transition, it also leads to the
coexistence curve of the higher order transition. We consider a
system whose free energy is a function of temperature $T$ and
magnetic field $B$, then the Clausius-Clapeyron equation becomes
$dB/dT=-\Delta{S}/\Delta{M}$. If the entropy and the magnetization
are continuous across the phase boundary, the transition is of
higher order. For the $p$th order phase transition, the vector
field is chosen as the $(p-1)$th derivative of $\delta{F}$,
$\phi^{1}=\partial^{p-1}_{T}\delta{F},\;\phi^{2}=\partial^{p-1}_{B}\delta{F}$.
Substituting $(\phi^{1}, \phi^{2})$ into Eq. (\ref{D(phi/x)}), we
arrive
\begin{eqnarray}\label{D(phi/B)}
D(\phi/q)=\frac{\partial^{p}\delta{F}}{\partial{T}^{p}}
\frac{\partial^{p}\delta{F}}{\partial{B}^{p}}-
\frac{\partial\partial^{p-1}\delta{F}}{\partial{B}\partial{T}^{p-1}}
\frac{\partial\partial^{p-1}\delta{F}}{\partial{T}\partial{B}^{p-1}}=0.
\end{eqnarray}
Considering the heat capacity
$\frac{{\partial}^{2}F}{{\partial}T^{2}}=-\frac{C_{B}}{T}$ and the
susceptibility $\frac{{\partial}^{2}F}{{\partial}B^{2}}=\chi$, the
bifurcation condition $D(\phi/q)=0$ is rewritten as
\begin{equation}\label{dB/dTp}
\left[\frac{dB}{dT}\right]^{p}=(-1)^{p}
\frac{\Delta\partial^{p-2}C/\partial{T}^{p-2}}{T_{c}\Delta\partial^{p-2}\chi/\partial{B}^{p-2}}.
\end{equation}
This equation is in perfect agreement with the equations in Ref.
\cite{kunmar}.

In mind of our holographic definition of phase transition Eq.
(\ref{Rp=/Rp}), we may also derive the holographic coexistence
equation using the fundamental order parameter field,
\begin{eqnarray}\label{Coexis-RFRA=RFRB}
&&\vec{\phi}=\frac{\delta^{p}[R(\theta)(F^{A}-F^{B})]}{\delta\theta^{p}}=\frac{\delta^{p}[R(\theta){\delta}F]}{\delta\theta^{p}},\nonumber\\
&&R(\theta){\delta}F=\sum_{0}^{n}\frac{1}{n!}(\hat{L}\theta)^{n}{\delta}F,\nonumber\\
&&\hat{L}={\gamma_{2}}\frac{\partial}{\partial{\gamma_{1}}}-{\gamma_{1}}\frac{\partial}{\partial{\gamma_{2}}}.
\end{eqnarray}
To study the $p$th order phase transition, one need to expand the
group element $R(\theta)$ to the $p$ the order, and split it into
the real part and imaginary part, i.e.,
\begin{eqnarray}\label{coexis-phi12}
&&\vec{\phi}={\frac{\delta^{p}[R(\theta){\delta}F]}{\delta\theta^{p}}}\bigr|_{\theta=0}=\phi^{1}+i\phi^{2}.\nonumber\\
\end{eqnarray}
Then the coexistence curve equation is
\begin{equation}\label{coexist-Qpoisson}
\{\phi^{1},\phi^{2}\}=\frac{\partial\phi^{1}}{\partial{\gamma_{1}}}\frac{\partial\phi^{2}}{\partial{\gamma_{2}}}
-\frac{\partial\phi^{1}}{\partial{\gamma_{2}}}\frac{\partial\phi^{2}}{\partial{\gamma_{1}}}=0.
\end{equation}
Under this definition, it is easy to see that Kunmar's result
(\ref{dB/dTp}) is only special case of a series of coexistence
equations for the $p$th order of phase transition, the complete
coexistence curve equations are given by
\begin{eqnarray}\label{coexis-dB/dTp}
&&\frac{\partial^{i}\delta{F}}{\partial{T}^{i}}\frac{\partial^{2p-i}\delta{F}}{\partial{B}^{2p-i}}
-\frac{\partial^{j}\partial^{p-j}\delta{F}}{\partial{B^{j}}\partial{T}^{p-j}}
\frac{\partial^{k}\partial^{p-k}\delta{F}}{\partial{T^{k}}\partial{B}^{p-k}}=0,\nonumber\\
&&(i,j,k=1,2,...,p).
\end{eqnarray}
Now we see, the universal coexistence equation not only reproduced
all the coexistence equations of classical phase transition in
physics, but also gave more equations that have never been
appeared before. This indicates that the game theory of
renormalization group transformation has very broad applications.
\begin{figure}
\begin{center}\label{channel2}
\includegraphics[width=0.4\textwidth]{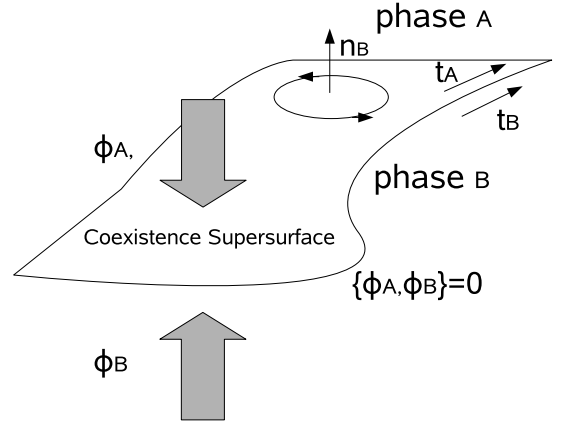}
\caption{In analogy with Newtonian mechanics, the difference of
output between two players is equivalent to gravitational
potential $V$, the first order phase transition means the two
phases have the same potential. For $p=2$, the order parameter
field,
$\phi^{A}=\partial^{p-1}_{\gamma_{A}}{O_{ut}},\;\phi^{B}=\partial^{p-1}_{\gamma_{B}}{O_{ut}}$,
represents the force of phase A and B. When they reach balance,
the two phase coexist on the hypersurface. This is the first order
phase transition. For $p=3$, $\phi^{A}$ and $\phi^{B}$ represents
the acceleration, although the force is not equal, but the
acceleration are the same, this is the second order phase
transition. On the two sides of the coexistence surface, the
tangent vector are continuous, but the normal unit vector has a
sudden jump.}
\end{center}
\end{figure}

For the magnetic field and temperature depended free energy $F(B,
T)$, the scaling laws were derived in Ref. \cite{kunmar}, the
exponents is defined as
\begin{equation}\label{scaling}
\frac{\partial^{p-2}C}{\partial{T}^{p-2}}=a^{-\mu},\;\;\;\frac{\partial^{p-2}\chi}{\partial{B}^{p-2}}=a^{-\kappa}.
\end{equation}
The equivalent expression in terms of the vector field
$\vec{\phi}$ is
\begin{equation}
\partial_{T}\phi^{1}=a^{-\mu},\;\;\;\partial_{P}\phi^{2}=a^{-\kappa}.
\end{equation}
This scaling law is only a special case, there were many other
scaling laws in various physical system, the same value of
critical exponents falls into the same universality class. In the
next section, we shall discuss the scaling laws based on the most
general definition of phase transition.

\subsection{Universal scaling laws and coexistence equation around Nash equilibrium point}

The phase transition of a physical system occurs at the critical
point where the correlation length between particles becomes
infinite\cite{stanley}. It is assumed that the free energy is a
generalized homogeneous function, i.e.,
$F(\lambda^{a}\gamma^{1},\lambda^{b}\gamma^{2})=\lambda{F}(\gamma^{1},\gamma^{2})$.
The quantity defined by the free energy obey power laws around the
critical point. From the two scaling exponent $a$ and $b$, one may
derive those critical exponents which obey some equalities. The
systems with the same scaling law fall into an universality class.

When it comes to the general phase transition defined as a war
game in this paper, all the critical phenomenons in physical
reappeared. During the war, the correlation between the all the
members of the participants becomes infinity, people may not know
each other, but every tiny work they do may cause great effect to
the final results. If we focus on the individual person in battle
field, one would see two opposite soldiers are fighting. Then we
go to a larger scale, we see two companies are fighting. We can
continue to magnify the scale, the participants who are fighting
range from hundreds of people to millions of people, range from a
small village to the whole world. No matter from which scale we
see it, it is the same war. At the critical point, the
participants of the war have equal strength, if any one of them
make a tiny mistake(the mistake may come from an unimportant
soldier), the whole army will lose the war, so the correlation
length between soldiers goes to infinity. In this sense, the
output field of the war should be a generalized homogeneous
function at the critical point, it obeys the relation
$\hat{O}_{ut}(\lambda^{i}\gamma_{i})=\lambda\hat{O}_{ut}(\gamma_{i})$.

According to the topological current of phase transition, we
established the tangent vector field of $\hat{O}_{ut}$. The
tangent vector field is the projection of physical field
configuration on the strategy manifold. The physical field
divergent at some singular points where the tangent vector field
vanished. It was proved that the nontrivial Riemannian curvature
just around these surviving strategy. The integral of the
Riemannian curvature is a topological invariant, the critical
exponent should bear a topological origin.

According to the universal definition of phase transition, the
phase transition point is a Nash equilibrium solution. A special
two dimensional output manifold is a saddle surface in the
vicinity of the critical point. The output manifold is maximum for
one parameter, but minimum for another parameter.

Suppose the topological dimension of the manifold around the
critical point is integer, i.e., $D=1,2,...,n$, we may introduce a
local coordinates to approximately express the output manifold in
the vicinity of the critical point as
\begin{equation}\label{F-cross-2}
\hat{O}_{ut}=\frac{\kappa_{1}}{2}\gamma_{1}^{2}+\frac{\kappa_{2}}{2}\gamma_{2}^{2},
\end{equation}
where $\gamma_{1}$ is the infinite small variable, based on which
the local coordination is
$r_{1}=(\delta{r}_{1}+\frac{1}{2}\Gamma_{ij}^{1}\gamma_{i}\gamma_{j})\sqrt{g_{ii}}+\cdots$.
We may abandon the quadratic term of $\gamma_{i}$. Eq.
(\ref{F-cross-2}) is the local approximation of the output
manifold. $\kappa_{1}$ and $\kappa_{2}$ are principal curvature.
When $\kappa_{1}>0$ and $\kappa_{2}>0$, it is a elliptic surface,
when $\kappa_{1}>0$ and $\kappa_{2}=0$, it is parabolic, for
$\kappa_{1}>0$ and $\kappa_{2}<0$, it is hyperbolic. Usually the
local manifold on a two dimensional output manifold is hyperbolic
at a phase transition point. In fact, if we carry out the
derivative of Eq. (\ref{F-cross-2}) to the seconde order, it
spontaneously leads to,
\begin{equation}\label{F-cross-2-2-2}
\partial^{2}_{\gamma_{1}}\hat{O}_{ut}={\kappa_{1}},\;\;\;\; \partial^{2}_{\gamma_{2}}\hat{O}_{ut}={\kappa_{2}},
\end{equation}
they are actually intrinsic geometric constant of the neighbor
manifold around the critical point.

Recall the game theory of renormalization group transformation in
the first section, one would see that the game operator is a
nonlinear operator, it defines an iterative map for the game
process. The dimension of the infinite small neighboring manifold
around the Nash equilibrium point can be exactly calculated from
the game operator. For most nonlinear game operators, the
dimension of the manifold around the Nash equilibrium is fractal
instead of integer. There are only a few very simple cases that
one can find an integer dimension. But the game operator in that
cases is too trivial to give us any interesting phenomena.
According to the experiments and numerical calculation of physics,
we can make a general hypothesis that the neighboring output
manifold around the critical point of phase transition, namely
around the Nash equilibrium point of a non-cooperative game, has
fractal dimension.

In the vicinity of the phase transition point,an approximation of
the scale invariant output manifold is a complex function in
fractal space,
\begin{equation}\label{F-cross-3}
\hat{O}_{ut}(\gamma)=\kappa_{i}\gamma_{i}^{d_{i}}+\kappa_{m}[f(\gamma_{j}^{d_{j}}+\gamma_{k}^{d_{k}}...)]^{d_{m}}
+\ldots,
\end{equation}
where $d_{1},d_{2},...,d_{n}$ are fractal dimensions wit respect
to different input parameters. Recall that most of the physical
observables in statistical mechanics are defined by the second
order derivative of free energy, we can define the observables of
a complex system by the second order derivative of the output
field. The free energy is only a special component of the output
manifold in physics. When we study the most general complex
system, as long as people can measure it, we can take any order of
derivative of the output field as observables. These observables
are just the components of the tangent vector field of the output
field. A simple example of the tangent vector field is
\begin{eqnarray}\label{F-cross-3-pp}
\phi_{ij}=\frac{\partial^{2}\hat{O}_{ut}(\gamma)}{\partial{\gamma_{i}}\partial{\gamma_{j}}}={d_{i}}\gamma_{i}^{d_{i}-1}+d_{j}\gamma_{j}^{d_{j}-1}+\cdots.
\end{eqnarray}
According to the topological phase transition theory, the tangent
vector field satisfy the phase coexistence equation
$\{\phi^{1},\phi^{2},...,\phi^{n}\}=0$ at the critical point, we
can derive a constrain on these vector field. We substitute the
explicit expansion of the observable quantities into the
coexistence equation, it would lead us to a constrain on the
fractal exponent index. These constrain relations are just the
scaling laws. Therefore scaling laws come from the coexistence
equation of the crossing defined physical quantities.

The scaling relations found in various physical system are
probably the simplest relations on the fractal space extended by
two parameters. One can reach all kinds of different scaling
relations\cite{stanley} in statistical mechanics by taking the
output field as free energy
$\hat{O}_{ut}(\gamma_{i})=F(\gamma_{i})$, and taking the input
$\gamma_{i}$ as physical parameters, such as temperature $T$,
pressure $P$, magnetic field $B$, and so on. In the vicinity of a
second order phase transition, the divergent physical quantity are
defined by the second order derivative of the free energy. Such as
the susceptibility $\chi=-\partial^{2}F/\partial{H}^{2}$, $H$ is
magnetic field. Each divergent quantity is characterized by a
critical exponent, this critical exponent comes from fractal
space. The two component coexistence $\{\phi^{1},\phi^{2}\}=0$
produced the scaling relations, such as the Fisher relation
$\nu{d}=2-\alpha$, Widom relation
$\hat{{\gamma}}=\beta(\delta-1)$, Rushbrooke relation
$\alpha+2\beta+\hat{\gamma}=2$, and so on. We take the Rushbrooke
relation as example to verify the coexistence equation. The Gibbs
free energy is $G=U-TS$, its differentiation is $dG=-SdT+VdP-MdH$.
Experimental and numerical calculation found that three
thermodynamic quantity obey the following scaling laws in the
vicinity of critical point,
\begin{eqnarray}\label{3}
M&=&-(\frac{\partial{G}}{\partial{H}})\sim{|T|}^{\beta},\nonumber\\
C_{P}&=&-T(\frac{\partial^{2}G}{\partial{T}^{2}})_{P,H}\sim{|T|}^{-\alpha},\nonumber\\
\chi&=&-(\frac{\partial{G}}{\partial{H}})\sim{|T|}^{-\gamma}.
\end{eqnarray}
The fundamental vector field can be taken as
\begin{equation}
\phi^{1}=(\frac{\partial{G}}{\partial{H}}),\;\;\;\;\;\phi^{2}=(\frac{\partial{G}}{\partial{T}}).
\end{equation}
Substituting the two vector field into the coexistence equation
$\{\phi^{1},\phi^{2}\}=0,$ one may derive
\begin{equation}
\frac{\partial^{2}{G}}{\partial{T}^{2}}\frac{\partial^{2}{G}}{\partial{H}^{2}}
-\frac{\partial^{2}{G}}{\partial{T}\partial{H}}\frac{\partial^{2}{G}}{\partial{H}\partial{T}}=0.
\end{equation}
Now we substitute the thermodynamic quantities (\ref{3}) into the
coexistence equation, it yields
\begin{equation}\label{4}
{|T|}^{-\alpha-\gamma}=\beta^{2}{|T|}^{2\beta-2}.
\end{equation}
When $T\rightarrow0$, we may ignore the coefficient $\beta^{2}$ at
the right hand side of Eq. (\ref{4}), then we obtained the
Rushbrooke relation $\alpha+2\beta+\hat{\gamma}=2$. Other scaling
relation can be verified following similar procedure. These
relations were firstly found by computational simulation and
experiments. Therefore the scaling law of universal phase
transition in a general complex system has solid numerical and
experimental foundation. Here it must be pointed out that the
commutable relation
$\partial_{T}\partial_{H}=\partial_{H}\partial_{T}$ have been used
in the calculation. This suggests that the partial differential
corresponding to different variables are commutable in the
vicinity of critical point. This is in consistent with our picture
of war game at the phase transition point. Further more, one may
choose different tangent vector field for the coexistence
equation, then one may obtain all different scaling relations in
the vicinity of critical point.

\subsection{The symmetry of Landau phase transition theory and
symmetry of game theory}

The Landau theory of continuous phase transition theory provides a
basic description to the phase transition characterized by
spontaneous symmetry breaking. Take its application to the
structure phase transitions as one example, it derives several
important features, namely, the change in the crystal's space
group, the dimension and symmetry properties of the transition's
order parameter, and the form of the free energy expansion. It has
the same range of validity as the mean-field approximation in
microscopic theories. A central assumption of the Landau theory is
that the free energy can be expanded as a Taylor series with
respect to the order parameter $\eta$:
\begin{equation}\label{freeenergy}
F(P,T,\eta)=F_{0}+A(P,T)\eta^{2}+B(P,T)\eta^{3}+C(P,T)\eta^{4}+...
\end{equation}
in which the phases are marked by the order parameter $\eta$. The
symmetry in Laudau theory talks about the invariance of this free
energy when we do some transformation on the order parameter field
$\eta'\rightarrow{U}\eta$. This symmetry is the same conception as
that in quantum field theory, such as the $\psi^{4}$ model,
$\mathcal{L}=\frac{1}{2}(\partial_{\mu}\psi)^{2}-\frac{1}{2}m^{2}\psi^{2}-\frac{\lambda}{4}\psi^{4}$.
If the equations of motion derived from this Lagrange is invariant
under some transformation $\psi\rightarrow\psi'$, we call such a
transformation a symmetry transformation.

But in this paper, the symmetry we are talking about in the game
theory of phase transition is a different conception.

The object of our research is a very general system
$\vec{O}_{ut}=F_{out}(\vec{x},\vec{\gamma})$, $\vec{\gamma}$ is
the input vector. In game theory, the input vector represent the
input strategies of different players. In physics, the input
vector are physical operational quantity. $\vec{O}_{ut}$ is output
vector, it encompasses all the information the observer can
received by sending different inputs. In physics, the outputs are
physical observables. The state $\vec{x}$ represent the inner
states of system, it plays a similar role as the order parameter
field $\eta$ in the free energy expansion equation
(\ref{freeenergy}). In our game theory of topological phase
transition theory, all the state vector have been integrated out,
the fundamental starting point is the output field
$\vec{O}_{ut}=F_{out}(\vec{\gamma})$, the symmetry we mentioned in
this theory is about the transformation invariant property of
$\vec{O}_{ut}$ under the transformation
$\vec{\gamma}^{\ast}\rightarrow{U}\vec{\gamma}$. For example, in
the free energy equation (\ref{freeenergy}), $F(P,T)$ is output,
$P$ and $T$ are the two players, we study the symmetry of $F(P,T)$
when $T\rightarrow{T}'$ and $P\rightarrow{P}'$. The order
parameter $\eta$ is not an operation quantity, it describes the
inner state of the system.

The Landau theory of phase transition can be summarized as a
differential game(see Appendix \ref{dif-game},\ref{inf-game}). The
order parameter is the state vector of the game. In the frame work
of our topological current theory of phase transition, the tangent
vector field $\phi^{i}$ consists of the complete set of phase
dynamics system. The vector field
\begin{equation}\label{deltaF}
\phi^{i}=\partial^{q-1}_{P}\partial^{p-1}_{T}(\frac{{\delta}F(P,T,\eta)}{\delta\eta})
\end{equation}
describes how the two player $T$ and $P$ behave across the state
vector space $\eta$. The free energy is the output function.
$F(P,T,\eta)$ should be written as an expansion in the even powers
in the spirit of Gintzburg-Landau formalism, for $F(P,T,\eta)$
must be gauge invariant with respect to the order parameter
$\eta$. Then the series of free energy in even powers truncated at
the fourth order is
\begin{equation}\label{F4}
F(P,T,\eta)=F_{0}+A(P,T)\eta^{2}+C(P,T)\eta^{4},
\end{equation}
A special phase vector field may be chosen as
\begin{eqnarray}\label{phi1=2=}
\phi^{1}=2\partial^{1}_{T}[A(P,T)+2C(P,T)\eta^{2}]\eta,\nonumber\\
\phi^{2}=2\partial^{1}_{P}[A(P,T)+2C(P,T)\eta^{2}]\eta.
\end{eqnarray}
If $\phi^{1}$ and $\phi^{1}$ do not commute with each other, both
the two players have surviving strategy which are the solutions of
$\phi^{1}=0,\;\phi^{2}=0$, these solution sit at some isolated
points
\begin{equation}\label{solut}
T_{k}=T_{k}(\eta),\;\;P_{k}=P_{k}(\eta),\;\; k=1,2,...,l,
\end{equation}
Each of these isolated solutions has a winding number $W_{k}$.
These surviving strategy pair ($T$,$P$) are changing according to
the state vector $\eta$. The surviving strategy of new phase and
old phase carry opposite winding number, they are annihilating and
generating at the phase coexistence region. The sum of these
winding number is a topological quantity which is determined by
the topology of the state vector manifold. .

Usually $A(T,P)$ has the form of $a(P)(T-T_{c})$ near the critical
temperature $T_{c}$, and C(P,T) is supposed to be weakly dependent
on the temperature, i.e., $\partial_{T}C\ll1$. Considering Eq.
(\ref{phi1=2=}), the surviving strategy for the old phase and new
phase can be derived from
\begin{eqnarray}\label{phi1=2=0}
\eta[a(P)+2\partial_{T}C(P,T)\eta^{2}]=0,\nonumber\\
\eta[\partial_{P}a(P)(T-T_{c})+2\partial_{P}C(P,T)\eta^{2}]=0.
\end{eqnarray}
One sees that $\eta=0$ corresponds to a stable phase solution of
the equation above. For the case $\eta\neq0$, there are two
solutions:
\begin{equation}\label{psi=aP}
\eta=\pm\left[\frac{-a(P)}{2\partial_{T}C}\right]^{1/2},\;\eta=\pm\left[\frac{\partial_{P}a(P)(T_{c}-T)}{2\partial_{P}C}\right]^{1/2},
\end{equation}
if we chose $C=exp({T+P})$ and $a=exp(P)$, it turns into a
familiar form
\begin{equation}\label{psi=aP2}
\eta_{1}=\pm\left[\frac{-a(P)}{2C}\right]^{1/2},\;\eta_{2}=\pm\left[\frac{a(P)(T_{c}-T)}{2C}\right]^{1/2}.
\end{equation}
The coexistence curve equation $\{\phi^{1},\phi^{2}\}=0$ is an
universal equation, it holds for this two player game. Consider
the two special vector field Eq. (\ref{phi1=2=}), we arrived a
sophisticate coexistence equation,
\begin{equation}\label{LD=0}
4\eta^{2}\partial^{2}_{T}C[\partial^{2}_{P}A(T,P)+2\partial^{2}_{P}C\eta^{2}]=[\partial_{P}a(P)+2\partial_{P}\partial_{T}C\eta^{2}]^{2}.\nonumber
\end{equation}
If we have obtained the explicit relation of $A(T,P)$ and
$C(T,P)$, then we can depict the phase diagram in $T-P$ plane for
different state vector $\eta$.

\subsection{Symmetry and evolution of phases in game theory}

In game theory, the output vector are payoff functions, the inputs
are strategies. A symmetry transformation in strategy space does
not always lead to different output, the Nash equilibrium solution
is invariant fixed point under continuous transformation. As shown
in Ising model, the renormalization group transformation keep
mapping a pair of strategy to another pair, no matter where it
starts, it always flows to the fixed point.

The phase of a system evolutes in the direction of renormalization
group transformation flow. The direction of renormalization group
transformation flow is determined by the Second Law of
Thermodynamics, which states all physical systems in thermal
equilibrium can be characterized by a quantity called entropy,
this entropy cannot decrease in any process in which the system
remains adiabatically isolated. We can ignore the thermodynamics,
and grab the central point that an isolated system only evolutes
in the direction of increasing entropy. This statement of the
Second Law of Thermodynamics may also hold in game theory, but
notice that the entropy here in our theory is not the conventional
conception defined in statistical physics, since we are not
manipulating the physical particles, it is the entropy of players
$\vec{\gamma}$ instead. In physics, the players are those physical
observables. An isolated system evolves spontaneously toward a
maximal symmetry.

The entropy of the multi-player game measures indistinguishability
of players. Let's look at an ancient battlefield with two armies
ready to fight, before the beginning of the combat, the two troops
are organized in ordered states separated by vacant glacis, every
part of them has special functioning. A bystander can tell which
is which. Once the combat begins, they rush at each other and
fused into one. In this case, the bystander is unable to
distinguish them, it is completely in chaos. When the war
finished, the loser surrendered, winner gathered, the two troops
translated into another ordered states. The surviving troops are
distinguishable again, but those dead can never be back to ordered
states again.

High symmetry leads to high entropy. If we do some transformation
to a system in chaos, it won't make any difference. As in the war,
the two armies become a mixture, it is an unstable equilibrium
state with the highest symmetry. The higher symmetry a system
owns, the more unstable it would be. The state before and after
the war are both of less symmetry states. Like the beginning of
universe, all different conflicting interactions confined in a
singular point, reach an unstable equilibrium state. It has the
highest space time symmetry. A minor imbalance results in the Big
Bang. The explosion breaks the singularity apart into all
different symmetry zone. Each symmetry group brings about a stable
phase.

The stable phases at different stage of history may be found
following the group chain of the original symmetry group. A high
symmetry group usually has a series of subgroup. A physical system
liken to stay in the most fundamental symmetry state. People love
peace, but hate war. They like to live at the most stable point of
the system. A war cost too much energy. The subgroup chain of a
given group is always finite,
\begin{equation}\label{Uchain}
U\supset{U_{1}}\supset{U_{2}}\supset{U_{3}}...\supset{U_{p}}.
\end{equation}
For example, $SO(4,2)$ constitutes a dynamical group for the
Hydrogen atom\cite{nambu}. It has many four subgroups\cite{brian},
\begin{equation}\label{chain}
SO(4,2)\supset{SO(4,1)}\supset{SO(4)}\supset{SO(3)}\supset{SO(2)}.
\end{equation}
This group chain may be followed by point group. For instance, the
$U(1)$ symmetry group may be decreased to $T_{n}$ which means a
physical system is invariant when it rotates $\frac{2\pi}{n}$
around one axel. So we can roughly find the stable phase according
to the group chain. Generally speaking, Lie group is highes rank
of group. The discrete group are imbedded in Lie groups. The most
stable phase focus on the lowest symmetry.

A game system evolutes following the first principal, the players
take different strategies to ground state. The $N$ players
represents $N$ interaction, or $N$ input vectors. In the
beginning, each of them occupies a small domain in phase diagram
and is the ruler in his own domain. There is temporary balance
between the different players. The war is going on all the time
between neighboring domains. Whenever a player loses, he dies out,
his enemies absorbed his domain. When a player fails, an old phase
died, and a new phase is born, this indicates a phase transition.
The balance on the boarder between different domains represent the
phase coexistence region, on which Nash equilibrium is reached.
But this state is an unstable equilibrium state, there is
constantly minor conflicting to destroy this balance. Then the war
goes on, the winners gain advantage, and becomes stronger and
stronger, the war will not stop until he unified the domains as
one.

The phase diagram of a physical system is depicted in the frame of
some static physical parameter. We can view the evolution of phase
as an continuous tuning of the physical parameter. For example, if
we continuously raise temperature, we will see single curve in the
phase diagram bifurcates at certain temperature, this is the
evolution of physical phase, although it is not the usual
evolution in the common sense of time.

This phase evolution with respect to temperature can be viewed as
a generalized dynamic process in parameter space. As shown in
section (\ref{exp(s)}), a multi-player game can be equivalently
expressed as a group of differential equations. The evolution of
output vector $\vec{O}_{ut}$ is governed by the differential
equations,
\begin{equation}\label{Out(t)}
\partial_{s}\vec{O}_{ut}=\hat{L}(\gamma,\partial_{\gamma})\vec{O}_{ut},
\;\;\;\;\;i=1,2,...,n.
\end{equation}
If we take the tuning parameter $s$ as temperature $T$, then we
understand the evolution of phase with respect to temperature. In
physical system, we can obtain the function of linear response by
perturbation theory. The response function $\vec{O}_{ut}$ may have
sophisticate relation with input parameters,
$\vec{O}_{ut}(\gamma)$. Some input parameters are function of
temperature $\gamma_{i}=\gamma_{i}(\gamma_{j})$. For example,
during the isothermal process, the product of pressure and volume
is a constant $PV=constant$, so $P$ and $V$ are related by this
equation. We take the $\gamma_{i}(\gamma_{j})$ as fundamental
variables, $\gamma_{j}$ is the tuning parameter. Then the
generalized momentum is defined as
\begin{equation}
P_{\gamma_i}=\frac{\partial{\phi(\gamma,\partial_{\gamma_{j}}\gamma,\gamma_{j})}}{\partial{(\partial_{\gamma_{j}}\gamma^{i})}},
\end{equation}
where $\phi(\gamma,\partial_{\gamma_{j}}\gamma,\gamma_{j})$ is a
reference function like Lagrangian function. We can further find a
vector field
$\phi_{H}(\gamma,\partial_{\gamma_{j}}\gamma,\gamma_{j})$ which is
similar to the Hamiltonian function in Lagrangian mechanics. For a
given field $\phi_{H}$, the game operator
$\hat{L}(\gamma,\partial_{\gamma})$ bear an expression in terms of
Poisson bracket,
\begin{equation}
\hat{L}(\gamma,\partial_{\gamma})=\frac{\partial\phi_{H}}{\partial{P_{\gamma_i}}}\frac{\partial}{\partial{\gamma}}
-\frac{\partial\phi_{H}}{\partial{\gamma}}\frac{\partial}{\partial{P_{\gamma_i}}}.
\end{equation}
This formula of game operator leads to another equivalent
representation of the differential equation \label{Out(t)},
\begin{equation}\label{Out(t)Poisson}
\partial_{\gamma_{j}}{O}_{ut}=\{\phi_{H}, {O}_{ut}\}.
\end{equation}
This equation actually describes the game process between
$\phi_{H} $ and ${O}_{ut}$, each of them navigates one stable
phase. If they are in the same complete set, $\{\phi_{H},
{O}_{ut}\}=0$, then the two phases coexist. In fact, the
Hamiltonian function does not play any special role in our
topological current theory of phase transition, it is just one of
the tangent vector field in Lie algebra space.

\subsection{Phase coexistence boundary and unstable vacuum state}

As we defined in the beginning, phase transition is a transition
from one stable state to another stable state, there is a critical
point at which the old stable state collapsed and the new state
arise from the shambles and grow to a stable state. A
oversimplified model is take two states as a vector of two
component $(x,y)$, the evolution operator is a diagonal 2 by 2
matrix $\hat{L}=\texttt{diag}(-1,1)$ at the coexistence region.
The evolution of the two phase follows the equation
$\partial_{\gamma}\vec{r}=\hat{L}\vec{r}$, which we established in
last section. Then in the vicinity of the coexistence region, we
would see the old phase decays following $x\propto{e^{-a\gamma}}$,
in the meantime the new phase blows up $y\propto{e^{+a\gamma}}$,
where $a>0$.

The mechanical simulation of the phase coexisting state is a ball
on the maximal tip of the parabola $f(z)=-z^{2}$. At Nash
equilibrium point, a player take the best strategy to minimize his
own damage, and obtain his maximal profit in the meantime. In
fact, it is the profit difference between them that they fight
for. When the two conflicting force reach a balance, the game
arrived at a Nash equilibrium.

The Nash equilibrium state is an unstable maximal point in the
mechanical potential of the output field. Let
$\hat{O}_{ut}^{i}(\vec{\gamma})$ be the output function of the
player $\gamma^{i}$, the effective potential for two players in a
game is
$\Delta{O}_{ij}=\hat{O}_{ut}^{i}(\vec{\gamma})-\hat{O}_{ut}^{j}(\vec{\gamma})$.
Nash equilibrium sits at the minimal point of $|\delta{O}_{ij}|$.
Notice that the player $\gamma^{i}$'s profit is the damage of
player $\gamma^{j}$, so $\Delta{O}_{ij}=-\Delta{O}_{ij}$. We may
establish a general equation of motion for player $\gamma^{i}$'s
profit function,
\begin{equation}\label{imotion}
\partial^{2}_{{\gamma}^{i}}\hat{O}_{ut}^{i}=\partial_{{\gamma}^{i}}V(\hat{O}_{ut}^{i}),
\end{equation}
$V(\hat{O}_{ut}^{i})$ is a general potential which is
self-consistently decided by other players, usually the player
$\gamma^{i}$ sits at some of its minimal points, while the other
players occupied its maximal point. The typical profit function
$\hat{O}_{ut}^{i}$ of player $\gamma^{i}$ is a kink solution,
\begin{equation}\label{kink}
\hat{O}_{ut}^{i}=\pm\tanh(\gamma^{i}-\gamma^{\ast})=\frac{e^{\gamma^{i}-\gamma^{\ast}}-e^{\gamma^{\ast}-\gamma^{i}}}{e^{\gamma^{i}-\gamma^{\ast}}+e^{\gamma^{\ast}-\gamma^{i}}}.
\end{equation}
where $\gamma^{\ast}$ is the optimal strategy. Eq. (\ref{kink}) is
the familiar kink solution of quantum tunnelling
problems\cite{kleinert}. Any minor deviation from the Nash
equilibrium solution would results in drastic increase or decrease
of the profit function. Therefore the Nash equilibrium solutions
of a game play the same role as the vacuum solution to quantum
tunnelling. A collection degenerate vacua means there are a series
of Nash equilibrium solutions with the same optimal value.

The game of phase transition has two different equilibrium states.
One is all different interactions reach an agreement to maintain
peace, this is the optimal strategy of cooperative game. It is the
trivial vacuum. The other is that no peace agreement is derived, a
war breaks out. At the critical point, different phases coexisted
but against each other. This is the Nash equilibrium state. It is
an unstable vacuum state.

The coexistence equation of $n$-different phases we obtained in
previous sections,
\begin{equation}
\{\phi^{1},\phi^{2},...,\phi^{n}\}=0,
\end{equation}
describes a Nash equilibrium state, or in other words, the
coexisting unstable vacuum state, here
$\vec{\phi}=(\phi^{1},\phi^{2},...,\phi^{n})=\frac{\delta^{p}[U(\theta){O}_{ut}]}{\delta\theta^{p}}{|}_{\theta=0}$.
For this equation is antisymmetric, exchanging any two of the
players would add a $(-1)$ to the output. As shown in the
topological current of phase transition, the sum of the winding
numbers around the surviving strategies is a topological
number----Chern number. The sign of each winding number is
determined by the sign of $\{\phi^{1},\phi^{2},...,\phi^{n}\}$.
Every $\phi^{i}$ field is a player, if exchange any two of them,
the winner becomes loser, the loser turns into winner. Therefore
if $\phi^{i}$ has a positive winding number, his opponent must has
a negative one. The strategy for $\phi^{i}$ to survive is the
anti-strategy of his opponents. During phase transition, the old
phase is the opponent of the new phase. The winding number of the
new phase's strategy  $\gamma^{j}$ is $W^{j}_{New}>0$, the winding
number around the surviving strategy $\gamma^{i}$ of the old phase
is $W^{i}_{Old}<0$. The topological constrain suggests
 \begin{equation}\label{chern}
N_{ChernNumber}=\sum_{i}W^{i}_{Old}+\sum_{j}W^{j}_{New}.
\end{equation}
We may view each strategy of the new phase as one particle with
topological charge $W_{New}$, and its corresponding anti-particle
is the anti-strategy of the old phase which carries a negative
winding number. The soldiers of the new phase are the strategies,
they carry positive winding number, rush at the coexist curve to
fight against the anti-strategy of the old phase. The particles
and antiparticles annihilate at the phase boundary, so the
coexisting phase boundary behaves as vacuum. This war is going on
under the constrain of Eq. (\ref{chern}).

The generator of translation group along $\gamma^{\mu}$ is
$i\partial_{\gamma^{\mu}}$. The evolution of the output on the
strategy space follows the classical Hamiltonian-Jacobi equation,
\begin{equation}\label{pt-heisenberg}
i{\partial_{\gamma^{\mu}}{\langle\hat{O}_{ut}(\theta)\rangle}}=\{\langle\hat{O}_{ut}(\theta)\rangle,\;L\},
\end{equation}
The second quantization of this equation is just the Heisenberg
equation $i\partial_{t}\hat{O}_{ut}=[\hat{O}_{ut},H]$, which has a
more general covariant form
\begin{equation}\label{pmu}
i\partial_{\gamma^{\mu}}{\hat{O}_{ut}}=[\hat{O}_{ut},\hat{P}_{\mu}],
\end{equation}
here ${\gamma}^{\mu}=(t,\vec{\gamma})$, $P_{\mu}=(H,\vec{P})$. In
four dimensional Minkoveski space time, the Hamiltonian is the
generator of translation group with respect to time $t$. While the
momentum operator $\hat{P}_{\mu}$ is the generator of space
translation group. Integrating the covariant Heisenberg equation
(\ref{pmu}), we arrive
\begin{equation}\label{O=uOu-1}
\hat{O}(\gamma)=e^{i\hat{P}\gamma_{\mu}}\hat{O}(0)e^{-i\hat{P}\gamma_{\mu}}.
\end{equation}
If the output is expressed by quantum operators, its evolution in
strategy space is governed by renormalization group transformation
\begin{widetext}
\begin{eqnarray}\label{elOe-l}
U(\theta)\langle\hat{O}\rangle{U}^{-1}(\theta)=e^{i\theta\vec{n}\cdot{\vec{L}}}\langle\hat{O}{\rangle}e^{-i\theta\vec{n}\cdot{\vec{L}}}=\langle\hat{O}\rangle+i\theta\;[\vec{n}\cdot{\vec{L}},\;\langle\hat{O}\rangle]+
\frac{(i\theta)^{2}}{2!}[\vec{n}\cdot{\vec{L}},\;[\vec{n}\cdot{\vec{L}},\;\langle\hat{O}\rangle]]+\cdots.
\end{eqnarray}
\end{widetext}
For the most simple Lie group $SO(2)$ whose Lie algebra has only
one generator $\hat{L}_{z}$, this transformation equation takes a
very simple form,
\begin{eqnarray}\label{SO(2)-elOe-l}
&&U\delta\langle\hat{O}\rangle{U}^{-1}\nonumber\\
&&=\delta\langle\hat{O}\rangle+i\theta\;[\hat{L}_{z},\;\delta\langle\hat{O}\rangle]+
\frac{(i\theta)^{2}}{2!}[\hat{L}_{z},\;[\hat{L}_{z},\;\delta\langle\hat{O}\rangle]]+\cdots.\nonumber
\end{eqnarray}
In quantum mechanics, if a group transformations $U$ commutes with
Hamiltonian $\hat{H}$, $[U,H]=\hat{H}U-U\hat{H}=0$, i.e.,
$H=H'=U\hat{H}U^{\dag}$, then $\hat{H}$ is invariant under
transformation of group $U$, they share the same eigenfunction. In
a general game theory, the Hamiltonian has nothing special. The
generators of Lie algebra plays the role of Hamiltonian operator.
The quantization of the coexistence equation reads
\begin{equation}\label{Qcoexist}
[\hat{\phi}^{1},\hat{\phi}^{2},...,\hat{\phi}^{n}]=0.
\end{equation}
Each operator $\hat{\phi}^{i}$ represents a quantum operator, if
they commute with each other, they share the same eigenspace. Eq.
(\ref{Qcoexist}) is the quantum coexistence equation.

If the $\hat{\phi}^{i}$ are vectors expanded in the tangent space
of Lie group around the identity, they are vectors of Lie algebra.
For Mathematician, this coexistence equation corresponds to an
invariant Cartan space of Lie algebra. The coexistence phase space
is the eigenspace of Cartan subalgebra. The Cartan subalgebra is
the maximal Abelian subalgebra. An arbitrary Lie algebra vector
commuting with this subalgebra is still a Lie vector in the same
space. The elements $C_{j}$ in this commutative subalgebra must
satisfies
\begin{equation}\label{cartan}
[C_{i},C_{j}]=0, \;\;\;\;(i,j=0,1,2,...,l).
\end{equation}
$l$ is the dimension of the Cartan subalgebra space. It is also
the number of coexistence phases. For the coexisting point of
three phases, we define the three operator commutator as
\begin{eqnarray}\label{3cartan}
&&\;\;\;[C_{i},C_{j},C_{k}]\nonumber\\
&&=[C_{i},C_{j}]C_{k}+[C_{j},C_{k}]C_{i}+[C_{k},C_{i}]C_{j}\nonumber\\
&&=0,\;\;\;\;(i,j,k=0,1,2,...,l).
\end{eqnarray}
It has a straight forward generalization for the points at which
$n$ phases intersects,
\begin{eqnarray}\label{ncartan}
[C_{i},C_{j},C_{k},...,C_{f}]=0,\;(i,j,k,...,f=0,1,2,...).
\end{eqnarray}
So we have a well defined quantum operator
\begin{equation}\label{DQcoexist}
\hat{D}^{n}=[\hat{\phi}^{1},\hat{\phi}^{2},...,\hat{\phi}^{n}],
\end{equation}
This operator is defined from by vectors of Lie algebra, we call
it phase coexistence operator. According to the group
representation theory\cite{chenjq}, we can always find the
representation of coexistence operator,
\begin{equation}\label{lpt=npt}
\hat{D}^{n}|\psi_{n}\rangle=D|\psi_{n}\rangle.
\end{equation}
If $n=2$, $\hat{D}^{n}$ is just the commutator of two operators
which represent two stable phases. The phase boundary between
phase $A$ and phase $B$ is given by the zero modes of the
commutator of the two phase operators. When $\hat{\phi}_{A}$ and
$\hat{\phi}_{B}$ are not commutable, the eigenfunctions of the
phase coexist operator may be divided into positive modes and
negative modes,
\begin{equation}\label{W+-}
\hat{D}^{n}|\psi_{n}\rangle=sign{W}|\psi_{n}\rangle.
\end{equation}
in which $W$ is the winding number around the surviving strategy.
$sign(W)=+1>0$ corresponds to stable phase A and
$sign(W_{pt})=-1<0$ corresponds phase B, while $(W_{pt})=0$
indicates the phase boundary. One of the most familiar example is
to take the phase operator as the three component of angular
momentum operator, $\hat{\phi}_{A}=L_{x}$ and
$\hat{\phi}_{B}=L_{y}$, then $\hat{D}^{2}=L_{z}$.
$L_{z}|\psi_{m}\rangle=m|\psi_{m}\rangle$, $m=\{0,\pm\}$. $m=0$ is
the coexistence eigenvalue, $m=\pm1$ represents the two stable
phases.

In topological quantum field theory, the topological information
of the zero mode is described by the Atiyah-Singer index theory.
The strategy space of the game of phase transition can be split
into positive eigenspace and negative eigenspace. The surviving
strategies of the new phase carry positive winding number, we call
them positive modes $\phi^{+}$. The strategies of the old phase is
the enemy of new phase, they carry negative winding number, so we
call them negative modes $\phi^{-}$. As shown by Eq.
(\ref{jidelta}), our topological current theory of phase
transition proved that the Euler characteristic number on two
dimensional strategy space is,
\begin{eqnarray}\label{jidelta+-}
Ch&=&\int\delta^2(\vec{\phi}_{+})\{\phi_{+}^{1},\phi_{+}^{2}\}-\int\delta^2(\vec{\phi}_{-})\{\phi_{-}^{1},\phi_{-}^{2}\},\nonumber\\
&=&\sum_{i}W^{i}_{+}-\sum_{j}W^{j}_{-}.
\end{eqnarray}
For the 2n-player game, the topological Chern number on this $2n$
dimensional strategy manifold is determined by Atiyah-Singer
theorem,
\begin{eqnarray}\label{B-i=+--}
Ch_{B_{i}}=Index \hat{D}^{n}=dim \hat{D}_{+}^{n}-dim
\hat{D}^{n}_{-},
\end{eqnarray}
where $\hat{D}^{n}$ is th quantized phase coexistence operator.
The positive modes behaves as particles, and the negative modes
are antiparticle, they coexist in vacuum. They are born by pairs
from vacuum, and annihilate by pairs to vacuum.

In the language of  Atiyah-Singer index theorem(see appendix
section \ref{Atiyah-Singer}), the topological index is the
difference between the number of positive eigen-modes and negative
eigen-modes. The positive modes is the surviving strategy for the
new phase, the negative modes is the surviving strategy for the
old phase. Thus the topological index counts how many extra
strategies the new phase have after his soldiers rushed at front
and annihilated at the coexistence phase boundary. We first assume
every surviving strategy has a topological $|W|=1$ for
convenience, the Euler number $Ch=+2$ on a sphere, this
topological constrain says that when all the soldier of the old
phase died, there are at least two positive soldiers of the new
phase left. The victory of the new phase has been determined by
topology of the strategy manifold. If it is on a torus, the Euler
number is zero, $Ch=0$. The new phase is not that lucky now, his
total number of soldiers is identical with that of the old phase.
When the war break out, the strategy-anti-strategy pair annihilate
on the phase coexistence boundary, the final result is a draw.
When the base manifold of strategy space is a torus with $n$-holes
, the Euler number is $Ch=2(1-n),n>1$, the final victory goes to
the old phase, he has at least $2(1-n)$ surviving soldiers when
the new phase is extinguished.

\section{Phase transition and entanglement in game theory}

\subsection{Phase transition and entanglement}

The participants of a war entangled with each other before the war
breaks out, otherwise there will not be a war at all. In fact, a
game is played by many interacting players. During the game, a
player choose strategy according to other players' strategy, so he
could never be a free particle unless he is not in the game.

The entanglement we are talking about here is a much more general
conception, it includes the relation, interaction or connection
between the elements of a system. The quantum entanglement in
physics is one special case.

When the players are at peace, the entanglement between them is
kept at a stable level. As the imbalance between different players
increase, the war is coming, their entanglement grows stronger and
stronger. The whole system becomes more and more unstable. When
the critical point arrived, a negligible event trigged the war,
which spread the whole system through the strong entanglement.
During the war, all the players summon up their internal strength
to collect information from all the other players, and make the
best strategy to win the war, the entanglement reach a climax. The
external response during the war also reached the strongest level.
When the war is over, everything is in order, their entanglement
gradually decay to another stable level.

Therefore, phase transition occurs when the entanglement between
different phases reaches a maximal point. Every local maximal
point of the entanglement indicates a transition, or a war. In
order to give an exact prediction on where or when the war arise,
we need to find some detectable quantity to measure the
entanglement. So that we can quantitatively determine the position
of the phase transition point.

The most familiar quantity for physicist is the von Neumann
entropy. Quantum statistics suggests that the von Neumann entropy
of a pure ensemble is zero. Entropy is a quantity to measure how
disorder a system is. More disorder means higher entropy. The von
Neumann entropy measure the disorder of the mixed states. The von
Neumann entropy is a relatively small quantity if the system is in
a stable phase. When the phase transition occurs, it would reach a
climax point.

In fact, any sensitive output function corresponding to a group of
inputs can be used as measure of entanglement. If the inputs have
strong relation between them, one minor change would definitely
change the others, and this would leads to the response of many
outputs. One can see this from a war, it is at the war that the
ignorable individuals began unite to work as a team, they are
strongly correlated and entangled.

In our topological current of phase transition, a quantity to
measure entanglement may be defined as
\begin{equation}  \label{Eent}
E_{ent}=\frac{1}{2\sqrt{\pi{h}}}\exp{[-{D(\phi/\gamma)}/(4h)]}
\end{equation}
where $D(\phi/\gamma)=\{\phi^{1},\phi^{2},...,\phi^{n}\}$ is the
coexistence equation, and $h$ is step size of renormalization
group transformation. As shown in the previous section, the phase
evolution equation in parameter space is
$\partial_{\gamma_{j}}{O}_{ut}=\{\phi_{H},
{O}_{ut}\}=\hat{L}{O}_{ut}.$ So Eq. (\ref{Eent}) is some kind of
propagator of phases similar to the propagator in physics
$U(t)=e^{-iHt}$. We rewrite the coexistence equation
$D(\phi/\gamma)$ as Det$\;\Phi$, where $\Phi$ is the matrix
extended by $\partial_{\gamma_{i}}\phi^{j}$. Recall the definition
of Pfaffian for matrix $M$, $[$Pf$\;{M}]^{2}=$Det$\;{M}$, we can
decompose the coexistence equation as
$D(\phi/\gamma)$=$[$Pf$\;{\Phi}]^{2}$. The entanglement equation
(\ref{Eent}) reads,
\begin{equation}  \label{Eent2}
E_{ent}=\frac{1}{2\sqrt{\pi{h}}}\exp{[-\frac{({\textsf{Pf}\;\Phi})^{2}}{4h}]}.
\end{equation}
A typical physical example of Pfaffian is the fermionic parity. In
the game theory here, a Pfaffian is a sum over all partition of
the players into pairs, exchanging any two of them contributes a
minus sign. For a bargain game between buyer and seller, the step
size $h$ is amplitude of the unfixed money they are fighting for.
The smaller $h$ they are negotiating on, the more information they
are exchanging so that they could persuade each other to accept
his offer. This means the entanglement between them increases when
$h$ approaches to zero. So the entanglement increases when the
renormalization group transformation goes on. The entropy
increases in the mean time. The maximal entanglement appears at
the phase coexistence state,
\begin{equation}  \label{Eent3}
E_{ent}=\lim_{h\rightarrow0}\frac{1}{2\sqrt{\pi{h}}}\exp{[-\frac{({\textsf{Pf}\;\Phi})^{2}}{4h}]}=\delta(\textsf{Pf}\;\Phi).
\end{equation}
Here ${\textsf{Pf}\;\Phi}=0$ is an equivalent expression of the
coexistence equation. The matrix $\Phi$ in $E_{ent}$ is
constructed by output vector field. The output vector $\vec{\phi}$
could be any physical parameters. For the most simple case of two
phases, $\vec{\phi}$ has two component $\phi^{A}$ and $\phi^{B}$,
the Pfaffian is just the Poisson bracket.

Besides the statistical observable or external response, the
output vector can also be chosen as the conventional order
parameter which is a quantity used to characterize the structural
and inner order changes of physical system at the phase transition
point. In the superconductor-insulator transition, upon tuning
some parameter in the Hamiltonian, a dramatic change in the
behavior of the electrons occurs, the order parameter of this
quantum phase transition $\phi=\Delta{e^{i\theta}}$ is the energy
gap function of cooper pair theory. For the ultracold Bosonic atom
gas confined in an optical lattice, the order parameter is the
mean field value of the operator of Bosons
$\phi=\langle{b}\rangle=\langle{b}^{\dag}\rangle$. There were some
numerical calculations in the Hubbard model, it shows the
entanglement follows different scaling with the size on the two
sides of the critical point denoting an incoherent quantum phase
transition\cite{kais}.

Entanglement state is the most important resource for quantum
information technology. The entanglement between the different
stable phases is a good candidate for quantum computer. The most
entangled states exist at the critical point of phase transition.
The critical point is where the war breaks out, one of the
fundamental character of war is chaos, thus any minor change of
parameter would leads to totally different results. Quantum
computation using entangled states is extracting information from
chaos, and control its output in a exact way. In fact, no matter
it is classical system or quantum system, entanglement is the
basic source of information manipulation, the best entanglement
states exist in a chaos state. The biggest problem for present
quantum computing schemes is decoherence of entanglement, the
quantum entanglement decays rapidly as time goes on.

However, the entanglement we present in this section is
independent of time, it only relies on the physical parameter. It
would be much easier to tune the physical parameters that to fight
against time. So the entanglement in the vicinity of phase
transition is very promising candidate for quantum computation.

\subsection{Quantum states and Nash equilibrium solution of games}\label{statistics-game}

We have shown that phase transition occurs at the Nash-equilibrium
point of a non-cooperative game. While the most entangled states
just arise from this Nash-equilibrium point. We need to get a
deeper understanding on how the entanglement grows stronger and
stronger in a game process.

The most fundamental principal for a physical system is the
Principle of Least Action.  Newton's mechanics is unified in
Hamilton's principle of least action as well as in Gauss's
principle of least constraint. Maxwell's equations can be derived
as conditions of least action for electromagnetic field
propagation. When light goes through optical systems, it always
take the path of least time, and takes short cuts in glass and
water where light travels slower. The most stable state of many
body system is the ground state, which possesses the least energy.
The ground state is the final game results of the particles in
many body system. Every particle wants to stay at the most stable
point, it is the medal all the other particles struggle to get,
this drives the particles in the game.

The Principle of Least Action of a cooperative game is to maximize
the profit of the whole group. The Principle of Least Action of a
non-cooperative game is that all players only take strategies to
maximize his payoff function or to minimize his loss function. Of
course his personal strategy must take into account of other
players' strategy, because if the whole group breaks down, he will
get nothing.

The players may be classified by their statistics in analogy with
the statistics of particles in physical system. As all know, there
are two types of elementary particles: Fermions and Bosons. There
are anyons whose statistics stands between fermions and bosons, we
first put that aside for convenience. The combination of fermions
may form boson. Particles obeying Bose statistics shows a
statistical attractive interaction. While the fermions
demonstrated a statistical repulsive interaction which comes from
the Pauli exclusion principle.

The predetermined physical environment is the game rule of the
players. Players choose their states according to its interaction
with external field and other players. We can alternatively
specify each strategy profile $(s_{1},s_{2},...,s_{N})$ by the
occupation number $(n_{1},n_{2},...,n_{N})$, which means there are
$n_{i}$ players take strategy $s_{i}$. Since all men are born free
and equal, there would be no payoff difference if they choose the
same strategy.

The players in a non-cooperative game behaves as fermions in
physical system. If we exchange any two of the players, the
difference of profit between them would change a sign. They do not
share the same occupation. So the expected number of players in
noncooperative game is fermi-dirac distribution,
\begin{equation}\label{bose-Einstein}
n_{i}=\frac{g_{i}}{e^{\beta(u_{0}-u_{i})}+1},
\end{equation}
where $u_{i}$ is the benefit the player could get by strategy
state $s_{i}$. The players in cooperative game are altruistic
players, they concern for the welfare of others to maximize the
benefit of the whole group. They do not care too much about their
own profit. so exchanging two of them does not make any
difference, they like to share with other players. In this case,
the distribution obey Bose-Einstein statistics,
\begin{equation}\label{bose-Einstein}
n_{i}=\frac{g_{i}}{e^{\beta(u_{0}-u_{i})}-1}.
\end{equation}
However, it is naive to say a player in reality is altruistic
player who only concerns about others or the egoistical player who
only concerns about himself. Every real player is in a mixed state
of the altruistic and the selfish. We consider a rational player,
each time he sets a step further by choosing a particular
strategy, he has to confront a wining result or a losing results.
When he loses, his benefit is transferred to other players, we
call him altruistic, on the contrary, we say he is selfish. We
expressed the altruistic state as $|1\rangle$ and the selfish
state as $|0\rangle$,
\begin{equation}\label{01}
|altruistic\rangle=|1\rangle,\;\;\;\;\;|selfish\rangle=|0\rangle.
\end{equation}
$|1\rangle$ and $|0\rangle$ form an orthogonal basis for the
self-state space of player, i.e., $\langle1|0\rangle=0$. Because
when he faces a particular pure strategy, there is only two
possibilities: take it or give it up. An arbitrary mixed
self-state vector is the linear combination of the two basis,
\begin{equation}\label{a0+b1}
|m\rangle=a|1\rangle+b|0\rangle,
\end{equation}
where $a$ and $b$ are complex number. $|m\rangle$ is a unit
vector, $\langle{m}|m\rangle=1$, it is equivalent to
$a^{2}+b^{2}=1$. $|m\rangle$ means the player has the possibility
of $a^{2}$ to be altruistic and $b^{2}$ to be selfish.

A game consists of $N$ players is physically equivalent to a many
body system with $N$ particles. Each player represents a particle,
the self state is the quantum states of the particle. A statistic
distribution of the $N$ particle states can be present in an
ensemble which is a collection of identically prepared physical
system. For a $n$ player game, the global strategy state space is
the product space of the $n$ players strategy space,
\begin{equation}\label{s=Sixxx}
S=S^{(1)}\otimes{S^{(2)}}\otimes\ldots\otimes{S^{(n)}}.
\end{equation}
The payoff function is a set value map from this Hilbert space to
a number in Euclidean space. The payoff function plays the role of
a negative Hamiltonian in may body physical system.

We first take the prisoner's dilemma to illustrate the basic
phenomena of quantum states in non-cooperative game theory.

In conventional text books about prisoner's dilemma, the two
players are rational players(see Appendix \ref{prisonerdilemma}).
When they are prevented from cooperation, both of them would
confess to minimize his own loss, both of them spend $b$ years in
jail. If Alice and Bob communicate and cooperate with each other,
they would not confess so that they only serve $a<b$ years in
prison.

Here we take a different angle to read the Prisoner's dilemma for
the sake of quantum statistics. The self-states of two players
have four different cases: (1) Alice and Bob are selfish players;
(2) Alice and Bob are altruistic players; (3) Alice is selfish
player and Bob is altruistic players; (3) Alice is altruistic
player and Bob is selfish players. Selfish player take strategy to
maximize his own benefit, and altruistic players tries to maximize
the other's benefit.

If we know which kind of players Alice and Bob are, we can find
the Nash equilibrium fixed point, the two players does not need to
commute with each other. For case (1), Alice knows Bob would
certainly choose confess that can maximize his benefit, the only
Nash equilibrium solution is that she also confess, so they reach
the Nash equilibrium point $(a,\;a)$. For case (2), the Nash
equilibrium is neither of them confess. In case (3), Alice does
not confess, Bob confess. In case (4), Alice confess, Bob does not
confess. The above is the case when the two players are in pure
states, Alice(bob) is either altruistic $|1\rangle$ or selfish
$|0\rangle$. We denotes the four mixed self-states as
\begin{equation}\label{00100111}
|1\rangle_{A}|1\rangle_{B},\;\;|1\rangle_{A}|0\rangle_{B},\;\;|0\rangle_{A}|1\rangle_{B},\;\;|0\rangle_{A}|0\rangle_{B}.
\end{equation}
The Nash equilibrium solution of the four states are summarized in
table (\ref{prisonertable})
 \begin{equation}\label{prisonertable}
\begin{tabular}{|c|c|c|}
  \hline
  Alice\; & \;Bob \;&\; equilibrium point \\
  \hline
 \; $|$0$\rangle$\;&\; \;$|$0$\rangle$ \;&\; (a,\;a) \\
   \hline
 \; $|$0$\rangle$\;& \;\;$|$1$\rangle$\;& \;(0,\;c)  \\
   \hline
 \; $|$1$\rangle$ \;& \;\;$|$0$\rangle$\;&\; (c,\;0)  \\
   \hline
 \;$|$1$\rangle$ \;&\; \;$|$1$\rangle$ \;&\; (b,\;b) \\
\end{tabular}
\end{equation}
This table is the density matrix of the prisoner dilemma. The
statistical weight of $|1\rangle_{A}|1\rangle_{B}$ is the Pareto
optimal point of collective payoff. The statistical weight of
self-state $|0\rangle_{A}|0\rangle_{B}$ is Nash equilibrium. For
the other two states $|1\rangle_{A}|0\rangle_{B}$ and
$|0\rangle_{A}|1\rangle_{B}$, the two equilibrium points appear in
the off diagonal elements of the payoff matrix.

The above is the simplest case of prisoner's dilemma, there are
only two available strategies: confess or non-confess. Now we
consider a more complex case: the available strategies of Alice is
$\{e_{A1},e_{A2},...,e_{Aq}\}$ and Bob's strategies are
$\{e_{B1},e_{B2},...,e_{Bp}\}$. The payoff matrix is now a
$q\times{p}$ bi-matrix.
 \begin{equation}
\begin{tabular}{|c|c|c|c|c|}
  \hline
   &  & Bob & $\cdots$ &Bob \\
  \hline
  & & \;\;$e_{B1}$ \;\;& \;\;$\cdots$\;\;& \;\;$e_{Bp}$\;\;   \\
   \hline
Alice  &  \;\;$e_{A1}$\;\;& ($A_{11}$,\;$B_{11}$) & $\cdots$\;& ($A_{1p}$,\;$B_{1p}$)   \\
   \hline
$\vdots$  & \;\;$\vdots$ \;\;& $\vdots$ & $\ddots$  \;& $\vdots$  \\
  \hline
Alice  & \;\;$e_{Aq}$ \;\;& ($A_{q1}$,\;$B_{q1}$) & $\cdots$  \;& ($A_{qp}$,\;$B_{qp}$) \\
  \hline
\end{tabular}
\end{equation}
The prisoner dilemma told us, the fixed point solution of a game
can be determined by self-state of players, namely, they are
altruistic or selfish. A player could be in a mixed state of
altruistic and selfish. For each pure strategy $|e^{k}\rangle$, we
introduce a fractional number $\rho_{k}$ to measure the altruistic
degree of a player, $\rho_{k}$ satisfies the constrain
$\sum_{k}\rho_{k}=1$. For example, we divided the altruistic
degree into $n$ stages,
\begin{equation}\label{1/p}
|0\rangle,\;\;|\frac{1}{n}\rangle,\;\;
|\frac{2}{n}\rangle,\;\;\ldots
,\;\;|\frac{n-1}{n}\rangle,\;\;|1\rangle.
\end{equation}
From $|0\rangle$ to $|1\rangle$ record how a player grows from a
selfish player to a altruistic player step by step. Once we know
the self-state of a player, we know the probability he would play
a certain strategy. As shown in Prisoner dilemma, the two player
is either altruistic or selfish, the probability of confess as
well as not-confess is either 0 or $1$. A vector of this
altruistic degree could be the self-characteristic vector of a
player, the equilibrium point of a game is utterly relies on this
self-characteristic vector. There exists a one-to-one
correspondence between the self-characteristic vector of players
and the fixed point of a game.
\begin{figure}
\begin{center}\label{phasechanel}
\includegraphics[width=0.4\textwidth]{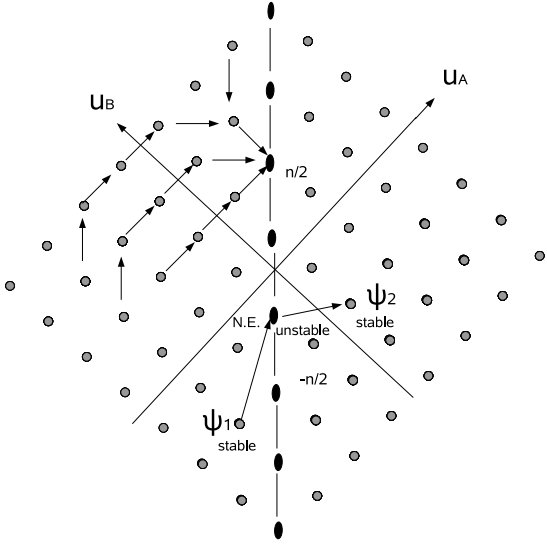}
\caption{This is the payoff table which is rotated by $45$ degrees
in clock wise direction. Each black dot represents a pair of
payoff value, the elliptic black point in the middle represent the
Nash-Equilibrium(N.E.) points. The game begins at an arbitrary
pair of strategies, there is an imbalance between Alice's payoff
and Bob's payoff in the begging. Then the losing player takes a
better strategy to eliminate the imbalance in the next step, the
winning player also choose his best response to keep his
predominance. This game process is actually a renormalization
group transformation, it finally converges at the Nash-Equilibrium
point. }
\end{center}
\end{figure}

This self-characteristic vector of one player uniquely determined
the probability he choose certain strategy. Thus when we introduce
the mixed strategy for two players Alice and Bob,
\begin{eqnarray}\label{s=e+e-fourier}
&&|s_{A}\rangle=\sum_{k=1}^{q}\rho_{Ak}|e^{k}_{A}\rangle,\;\;\;\sum_{k}\rho_{Ak}=1,\\
&&|s_{B}\rangle=\sum_{k=1}^{p}\rho_{Bk}|e^{k}_{B}\rangle,\;\;\;\sum_{k}\rho_{Bk}=1.
\end{eqnarray}
the self-characteristic vector of Alice and Bob have been uniquely
defined by $\{\rho_{Ak}\}$ and $\{\rho_{Bk}\}$. So every mixed
state has an unique fixed point in the payoff matrix. In physics,
payoff function or loss function corresponds to physical
observables, such as Hamiltonian, angular momentum, the given
mixed strategy is just the eigenfunction with the fixed point
solution as the eigenvalue. We can define the density matrix using
the strategy vector $|s\rangle$, $\rho=Tr(|s\rangle\langle{s}|)$.
The von Neuman entropy
$S(\rho(\gamma))=-Tr(\rho(\gamma)\lg{\rho(\gamma)})$ measures the
entanglement between the strategy vectors.
\\
\\
\noindent{{\emph{ Game theory of Hohenberg-Kohn theorem}}}
\\
\\
A many electron system can be mapped into a $n$ player game, each
player carries $\pm\frac{1}{2}$ spin. The ground state energy of
an electronic system is completely determined by the minimization
of the total energy as a functional of the density function. The
external potential together with the number of electrons
completely determines the Hamiltonian, these two quantities
determine all properties of the ground state. Hohenberg-Kohn
theorem states: the external potential $V$ is determined, within a
trivial additive constant, by the electron density. This theorem
insures the that there can not exist more than one external
potential for any given density.

The electrons are players, the many body wave function are the
strategy space. The external potential comes from external
constrain. For example, in the prisoner dilemma, the police could
put the two prisoners together or separate them. If the two
prisoner are put together, they would conspire to remain silent so
that both of spend less time in jail. If the police put them in
separated rooms, they can not communicate with each other. Since
they do not trust each other, they would confess. So we see
external potential from  police decided the behavior of the
player, or in another point of view, the external potential
transform the altruistic into the selfish by separating them, and
transform the selfish into the altruistic by putting them
together. Therefore there is only one self-characteristic vectors
for a given external potential. This is Hohenberg-Kohn theorem in
game theory.

\subsection{{{ Renormalization group transformation and quantum entanglement}}}

In fact, the von Neuman entropy measures the entanglement between
the self-characteristics vectors. Every self-characteristics
states vector is a mixed states of altruistic and selfish. If the
player is in a complex mixed states of altruistic and selfish, it
is hard to predict where he ends. The Hohenberg-Kohn theorem told
us the external potential uniquely determined the state vector of
players, as shown in prisoner dilemma. This suggests us a possible
way to operate entanglement state using external potential.

The entanglement states grows stronger and stronger following the
step of renormalization group transformation. For example, in the
bargain game, the seller first gave a price to see if the buyer
take it. The buyer thought it too expensive, he feedback his price
to the seller. Then the buyer and seller both know each other at
the first step. As this bargain goes on, they know each other
better and better. In other words, the entanglement between them
grows stronger and stronger. When this entanglement reach a
maximal point, a Nash equilibrium state is arrived. At this point,
if the seller raise one penney, the buyer won't buy it, on the
other hand, if the buyer lower the price one penney, the seller
will not sell his product. When the game passed over the Nash
equilibrium, the entanglement decreased dramatically.

The entanglement in non-cooperative game is much stronger than the
entanglement in cooperative game. In last section, we know if the
self-character of players is determined, we can uniquely find an
equilibrium solution. We call the altruistic player an angle
player, while the selfish a devil player. The angel player who
manifests goodness, purity, and selflessness, behaves like bosons.
While the devil player are fermions.

The devil player do not trust each other, they most likely to
betray in the game. So two devil players would try their best to
bound them together and form a pair, they increase communication
and cooperation to prevent his companion from betray. The
entanglement between them is very strong, this entanglement would
reach a maximal when the trial is around the corner.

But angle players always trust their companions. Everything they
do is to increase the welfare of other players. The angel player
does not need to communicate too much. They behave much like
non-interacting, indistinguishable particles. So angle player is
boson. An unlimited number of bosons may occupy the same state at
the same time. At low temperatures, bosons can behave very
differently than fermions; all the particles will tend to
congregate together at the same lowest-energy state to warm each
other, this is Bose-Einstein condensate.

For a given many particle system, the entanglement can be measured
by the particle-particle correlation length. the longer
correlation length there exist between particles, the stronger
entanglement they have. we can read the entanglement out following
Kadanoff block transformation. For example, in the two dimensional
Ising model, the spins are players of game. They are mixed type of
players between angle and devil. Their self-characteristic state
vector $|\psi\rangle$ can be expanded by four entangled
eigenstates of the Bell operators:
$\psi^{\pm}=\frac{1}{\sqrt{2}}(|\downarrow\rangle|\uparrow\rangle\pm|
\uparrow\rangle|\downarrow\rangle),\;\phi^{\pm}=\frac{1}{\sqrt{2}}
(|\downarrow\rangle|\downarrow\rangle\pm|\uparrow\rangle|\uparrow\rangle).$
The density matrix, describing all the physical variables
accessible to entanglement state $|\psi\rangle$, is given by
$\rho=Tr(|\psi\rangle\langle\psi|)$. Then we can calculate von
Neuman entropy
$S^{0}(\rho(\gamma))=-Tr(\rho(\gamma)\lg{\rho(\gamma)}) $ for the
first order Kadanoff block transformation. Then we construct the
entanglement states between two blocks and obtained the second
order von Neuman entropy $S^{1}$. We can continue this
renormalization group transformation, and finally derive an exact
von Neuman entropy.

\subsection{Topological phase in strategy space for multi-player game}

We study the topological quantity in the strategy space of
n-player game in this section. The total payoff function of the
$n$ players could be view ed as Hamiltonian matrix.
\begin{equation}\label{Hami-payoff}
\mathbb{H}=\{\hat{u}_{1},\hat{u}_{2},...,\hat{u}_{n}\}.
\end{equation}
A strategy vector of the many players's strategy space reads
\begin{equation}\label{s=sixxx}
s=\{s^{(1)}_{i},{s^{(2)}_{j}}\ldots{s^{(n)}_{k}}\},
\end{equation}
$s^{(n)}_{k}$ is the $k$th strategy of the $n$th player.
$\hat{u}_{n}$ maps this strategy vector into the payoff value of
the $n$th player, i.e., $\hat{u}_{n}s=u_{n}$. When the Hamiltonian
matrix of the payoff function operates on this vector, it produces
the eigenvalues
\begin{equation}\label{uns=u}
\mathbb{H}\;s=\{{u}_{1},{u}_{2},...,{u}_{n}\}.
\end{equation}
The players fight against each other to reduce the difference
between their payoffs
 \begin{equation}\label{delat1--n}
\Delta_{1}=u_{1}-u_{2},
\;\Delta_{2}=u_{2}-u_{3},\cdot\cdot,\;\Delta_{n}=u_{n}-u_{1}.
\end{equation}
In a quantum system, $u_{p}$ could be interpreted as the $p$th
energy level. The Nash equilibrium solution is governed by the
coexistence equation
\begin{eqnarray}\label{berry-del}
D({\Delta}/{s})=\{\Delta_{1},\Delta_{2},...,\Delta_{n}\}=0.
\end{eqnarray}
The war between energy levels breaks out in the degenerated
eigenspace. The players take their strategy to decrease the energy
gap between them. Those points at which the energy gap vanishes
are the core center of vortex, they are the energy level crossing
point. The $i$th player choose strategies according to other
players' strategy, thus his strategy vector is a map from the
other players's eigen-strategy to his own space
$\psi_{i}(\gamma)$. In analogy with conventional Berry phase of
quantum mechanics, we proposed a general topological quantity(See
Appendix \ref{topologicalphase}). The topological phase of many
players reads
\begin{equation}\label{dual-n-berry}
\Omega_{i}=i\epsilon_{tj...kl}\langle\partial_{\gamma_t}\psi_{i}({\gamma})|
\partial_{{\gamma}_j}\psi_{i}({\gamma})\rangle...\langle\partial_{{\gamma}_k}\psi_{i}({\gamma})|\partial_{{\gamma}_l}\psi_{i}({\gamma})\rangle.\nonumber
\end{equation}
$\Omega_{i}$ is actually the Riemannian curvature tensor in the
strategy vector bundle. This quantity is equivalent to a
topological current of the payoff functions
\begin{eqnarray}\label{Dup}
\Omega_{i}^{p}=\prod_{j\neq{i}}\delta(\Delta_{i}-\Delta_{j})d\Delta_{1}\wedge{d\Delta_{2}}\wedge\cdots{d\Delta_{n-1}}\wedge{d\Delta_{n}}.\nonumber
\end{eqnarray}
From our general generalization of conventional Berry phase, the
most general topological quantity is the Chern character. In this
strategy vector bundle, the Chern character can be derived from
the exponential map,
\begin{eqnarray}\label{mo}
&&Ch(x)=Exp({d\langle\psi(\gamma)|\wedge{d|\psi(\gamma)\rangle}}).
\end{eqnarray}
These topological quantities are good candidate to measure the
entanglement between different players in strategy space. One can
see from Eq. (\ref{Dup}), topological charges focused on the
equilibrium solutions where all players get the same payoff, there
is no difference among any two players. But remember these
equilibrium point are isolated point, they are the center of
vortex. A tiny deviation from these point would break the
equilibrium.

\section{game theory of many body physics}

\subsection{Cooperative game and classical many body system}

The particles of a physical is the players of a game. They act
following the least principal, the aim of the players is to
decrease the total energy. Some of them may united to form a
subgroup due to local potential. Two subgroups may fuse into one
when this fusion can make both of them more stable. It will be
shown the many player cooperative game(see Appendix
\ref{cooperative}) is just a physics system.

We consider a 3-person cooperative game $\{1,2,3\}$. We choose the
payoff function of this game as the conventional Hamiltonian, the
first step of the cooperative game is to find out all subgroups
$\{\emptyset,1,2,3,(1,2),(1,3),(3,2),(1,2,3)\}$. The energy
function maps each subgroup to a real number,
\begin{equation}\label{vmap-8}
\varepsilon_{\emptyset}=0,\;\varepsilon_{1},\;\varepsilon_{2},\;\varepsilon_{3},
\;\varepsilon_{12},\;\varepsilon_{23},\;\varepsilon_{13},\;\varepsilon_{123}.
\end{equation}
This energy mapping may have various formulas. The total energy of
the three particle is
$E(N)=\varepsilon_{1}+\varepsilon_{2}+\varepsilon_{3}+\varepsilon_{12}+\varepsilon_{13}
+\varepsilon_{23}+\varepsilon_{123}.$ The super-additivity
requires $\varepsilon_{2}+\varepsilon_{3}\leq\varepsilon_{23}$.
$e(23)=\varepsilon_{23}-\varepsilon_{2}-\varepsilon_{3}$ is the
difference of payoffs called excess. In fact, this is a nuclear
reaction in physics. In game theory, people are looking for
optimal self-content energy vector of the n particles
${\varepsilon}^{\ast}$ and the excess value pair
$\epsilon_{\ast}$. In the eyes of physicist, this is group
expansion algorithm.

For a standard Hamiltonian system,
\begin{equation}\label{Hc00}
H=\sum_{i=1}^{N}\frac{1}{2}p^{2}_{i}+V(q_{1},...,q_{N}),
\end{equation}
we take $N$ particles as players, their position $q_{i}$ and
momentum $P^{i}$ as their strategy, the energy is the payoff. The
target of this game is to find the stable ground states.

To investigate the phase transition of this game in the frame work
of our topological phase transition theory, one only need to
replace the output function $\hat{O}_{ut}(\gamma)$ with the
Hamiltonian $H$, or any other independent integrals in the
complete set $(\phi_1 = H, \phi_2, . . . , \phi_n)$ which are in
involution. The strategy vectors is taken as
$\gamma=(q_{1},...,q_{N},p_{1},...,p_{N})$. Then all the
conclusion on topological phase transition we obtained in the
previous section holds here.

However the physical system we encountered always have a very
larger number of particles, usually $N\sim10^{23}$. It is
impractical to control the position and momentum of particles.
Therefore we take a different way to model the many body system.
We focus on a few interaction parameters and integrate the
configuration space out. For example, we choose the Hamiltonian,
\begin{equation}\label{Hc00}
H=\sum_{i=1}^{N}\frac{1}{2}p^{2}_{i}+\gamma^{1}V_{1}(q_{1},...,q_{N})+\gamma^{2}V_{2}(q_{1},...,q_{N})+....\nonumber
\end{equation}
The interaction potential relies on $n$ players above
configuration space,
 \begin{equation}
V(\vec{q})=\gamma^{1}V_{1}+\gamma^{2}V_{2}+...+\gamma^{n}V_{n}.
\end{equation}
where $\{\gamma_{1},\;\gamma_{2},...,\}$ are physical parameters,
and $V_{i}=V_{i}(q_{1},...,q_{N})$ is the potential functions. We
integrate out all the uncontrollable position and momentum
variables in the Helmoltz free energy on configuration space, then
the effective output function for a finite player game is derived,
$F(\gamma_{1},\gamma_{2},...)=-(2\beta)^{-1}Log(\pi/\beta)-f(\beta,\gamma_{1},\gamma_{2},...)/\beta$.
with
$f(\beta,\gamma_{1},\gamma_{2},...)=\frac{1}{N}Log\int{d^{N}q\;exp[-\beta{V(\vec{q},\gamma_{1},\gamma_{2},...)}]}$.

The free energy function is merely one special component of output
vector field corresponding to Hamiltonian function. We can choose
any output function $O_{ut}$ on the strategy manifold, and choose
a arbitrary component of the compete set whose members are in
involution as Hamiltonian function $\phi_{H}$, then we introduce
the renormalization group transformation upon the output vector
field. The basic tangent vector field $\vec{\phi}$ to characterize
the topology of the extended strategy space is given by the
variational principal
\begin{equation}
\frac{\delta^{p}[\hat{U}{O_{ut}}(\gamma_{1},\gamma_{2},...)]}{\delta{\theta}^{p}}=
\frac{\int{d^{N}q{d^{N}p}\;\exp[-\beta{\phi_{H}}]}\delta^{p}[\hat{U}{O_{ut}}]}
{N\int{d^{N}q{d^{N}p}\;\exp[-\beta{\phi_{H}}]}}.\nonumber
\end{equation}
A special subset of renormalization group transformation is Lie
group $\hat{U}=e^{i\theta\hat{L}}$. Now we can apply the
topological phase transition theory to study the phase transition
on the strategy space extended by the interaction parameters
$\gamma=(\gamma_{1},\gamma_{2},...)$. The conventional conjugate
momentum and position variables have been integrated out, they are
out of the set of players for the phase transition game. Of
course, people can bring them in the game, in that case, the
number of players is too large to manipulate, it does not tell us
any practical information.

Generally a general dynamic system can be viewed as a
non-cooperative game between players
$\gamma_{1},\;\gamma_{2},...\;\gamma_{n}$. If we view the
classical system as a cooperative game, we can divided the
effective interaction parameters
$\{\gamma_{1},\gamma_{2},...,\gamma_{n}\}$ into several groups
$\{(\gamma_{1},\gamma_{2}),(...\gamma_{i}),...,\gamma_{n}\}$, and
consider the interaction between these groups. Then we analyze
their coalition, their excess value pair, their payoff, etc. We
could get the most effective information about the system through
the phase coexistence equation.

\subsection{Topological phase transition of quantum many body system}

The topological phase transition theory in this paper aims at the
most general systems, no matter they are classical physical
systems or quantum physical systems, or biological system, or
social system. For a quantum many body system, any output that
responses corresponding to certain input can be used to detect a
phase transition. The output $\hat{O}_{ut}$ could be any physical
observable, such as the correction to ground state energy
$\delta{E_{g}}$, the correction to thermal potential
$\delta{\Omega}$,  the single-particle current operator
$\hat{J}=\int{d^{3}x}j$, the number density operator
$<\hat{\rho}(x)>$, the spin density operator $<\hat{\sigma}(x)>$,
the free energy $F$, susceptibility $\chi$, specific heat $C_{H}$,
correlation length $\xi$, compressibility $\kappa_{T}$, $\cdots$,
and so on. The input are any physical parameters, such as magnetic
field, electric field, radiation, temperature, pressure, on-site
repulsive interaction, chemical potential, density, potential,
scattering length, neutral current, electron wave, probing laser
beams, enzymes, ..., and so forth. Our topological phase
transition theory can be successfully applied to explain the
quantum tunnelling of the magnetization vector in ferromagnetic
nano-particles\cite{tieyan}.

We take a specific model to demonstrate the application in the
following. A gas of ultracold atoms in an optical lattice has
provided us a very good experimental observation of
superfluid-Mott-insulator phase transition\cite{greiner}. The
system is described by the Bose-Hubbard model
\begin{equation}\label{Bose-Hubbard}
H=-J\sum_{{\langle}i,j\rangle}a_{i}^{\dag}a_{j}
+\frac{U}{2}\sum_{i}a_{i}^{\dag}a^{\dag}_{i}a_{i}a_{i}-\sum_{i}{\mu}a^{\dag}_{i}a_{i},
\end{equation}
where the sum in the first term of the right-hand side is
restricted to nearest neighbors and $a^{\dag}_{i}$ and $a_{i}$ are
the creation and annihilation operators of an atom at site $i$
respectively. $J$ is the hopping parameter. $U$ corresponds to the
on site repulsion between atoms, and $\mu$ is the chemical
potential. This Hamiltonian admits two conflicting forces, $J$ and
$U$, player $J$ drive the particles hoping from one site to
another, but $U$ repulse any particles jumps to his sits, this
prevent the particles from moving site by site. The quantum phase
transition is governed by the two players, when the hoping player
is dominated, the ultralcold atoms can easily hop in the optical
lattice, this is the superfluid phase.When the on-site repulsive
force is dominated, the particles is strongly repulsed by its
neighbors, so it would be difficult for him to move to other
house, then the ultralcold atoms have to stay at home, this is the
Mott-insulator phase.

Using mean-field approaches, the ground state energy of
Bose-Hubbard model can be generally expressed as the functional of
$\varepsilon_{i}$, $J$ and $U$, i.e.,
$E_{g}=E_{g}(\varepsilon_{i}, J,U).$ Following perturbation theory
up to the second order\cite{oosten}, the variation of ground state
energy $\delta{E_{g}}$ is
\begin{equation}\label{c2}
\delta{E_{g}}=[\frac{g}{{U}(g-1)-J}+\frac{g+1}{J-{U}g}+1],
\end{equation}
From our definition of the first order phase transition, the
boundary between the superfluid and the Mott insulator phases
should be decided by the equation $\delta{E_{g}}=0$. We have
plotted the phase diagram(FIG. 6).
\begin{figure}
\begin{center}\label{no1}
\includegraphics[width=0.26\textwidth]{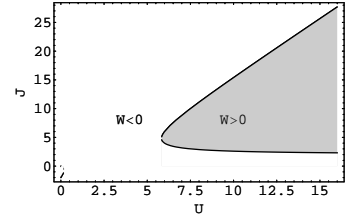}
\caption{The phase diagram of the Bose-Hubbard Hamiltonian for
g=1. Inside the curve is the Mott-insulating phase, outside the
curve is the superfluid phase.}
\end{center}
\end{figure}
This phase diagram is in agrement with the familiar phase diagram
of superfluid-Mott-insulator phase transition\cite{oosten}. It is
also the same as the phase boundary obtained from the minimizing
the free energy\cite{chen}, where they take the hopping term as
perturbation.

In this phase diagram, each phase is assigned with a winding
number. The winding number $\beta_{k}\eta_{k}=W_{k}$ is the
generalization of the Morse index in {Morse} theory. Its absolute
value $\beta_{k}$ measures the strength of the phase.
$\eta_{k}=$sign$\{\phi^{1},\phi^{2}\}=\pm1$, represents different
phases. In the dark area circulated by the curve,
$\eta_{k}=$sign$\{\phi^{1},\phi^{2}\}=+1$, it represents the
Mott-Insulator phase, while outside the curve,
$\eta_{k}=$sign$\{\phi^{1},\phi^{2}\}=-1$, it represents the
superfluid phase. On the curve, $\{\phi^{1},\phi^{2}\}=0$, that is
where a quantum phase transition takes place.

Most of the recent phase diagram of the quantum phase transition
are focused on the first order. As for the $p$th-order quantum
phase transition, it is characterized by a discontinuity in the
$p$th derivative of the variation of ground state energy. The
order parameter field of the $p$th order Quantum phase transition
can be chosen as the (p-1)th derivative of $\delta{E}_{g}$ with
respect to $U$ and $J$, i.e.,
$\phi^{1}=\partial^{p-1}_{J}\delta{E}_{g},\;\phi^{2}=\partial^{p-1}_{U}\delta{E}_{g}$.
The equation of coexistence curve is given by
$\{\phi^{1},\phi^{2}\}=D(\frac{\phi}{\gamma})=0$,
\begin{eqnarray}\label{D(E/JU)}
\{\phi^{1},\phi^{2}\}=\frac{\partial^{p}{\delta{E}_{g}}}{\partial{J}^{p}}
\frac{\partial^{p}{\delta{E}_{g}}}{\partial{U}^{p}}-
\frac{\partial\partial^{p-1}{\delta{E}_{g}}}{\partial{U}\partial{J}^{p-1}}
\frac{\partial\partial^{p-1}{\delta{E}_{g}}}{\partial{J}\partial{U}^{p-1}}=0,
\end{eqnarray}
here we have taken $\gamma_{1}=U,\;\gamma_{2}=J$.  From this
equation, one can arrive the phase diagram of the $p$th order
quantum phase transition from the equation above, and find some
new quantum phases.

The above discussion is based on the correction to ground state
energy, now let's choose the output field as thermal potential,
$\hat{O}_{ut}=\Delta{\Omega}$. One can do some perturbation
calculation up to the second order, and derive the Green function,
\begin{eqnarray}\label{G=u-U}
G^{-1}(p,i\omega_{n})=\epsilon-\epsilon^{2}\sum_{\alpha}(\alpha+1)\{\frac{n_{\alpha}-n_{\alpha+1}}{i\omega_{n}-\alpha{U}+\mu}\}
\end{eqnarray}
The two parameters are $\{\mu.U\}$. The correction of the
thermodynamic potential is
\begin{eqnarray}\label{delOmega2-yu}
\Delta{\Omega}={\Omega}-\Omega_{0}=\frac{1}{\beta}\sum_{p,n}\ln[{-{\beta}G^{-1}(p,i\omega_{n})}].
\end{eqnarray}
The first order phase transition is given by $\Delta{\Omega}=0$.
Put Eq. (\ref{G=u-U}) into $\Delta{\Omega}=0$, we have
$G^{-1}=-T.$ At zero temperature, $T=0$, it leads to $G^{-1}=0$.
This is exactly the condition for the superfluid-mott-insulator
phase transition in Ref. \cite{yuyue}.

Now we see different physical observables are only different
outputs, they lead to the same phase boundary.

\subsection{Elementary excitations and momentum space}

One of the best merits of quantum many body theory on lattice is
we do not need to consider canonical position coordination and its
conjugate canonical momentum. The wave vector has only three
spatial components. Although we can not manipulate them, they are
very good candidates for the player of a game.

The inverse of the wave vector is wave length which is quantized
due to boundaries in a finite solid state lattice. Like the string
fixed at both ends, the wave vector of the standing wave modes
only take a few discrete values. The standing wave patterns of the
electron wave and atomic wave in lattice is viewed as elementary
excitations, or quasi-particles. If we think the electrons as
soldiers, the elementary excitation is the collective dancing of
millions of soldiers. When the soldiers are confined in a square
battlefield, they self-reorganized into certain dancing modes to
avoid crashing each other. In fact, the independent spatial
components of wave vector are their commanders, this is a war game
between wave vectors. These oscillation modes, or elementary
excitations, are the surviving strategies of the commanders in
this battle.

In quantum many body system, the elementary excitations sit at the
singular points of the green function. The external response can
be derived from green function by linear response theory(see
Appendix \ref{Lresopnse }). These responses are the output vector
field of a quantum many body game. For the most general case, we
set the Green function on a $m$ dimensional momentum space,
$p\equiv(p_{0}=E, p_{1}, p_{2}, ... , p_{m})$. In analogy the
topological current of Riemannian curvature tensor on $m$
dimensional manifold, we introduce the topological current of
$N$-point Green function $G$ in momentum space,
\begin{eqnarray}\label{k-E-n-berry}
B_{G}=\epsilon_{tj...kl}\langle\partial_{p_t}G|\partial_{p_j}G
\rangle...\langle\partial_{p_k}G|\partial_{p_l}G\rangle,
\end{eqnarray}
where we denote $|\partial_{p_l}G\rangle=\partial_{p_l}G(p)$ and
$\langle\partial_{p_k}G|=\partial_{p_l}G^{\dag}(p)$. The inner
product between Bra $\langle|$and Ket $|\rangle$ is an integral,
i.e., $\langle\partial_{p_t}G|\partial_{p_j}G
\rangle=\int{dp}\partial_{p_t}G^{\dag}_{N}(p)\partial_{p_j}G_{i}(p)$.
Use this topological current, we can construct a more general
topological action
\begin{eqnarray}\label{Holo-Chern-G}
Ch(M)=Tr(\exp\frac{i}{\pi}d{\langle}G(p)|\wedge{d}|G(p)\rangle).
\end{eqnarray}
This topological action measures the non-trivial topology of Green
function which maps the momentum space to an external response.
The Green function corresponding to the exactly solvable terms of
hamiltonian can be exactly derived, the non-exactly solvable part
is always calculated using perturbation theory. According to Dyson
equation, the difference between the inverse of exact Green
function and the exactly solvable Green function gave us the
correction to energy,
\begin{equation}\label{G-G0+=sigma}
\delta{{E}}=G(p,E)^{-1}-{G_{0}(P,E)}^{-1}=\Sigma(p,E).
\end{equation}
usually people call it self-energy $\Sigma(p,E)$. In this case,
the nontrivial topological current of the N-point Green function
focus on
\begin{eqnarray}\label{B=p*sigma}
B_{p}=Tr[{d\langle{\Sigma}|\wedge{d|{\Sigma}\rangle}
\cdots{\wedge}d\langle{\Sigma}|\wedge{d|{\Sigma}\rangle}}_{2p}],
\end{eqnarray}
again we have the holographic topological action
\begin{eqnarray}\label{Holo-Chern-sigma}
Ch(M)_{\Sigma}=Tr(\exp\frac{i}{\pi}d{\langle}{\Sigma(p,E)}|\wedge{d}|{\Sigma(p,E)}\rangle).
\end{eqnarray}
The self-energy here is the output vector field. The momentum
vector and frequency are the strategies. The surviving strategies
are where the stable elementary excitations arise. One may apply a
Lie group transformation to self-energy to check if there the
discontinuity exist. Usually we take $SO(4)$ whose element can be
written as
$U(\theta)=e^{\theta{\hat{L}}}=\sum_{0}^{n}\frac{1}{n!}(\hat{L}\theta)^{n},$
where $\hat{L}$ is the generator of $SO(4)$, it is operator
constructed from the momentum vector $P$ and energy $E$. If there
is discontinuity, we introduce the vector field
$\vec{\phi}=\frac{\delta{U(\theta)\Sigma(p,E)}}{\delta\theta}$.
Then we can use the phase coexistence equation
$\{\phi^{i},\phi^{j},...,\phi^{k}\}=0$ to decide the phase
boundary.

We always assume the periodic boundary condition in solid state
physics, this boundary condition created a torus lattice manifold.
This naturally leads to a topological constrain in momentum space.
If we take the components of momentum as the player of a game,
their strategy space is the momentum space. If the momentum space
is noncompact, there is no topological constrain on strategies. If
the momentum space is a compact manifold like sphere or torus,
there is a nontrivial topological constrain on strategy manifold.

To avoid the non-figurative reasoning, we consider the
superconducting pairing Hamiltonian \cite{schrieffer},
\begin{equation}\label{EH}
H_{e}=\sum_{ij}(t_{ij}^{\sigma}c_{i\sigma}^{\dag}c_{j\sigma}+\Delta_{ij}c^{\dag}_{i\uparrow}c^{\dag}_{j\downarrow}
+\Delta_{ij}^{\ast}c_{j\downarrow}c_{i\uparrow})-\mu\sum_{i\sigma}c_{i\sigma}^{\dag}c_{i\sigma}.
\end{equation}
If the system is translationally invariant, then
$t_{ij}^{\sigma}=t^{\sigma}(i-j)$ and $\Delta_{ij}=\Delta(i-j)$.
We further require that hopping rate for the up-spin and down spin
is the same, $t_{ij}^{\uparrow}=t_{ji}^{\downarrow}$, then the
up-spin and down-spin have the same kinetic energy
$\xi^{\uparrow}(k)=\xi^{\downarrow}(k)=\xi(k)=\sum_{j}e^{-ik\cdot{r_{j}}}t^{\sigma}(j)-\mu$.
The momentum representation of Hamiltonian $H_{e}$ is simplified
as
\begin{equation}\label{H(k)}
H_{e}(k)=\vec{R}(k)\cdot{\vec{\sigma}},
\end{equation}
where $\vec{\sigma}=(\sigma_{x},\sigma_{y},\sigma_{z})$ are the
Pauli matrices, and
$\vec{R}=(R^{x},R^{y},R^{z})=[Re\Delta(k),-Im\Delta(k),\xi(k)]$
with $\Delta(k)=\sum_{j}e^{-ik\cdot{r_{j}}}\Delta(j)$ as the gap
function. Using the Bogoliubov transformation,
$\alpha_{k}=u_{k}c_{k}-v_{k}c_{-k}^{\dag},\;
\alpha_{k}^{\dag}=u_{k}^{\ast}c_{k}^{\dag}-v_{k}^{\ast}c_{-k}$ and
considering $[\alpha_{k}, H_{e}]=E_{k}\alpha_{k}$ for all $k$, the
effective Hamiltonian becomes
$H_{e}=\sum_{k}E_{k}\alpha_{k}^{\dag}\alpha_{k}+const$ with
$E_{k}\geq0$. The effective energy of quasi-particle is
$E_{k}=\sqrt{\xi_{k}^{2}+|\Delta_{k}|^{2}}.$ The Anderson's
pseudospin vector\cite{anderson}
\begin{equation}\label{n=xi/E}
\vec{n}=\frac{(Re\Delta_{k},\;-Im\Delta_{k},\;\xi_{k})}{E_{k}}.
\end{equation}
The pseudospin vector $\vec{n}$ describes a mapping from
\textbf{k}-space to a sphere $S^{2}$ in the pseudospin
space\cite{readprb}. This map defined a topological number to
classify the topology of the momentum space.

As observed in the calculated results for the berry phase
curvature in SrRu$O_{3}$. It has a very sharp peak near the
$\Gamma$-point and ridges along the diagonals\cite{monom}. The
origin for this sharp structure is the degeneracy and band
crossing. The transverse conductivity $\sigma_{xy}$ is given
by\cite{thouless}
\begin{equation}\label{sigmaxy}
\sigma_{xy}=\sum_{k}\frac{1}{e^{{\beta}(\varepsilon_{k}-\mu)}+1}\Omega_{xy}(k).
\end{equation}
We applied the topological current theory\cite{DuanSLAC} to the
momentum space of this pairing Hamiltonian, and proved that the
curvature only exist at the solutions of $\vec{R}=0$, i.e.,
\begin{equation}\label{sigmaQ}
\Omega_{xy}=\sum_{k}W_{l}\delta(k^{\mu}-k^{\mu}_{l})
\end{equation}
While $W_{l}$ is just the winding number of $\vec{{R}}$ around the
$l$-th solution $\vec{R}(k_{l})=0$.

If we take the whole two dimensional system as a war, the soldiers
are the two momentum components $\vec{k}=(p_{l},p_{2})$. Each
$p_{i}$ has a strategy space which is his available value across
the whole momentum space. $p_{i}$ fights against each other to
steer the navigation of quasi-particle. If the whole system is
homogeneous and external potential is not anisotropic, $p_{l}$ and
$p_{2}$ are exchangeable. In real space, the electrons in a
homogeneous system has no bias direction, they circular around
each other. Two dimensional system is rather special, the
electrons have to dance around each other to avoid crashing. When
the external magnetic field is present, the electrons change their
pattern of motion in real space, they are adjusting their
wavelength and oscillation modes in the mean time. The electrons
are spontaneous organized into collective oscillating modes in
which any two of them form pairs.

The interference of the electron wave forms periodic distribution
in the two dimensional momentum plane. Generally speaking, $p_{1}$
and $p_{2}$ are much like coworkers, they balance their inner
conflicting profits to fit the environment. The best surviving
strategy focused on the extremal point of the external potential,
at these points, the quasi-particle or collective oscillation
modes do not cost energy any more. Those solutions are given by
$\vec{R}=0$. These periodically distributed extremal points are
the basic channels through which the electron like to travel. Each
of these channels may be view as the center of cyclone, they carry
a winding number. The sum of the winding numbers of all these
channels is under a topological constrain which comes form the
topology of momentum manifold.

The topology of momentum manifold strongly depend on boundary
condition. It seems strange for a brick of material with a rough
surface to have some periodic or other boundary condition for the
wave function of electrons. In fact, like the droplet of water in
vacuum without gravitational field, their surface are
spontaneously compact sphere. Compact boundary is one way for a
system to fit the environment under certain environment. In other
worlds, compact boundary condition is the result of evolution
following the Principal of Least Action(The principal we talked
bout here is the generalized principal of least action in the
first several sections of this paper).

Therefore he electrons are organized into certain compact boundary
condition by themselves in order to save energy or to form a more
stable state. The topology of their momentum space is modified
according to external inputs, such as chemical potential, magnetic
field, et al. Anyway, if the Hamiltonian relies on many physical
parameter $\{\gamma^{i},i=1,2,...\}$, i.e.,
$H=H(\gamma^{1},\gamma^{2},...)$, the coexisting boundary at which
they reach a balance is given by
\begin{eqnarray}\label{berry-del}
D(\hat{\phi}/{\gamma})=[\hat{\phi}_{1},\hat{\phi}_{2},...,\hat{\phi}_{n}]=0.
\end{eqnarray}
If one of the operator is taken as Hamiltonian itself,
$\hat{\phi}=\hat{H}$, this is nothing but the complete set of
dynamic system in parameter space.

\subsection{Quantum many body theory of game}

The player of war consists of a large number of soldiers which act
like identical particles. The soldiers split into several
subgroups, inside every subgroup there is a common agreement
constituted and obeyed by all the members of the association. Thus
we may take mean field approach inside the subgroup whose member
move in a mean-field potential formed by all the other members.

Generally speaking, the subgroup consists of angel players is
stronger than that of devil players. Angle players are bosonic
players. The collective strategy is the symmetric combination of
the personal strategy,
\begin{equation}\label{boson-ijk}
|S_{Bos}(1,2,...,n)\rangle=\sqrt{\frac{\prod{m!}}{n!}}P(|s^{(1)}\rangle|s^{(2)}\rangle....|s^{(n)}\rangle).\nonumber
\end{equation}
The devil payer are fermionic players. Their collective strategy
is a totally anti-symmetric strategy, which can be written as
Slater determinant,
\begin{equation}\label{fermion-ijk}
|S_{Fer}\rangle=\sum\frac{1}{n!}\varepsilon_{ij...k}
\varepsilon^{\alpha_{1}\alpha_{2}...\alpha_{n}}(|s_{\alpha_{1}}^{(i)}
\rangle|s_{\alpha_{2}}^{(j)}\rangle....|s_{\alpha_{n}}^{(n)}\rangle).\nonumber
\end{equation}

The above discussion is only suitable for some oversimplified
system. When it comes to a general multi player game, we do not
distinguish which one is angel player and which one is devil
player. They are all rational players. In every round of the game,
the player choose one strategy from his strategy space, so do
other players. When they check their payoffs, what someone win is
what someone lose. We may call the individual loser an angel
player, and call the individual winner a devil player.

In fact, the players share the same strategy space. A wining
player takes good strategies, a losing player takes those bad
ones. It does not make any sense to talk about the good or the bad
of the strategy itself, the same strategy may bring different
results in different situations for a certain layer. Every
combinatorial series of strategies corresponds to a payoff
distribution to individual players. If we take particles as
player, a physics system always contain millions of particles, it
is impossible to calculate every particle's payoff. That is why we
need statistics physics through which we summarize the information
of payoffs into a few statistical observables, but we lost the
information on individual payoffs. We choose a more practical way
to grab the key information of many particle system. The player of
game is taken as the different interactions that govern the
microscopic behavior of particle.

A vacuum state $|0\rangle$ is defined by a state that no player
confirm his strategy. When the $i$th player has chosen his
strategy $|s_{\alpha_{1}}^{(i)}\rangle$, and place his card on the
table, it means that a strategy $s_{\alpha}$ is generated from
vacuum, i.e.,
$|s_{\alpha}^{(i)}\rangle=\hat{s}_{i\alpha}^{\dag}|0\rangle$. One
player's strategy is not isolated, it is always entangled with
other players's strategy, since whenever he makes the decision he
must reflects other player's choice, the collective strategy is
highly entangled states. The anti-strategy of the $i$th player
comes from the combination strategy he received from the rest
players, we denote it as
\begin{equation}\label{s+0=a}
\langle{S}^{i}|=\langle0|\hat{F}(\hat{s}_{\alpha_{1}},\hat{s}_{\alpha_{2}},...,\hat{s}_{\alpha_{k}},...\hat{s}_{\alpha_{n}}),\;(k\neq{i}).\nonumber
\end{equation}
When the $i$th player encounter the $j$th player, his payoff is
$u_{ij}=Tr_{k\neq{j}}[\langle{S}^{i}|s_{\alpha}^{(i)}\rangle]$. We
can further define the density matrix
$\rho=\sum_{j}p_{j}|S_{j}\rangle\langle{S}_{j}|$, and introduce
the familiar Von Neumann entropy to measure the entanglement.
\\
\\
\textsf{{{\textsl{Pairing mechanism in the war game}}} }
\\
\\
As shown in the prisoner dilemma, two altruistic players can get
their best payoff without communication and cooperate, they trust
each other and both choose the strategy for the welfare of other
people. But two selfish player can not get the best payoffs unless
they communicate and cooperate, in order to prevent the other
player's betray, they would like to bound into a pair so that any
betray would damage himself, this increases the entanglement
between them. in physics, two spin in magnetic field is a perfect
physical system to realize a prisoner dilemma. A direct physical
observation is that the particles with spin is lean to form pairs,
and the magnetic field would strengthen the entanglement between
two particles with opposite spins.

\begin{figure}
\begin{center}\label{saddle3}
\includegraphics[width=0.15\textwidth]{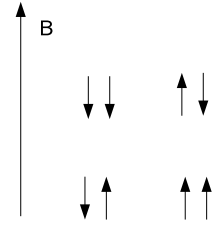}
\caption{Two spins in magnetic field is a good demonstration of
prisoner dilemma in physics.}
\end{center}
\end{figure}

A physical system to demonstrate the Prisoner's dilemma is two
spins in magnetic field. Two players are the two particles: Alice
and Bob, they both carry magnetic dipole momentum $\mu$. The
strategies: confession and accusation corresponds to spin up and
spin down, the external magnetic field is external law. Their loss
function is their energy in this system. They fight to minimize
their loss function. The payoff matrix is
\begin{equation}
\begin{tabular}{|c|c|c|c|}
  \hline
   &  & Bob & Bob \\
  \hline
  & & \;\;$\downarrow$ \;\;& \;\;$\uparrow$\;\;   \\
   \hline
Alice  &  \;\;$\downarrow$ \;\;& $(-\mu{B},\;-\mu{B})$ &$(\mu{B}-\Delta,\;\mu{B}+\Delta)$   \\
   \hline
Alice  & \;\;$\uparrow$ \;\;& $(\mu{B}+\Delta,\;\mu{B}-\Delta)$ & $(\mu{B},\;\mu{B})$   \\
  \hline
\end{tabular}\nonumber
\end{equation}
$\Delta$ is the energy shift due to the interaction between the
two particles. This is the classical picture of two-particle game.
There was experimental realization of prisoner dilemma on nuclear
magnetic resonance quantum computer \cite{jfdu}.

A general $n$-player prisoner dilemma may be summarized into a
$n$-spin Hamiltonian. A player has two strategies: spin-up and
spin-down. The payoff function is the Hamiltonian,
\begin{equation}\label{ij-ising}
H=\sum_{i}h_{i}\sigma_{z}^{i}+J^{ij}\sum_{ij}\sigma_{z}^{i}\sigma_{z}^{j}.
\end{equation}
If we take the local coupling $J^{ij}=J$ and consider only the
nearest neighbor interaction, this Hamiltonian becomes traditional
Ising model. The payoff matrix of this $n$-player prisoner dilemma
game for Ising model is diagonal block. A group of soldiers in a
war also encounter prisoner dilemma. Two soldiers helping each
other is stronger than the two that they do not help each other.

We consider a war on two dimensional lattice. The soldiers are
particles, they are divided into two equal groups, ${A}$ and
${B}$. The particles of $A$ is the anti-particle of $B$ and vice
versa. In the beginning, $A$ and $B$ are separated two parts on
the lattice. When the war break out, they rush to each other and
began the combat. A particle only fight against his nearest
antiparticles and help his nearest friend neighbors. If the
soldiers are identical particles at the same level, the
interaction between particles does not go too far, it is confined
in a local region. If we assume all the soldiers only fight
against enemy soldiers for his own survival, they do help friend
neighbors, the dynamic process only includes
\begin{equation}
\{\varepsilon_{ij}\;\hat{s}_{A_i}^{\dag}\hat{s}_{B_j},\;\;\;\varepsilon_{ji}\;\hat{s}_{B_j}^{\dag}\hat{s}_{A_i},\;\;\;A_i\in{A},B_j\in{{\bar{B}}}\},
\end{equation}
where $s$ represents the soldiers. This is a system composed of
free particle and antiparticles.  In fact, some soldiers of $G$
may help each other, they grouped into pairs to fight. So does the
soldiers of ${\bar{G}}$. Then we have to consider the pair
interactions. The war now contains three solders interaction and
four soldiers interaction,
\begin{eqnarray}
&&\varepsilon_{ij}\;\hat{s}_{A_i}^{\dag}\hat{s}_{B_j},\;\;\;\varepsilon_{ji}\;\hat{s}_{B_j}^{\dag}\hat{s}_{A_i},\nonumber\\
&&V^{3}\;\hat{s}_{A_i}^{\dag}\hat{s}_{A_k}^{\dag}\hat{s}_{B_j},\;...\;\;,\varepsilon_{ji}\;\hat{s}_{B_j}^{\dag}\hat{s}_{B_k}^{\dag}\hat{s}_{A_i},\nonumber\\
&&V^{4}\;\hat{s}_{{A_i}}^{\dag}\hat{s}_{{A_j}}^{\dag}\hat{s}_{{B_k}}\hat{s}_{{B_l}},\;\;
V^{4}\;\hat{s}_{{B_i}}^{\dag}\hat{s}_{{B_j}}^{\dag}\hat{s}_{{A_k}}\hat{s}_{{A_l}}.\nonumber\\
\end{eqnarray}
If we recall the BCS pairing mechanism in superconductor theory,
the war between electrons does not have three-particle
interaction. Even two electrons separated far from each other,
they can act as one pair. They do not even know each other, since
identical particle are indistinguishable. This interesting
phenomena in the quantum world may found some naive analogy in
war. During the war game, we can assume a pair of soldier are
confined in the lattice, they kept running from one place to
anther, but helping each other all the time. They shot any enemy
that try to kill his partner. This is a long range interaction.

There are two types of pairs, angle pair and devil pair. The angle
pair is more powerful than the devil pair as a whole. They love
each other and trust each other, so they can separated from each
other in a longer distance. But devil player have to stay closer
to prevent his partner from betraying him for his own survival,
their correlation length is shorter.

There are hierarchical structure in army. The soldiers are
organized into different groups in which a large number of them
fight as a whole. These hierarchical structure are well kept in
the region far from the coexistence boundary, but when combat
begins in one battlefield, all these hierarchical structure
becomes meaningless, the general or the captain at high plays no
different role as an ordinary soldier at the lowest level, they
are just the same individual fighters with arm. It is the
hierarchical structure or inner structure of a system that differs
it from others. When the phase transition occurs, all these
hierarchical structure are broken, men are born to equal, men are
also died to equal. This is the origin of universality class.

For a further consideration of war game, we have to take into
count of different hierarchical structures. The commanders are the
critical players, they are dressed up by a group soldiers, and
fight as a whole around the battlefield. We may take Kadanoff
block procedure to summarize all the many player interaction to
the commander. The many body interactions, such as
\begin{equation}
V^{n}\;\hat{s}_{{B_i}}^{\dag}\;...\;\hat{s}_{{B_j}}^{\dag}\;...\;\hat{s}_{{A_k}}\;...\;\hat{s}_{{A_l}},\nonumber\\
\end{equation}
are renormalized into partition function. The war is going on
between the renormalized commanders, ${S}_{B}$ and ${S}_{A}$,
where
\begin{eqnarray}
(\hat{s}_{{B_i}}^{\dag}\;...\;\hat{s}_{{B_j}}^{\dag})\rightarrow{\hat{S}}_{B},
\;\;(\hat{s}_{{A_i}}^{\dag}\;...\;\hat{s}_{{A_j}}^{\dag})\rightarrow{\hat{S}}_{A}.
\end{eqnarray}
Again we may consider the pairing interactions among these
commanders,
\begin{eqnarray}
&&V^{3}_{S}\;\hat{S}_{A_i}^{\dag}\hat{S}_{A_k}^{\dag}\hat{S}_{B_j},\;...\;\;,\varepsilon_{ji}\;\hat{S}_{B_j}^{\dag}\hat{S}_{B_k}^{\dag}\hat{S}_{A_i},\nonumber\\
&&V^{4}_{S}\;\hat{S}_{{A_i}}^{\dag}\hat{S}_{{A_j}}^{\dag}\hat{S}_{{B_k}}\hat{S}_{{B_l}},\;\;
V^{4}\;\hat{S}_{{B_i}}^{\dag}\hat{S}_{{B_j}}^{\dag}\hat{S}_{{A_k}}\hat{S}_{{A_l}}.\nonumber\\
\end{eqnarray}
Repeating this renormalization group transformation, we transform
the sophisticate many body problems into something we can handle.
This is the fundamental spirit of various numerical
renormalization group method developed in quantum many body
system.

The phase boundary appears when the fighting groups reach a
balance at certain frontiers regions. Friend and enemy are
maximally entangled in the coexistence state. One can not tell who
is who. In order to study the symmetry transformation at the
coexistence boundary, we introduce the position displacement
operator and its conjugate momentum operator,
\begin{equation}\label{xp}
X_{j}=\frac{\hat{S}_{j}^{\dag}+\hat{S}_{j}}{\sqrt{2}},\;\;\;
P_{j}=\frac{i(\hat{S}_{j}^{\dag}-\hat{S}_{j})}{\sqrt{2}}.
\end{equation}
Then the angular momentum operator is given by
$L_{ij}=(X_{i}P_{j}-X_{j}P_{i})$. The generator of the symmetry
transformation may be denoted as
$\hat{L}={i\sum_{ij}\theta{L}_{ij}}$, through which we derived the
Lie group element
\begin{equation}\label{Uxp}
U=e^{i\theta\hat{L}}=e^{{-i\sum_{ij}\theta(\hat{S}_{i}^{\dag}\hat{S}_{j}^{\dag}
-\hat{S}_{i}\hat{S}_{j})}}.
\end{equation}
The quantum output vector
$\hat{O}_{ut}(\hat{S}_{i},\hat{S}_{j},...)$ is expressed by the
operator of renormalized soldiers. Then we can detect the phase
transition by the transformation
\begin{eqnarray}\label{game-l-Uexp-p}
U_{(p)}\hat{O}_{ut}{U}_{(p)}^{-1}=e^{i\theta\cdot{\hat{L}}}\hat{O}_{ut}e^{-i\theta\cdot{\hat{L}}}
\end{eqnarray}
This transformation equation determined the evolution of the
quantum output in the vicinity of phase coexistence region. Inside
one phase, or inside one army, this transformation is continuous
up to infinite order. When never the symmetry lost, an enemy
popped out and initiated the revolution.

The uncertainty principal shows up across the whole phase diagram.
In the phase diagram, the war game is in different squeeze state
for different stage of combat. The transformation operator
(\ref{Uxp}) is a special squeeze operator. A general squeeze
state\cite{scully} generated from vacuum is
\begin{equation}\label{Uxpstate}
|\psi\rangle=e^{{-i\sum_{ij}(\theta\hat{S}_{i}^{\dag}\hat{S}_{j}^{\dag}
-\theta^{\ast}\hat{S}_{i}\hat{S}_{j})}}|0\rangle.
\end{equation}
When the two players began rushing toward each other but still do
not touch each other, their fluctuation of spacial position is
much higher than that of their force. When they meet in the phase
coexistence boundary, the fluctuation of spatial position is
around the boundary, it is much lower than the conflicting force.
At the Nash equilibrium state, the phase boundary is almost
stationary, but both of the two players summon up their maximal
force to against each other, they reach a dangerous equilibrium
state on the boundary. This is an extremal squeezed state, the
fluctuation of position is highly squeezed, but the fluctuation of
conflicting force is very strong.

So we may take two Hermitian operators $\gamma$ and $P_{\gamma}$,
$\gamma$ denotes the position in phase diagram, $P_{\gamma}$
represents the corresponding force. The commutation relation
derive from the second quantization is $[\gamma,P_{\gamma}]=iC$.
We can calculate the uncertainty of an operator $A$ by
$(\Delta{A})^{2}={\langle\psi|A^{2}|\psi\rangle-(\langle\psi|A|\psi\rangle)^{2}}$.
According to Heisenberg uncertainty principal,
$\Delta{\gamma}\Delta{P_{\gamma}}\geq\frac{1}{2}|C|$. The
uncertainty of the position operator is
\begin{equation}
\Delta{\gamma}\ll\frac{1}{2}|C|
\end{equation}
at the coexistence state. While the uncertainty of corresponding
momentum is much stronger $\Delta{P_\gamma}\gg\frac{1}{2}|C|$.

The angel pairing and devil pairing are two stable squeezed
states. The angle pairing has less uncertainty about their
trustworthiness, so they have larger uncertainty on their spatial
connection. But the devil pair has larger uncertainty on his
partner's trustworthiness, so they strengthen their spatial
connection, this reduced their spatial uncertainty. No matter
which kind of pairing it is, the total information in the two
pairing states should be conserved. If we can make sure a system
doe not lose information, we can maintain the pairing to a stable
level. Superconductor is such a system, below the critical
temperature, no chaos invades into it, no information escapes out
of it. There is a constant electric flow.

According to quantum field theory, symmetry means conserved
quantity, a loss of symmetry indicates a loss of conserved
quantity. What the old phase lost becomes the generator of the new
phase. The squeeze transformation on output field is a way to
detect the loss of information. Imbalance is the original engine
of development. For physicist, the physical measurement always
affect the output state itself. We can not detect a state without
changing it, this minor change is neglectable when it is done in a
stable phase. But at critical region, the minor effect of
detection might induce significant effect. In other words, when a
war is going on, once the spy step into the frontier, he is force
to fight for his own survival, it would be too hard for him to
sent out any information.

\section{summary}

{\noindent}(1){\texttt{game theory of general phase transition and
   $\;$Renormaization group transformation}}
\\
\\
Phase transition is a much more universal phenomena than the
conventional phase transitions in physics. Similar sudden changes
arise from evolution of biomolecule, self-organization in
nanoscale systems, cosmic evolution, and so forth. The general
phase transition breaking the envelop of physics can be viewed as
a non-cooperative game of many players. The players try different
strategies to win. Whenever a winning player fails and becomes a
loser, a phase transition would occur.

When the players represent different interactions between the
elements of a complex system, a winning player is a dominant
interaction which governs one stable phase of the system, the
other phases represented by loser's interaction are suppressed.
When we tune the interaction parameters, we are changing the
strategies of the players, the winning interaction may becomes
weaker and weaker. There is a critical point at which the losing
players grows stronger enough to balance the previous winning
player, then we arrived the Nash equilibrium point. At this point
, no one wins, and no one losses, it is the coexistence point of
the new phase and old phase. The Nash equilibrium point is an
unstable saddle point, any tiny deviation would decided the fate
of the coexisting new phase and old phase, there is only one
winner left to dominate the phase structure of the system.

Usually the Nash equilibrium of a war game is not derived in one
round of combat. The players have to play many rounds of
negotiation to find the optimal strategy. The renormalization
group transformation theory is actually one theoretical
description of this kind of game process in physics. In physics,
the players are the interaction parameter or physical parameter.
For example, we take the Ising model as a game, the two players
are the spin coupling interaction and external magnetic field. The
stable phase is determined by the dominant interaction.

At the first round of renormalization group transformation, the
two player accomplished the first round of combat, they made an
agreement on the most part of boundary between their domains. But
the unsettled part is occupied by the winner. So the loser
initiated the second round of renormalization group transformation
to get some of his losing back. As this transformation goes on,
the unsettled boundary between the players becomes less and less.
Finally, they reach the Nash equilibrium point, that is the fixed
point of the renormalization group transformation. This point is
where the phase transition occurs.

Usually to obtain the exact critical point, one has to perform the
renormalization group transformation to infinite order. However,
it is too far to reach in reality, we always cut it off at certain
order. It inevitably gives birth to a loser and a winner to
certain order. Although the players at that order maybe are
fighting for only one thousandth of a penny, but the loser is
loser, winner is winner, the title is  fixed. The order of cutoff
is the order of phase transition. we can take renormalization
group transformation on output function to observe different order
of phase transition. When we apply this definition to
thermodynamics physics, it naturally unified Ehrenfest definition.
\\
\\
{\noindent}(2)\texttt{{Universal coexistence equation and
topological conjecture of scaling laws in fractal space}}
\\
\\
The game is played on the strategy base manifold, so a phase
transition is related to the topology of the strategy base
manifold. If the strategy space is noncompact, we do not need to
consider topology at all, in that cases, the phase transition
governed by game dynamics. We may call it a non-topological phase
transition. When the strategy manifold is compact, there would be
a topological constrain on the player's strategies, then the
topological effect would appear in phase transition, we call it a
topological phase transition.

The topology is intimately related to the fixed points of
renormalization group transformation flow on strategy manifold. We
introduced the general flow vector field, and showed that the sum
of the winding number around the surviving strategies is a
topological number. These surviving strategies with opposite
charges annihilate at the universal coexistence curve, whose
equation is $\{\phi^{1},\phi^{2},...,\phi^{2}\}=0$, $\phi^{i}$ are
the component of the flow vector field of output field. In the
thermodynamics physics, the output field may be chosen as free
energy, the players are temperature and pressure, one can verify
that this coexistence equation unified all the coexistence
equations at different order classical phase transition. The
phases separated by the coexistence curve are assigned with
different winding number of opposite sign.

The strategy space is a fractal dimensional space. It is easy to
see this point from a war game. We focus on the battlefield, it is
a war between two armies. If we look closer, it is the war between
corps. We may continue the magnify the battle field, one would
see, it is a war at all different scales, it is the combat between
two individual soldiers at any local region. A war game may be
summarized into the war between two commanders by Kadanoff-block
procedure. There is an intrinsic fractal structure around the
critical point. So we make the hypothesis that a local
neighborhood on the strategy base manifold is homeomorphic to a
fractal dimensional space. Then the output field may be expanded
by the polynomial of power functions with fractal dimension index.
In analogy with the observables defined in the statistical
mechanics, we introduce the tangent vector field of the output
field as observables. These tangent vector field may be
approximated by fractal polynomials in the vicinity of Nash
equilibrium solution. By substituting these fractal polynomials
into the universal coexistence equation, one can derive the
scaling relations. As we shown, the coexistence operator has a
degenerated subspace, the coexistence equation bear a topological
origin, so these scaling relations are universal.
\\
\\
{\noindent}(3)\texttt{{Many body physics of games}}
\\
\\
Game theory is actually one different way to see many body system.
For example, a collection of millions of ultralcold atoms trapped
in optical lattice can be described by Bose-Hubbard model. There
are two key parameter, the hopping parameter $J$ and the on site
repulsive interaction $U$. We can model this system as a game
between $U$ and $J$. Player $J$ command the particles hoping from
one site to another, drive them into superfluid phase. While $U$
direct the soldiers to stick to the lattice site. If $J$ wins, the
ultracold atoms becomes superfluid. If $U$ wins, the ultracold
atoms are confined in the lattice, they form Mott-insulator phase.
The transition point is the Nash equilibrium point of this game.
The two phases coexist at the critical point. Another way to model
the millions of ultracold atoms is to take every atom as a player,
and their states at lattice sites are strategy space, this
reproduced the standard quantum many body theory.

A practical approach to a given many body system is first to find
out the main different interactions that are players of the game.
Secondly, we choose some output quantity to measure the states.
Physicist usually take the external response function, these
output field are the payoff function of the game. The tangent
vector field established on the hypersurface of these response
function is the fundamental vector field to study the topological
effect of the strategy space. Applying the universal coexistence
equation, we can find out the basic structure of different phases.
This scheme holds both for classical systems and quantum systems.

Game theory only provide us a mathematical frame work to
understanding the general property of a system. The topological
phase transition theory developed in this paper is a general
mathematical results, it does not depend on any specific model in
physics, biology or social system. When it comes to some specific
physical system, we may apply the conventional theory for that
special system to derive the specific relation between input
parameters and output field, or one may directly conduct some
experiments to find some approximated relation. As long as one
derived the specific relation between output and input, just
substitute it into the universal coexistence equation, one can get
basic phase diagram.

In fact, quantum many body theory may provide us new understanding
to many player games. We developed the density matrix theory of
many player game, it was shown there is a one-to-one
correspondence between fixed point of many player game and the
self-character vector of the player. Every quantum phase may be
represented by a state function, these state functions are the
players of multi-player game. A player's strategy must take into
account of the other players's strategy. So game theory is born to
be an entangled theory. This may help us to understand the
entanglement among coexistence phases. We find a new quantity to
measure the entanglement of different phases. The maximal point of
entanglement is the Nash equilibrium point at which different
phases coexist. The maximal entanglement between phases is
controlled by interaction parameter instead of time, so the
quantum entanglement between different quantum phases are good
candidates for quantum computation.

Further more, if we take two particles as the two players of
prisoner dilemma, it would be found that there are two types of
entangled pairs: the angel pair and devil pair, the angel pair
love each other, their entanglement is a little bit weaker. While
the devil players do not trust each other, they inclined to
bounded together to prevent the other from betray, it is a strong
entangled pair. This may help us to understand pairing mechanism
in superconductor.

\section{Acknowledgment}

The author expresses his gratitude to professor Yue Yu for his
support. This work was supported by the National Natural Science
Foundation of China.

\appendix

\section{Ehrenfest's definition about order of phase transition}
The zeroth order phase transition defines that the free energy of
the two phases $F(T, P)$ are not equal,
\begin{equation}
F_{A}(T,P){\neq}F_{B}(T,P).
\end{equation}
For the first order phase transition, the free energy of the two
phases $F_{A}$ and $F_{B}$ is continuous, but the first order
derivative is not continuous,
\begin{equation}
F_{A}(T,P)=F_{B}(T,P),\;\;\;\frac{dF_{A}}{dT}\neq\frac{dF_{B}}{dT},\;\;\frac{dF_{A}}{dP}\neq\frac{dF_{B}}{dP}.
\end{equation}
The second order phase transition is defined by the discontinuity
of the compressibility, susceptibility,
\begin{equation}
C_{p}^{A}\neq{C_{p}^{B}},\;\;\alpha^{A}\neq\alpha^{B},\;\;\kappa^{A}\neq\kappa^{B},
\end{equation}
in which the specific heat $C_{p}$, $\alpha,\kappa$ are defined
as:
\begin{eqnarray}
&&C_{p}=T(\frac{dS}{dT})_{P}=-T\frac{\partial^{2}F}{dT^{2}},\nonumber\\
&&\alpha=\frac{1}{V}(\frac{\partial{V}}{\partial{T}})_{P}=\frac{1}{V}\frac{\partial^{2}F}{\partial{T}\partial{P}},\nonumber\\
&&\kappa=-\frac{1}{V}(\frac{\partial{V}}{\partial{P}})_{P}=-\frac{1}{V}\frac{\partial^{2}F}{\partial{P}^{2}}.
\end{eqnarray}
The higher order of phase transition is defined from the
discontinuity of higher order derivative of free energy. We
rewrite Ehrenfest's definition into more compact formulism. The
$p$th-order quantum phase transition is characterized by a
discontinuity in the $p$th derivative of difference of free energy
$\delta{F}$,
\begin{eqnarray}\label{00-F-Qporder}
&&\partial^{p-1}_{\gamma_{1}}\delta{F=0},\;\;\partial^{m}_{\gamma_{1}}\partial^{p-m-1}_{\gamma_{2}}\delta{F=0},(m=1,2,...,p-1),\nonumber\\
&&\partial^{p-1}_{\gamma_{2}}\delta{F=0}.\nonumber\\
&&\partial^{p}_{\gamma_{1}}\delta{F\neq0},\;\;\partial^{m}_{\gamma_{1}}\partial^{p-m}_{\gamma_{2}}\delta{F\neq0},(m=1,2,...,p),\nonumber\\
&&\partial^{p}_{\gamma_{2}}\delta{F\neq0}.
\end{eqnarray}
If the $p$th derivative of $\delta{F}$ becomes continuous, the
phase transition jumps to the $p+1$th order.

\section{Differential geometry of free energy}

We demonstrate the basic conception of differential geometry on
the free energy manifold expanded by two thermal variables
temperature $T$ and pressure $P$.

In differential geometry, the open set $\delta{F}(T,P)$ in a
manifold may be mapped to an open set in three dimensional
Euclidean space through the homeomorphic mapping
$f_{\delta{F}}:\delta{F}(T,P)\rightarrow\textbf{X}_{\delta{F}}$,
$\textbf{{X}}=X_{1}(\gamma_{1},\gamma_{2})i+X_{2}
(\gamma_{1},\gamma_{2})j+X_{3}(\gamma_{1},\gamma_{2})k$, At each
point $\gamma^{0}$ in $\delta{F}$ there exist a set of tangent
vectors. The two basis of the tangent vector space on this
manifold are
\begin{equation}\label{e1e2}
e_{1}=\frac{\partial}{\partial{\gamma_{1}}}=\frac{\partial}{\partial{T}},\;\;
\;\;e_{2}=\frac{\partial}{\partial{\gamma_{2}}}=\frac{\partial}{\partial{T}},
\end{equation}
here we have defined $\gamma_{1}=T$(temperature) and
$\gamma_{1}=P$(pressure). An arbitrary tangent vectors may be
expressed as
\begin{equation}\label{00-tangent}
{\textbf{X}}_{\gamma_{i}}=\frac{\partial{\textbf{{X}}}}{\partial{\gamma_{i}}},\;\;
{\textbf{X}}_{\gamma_{i}\gamma_{j}}=\frac{\partial^{2}\textbf{{X}}}{\partial{\gamma_{i}}\partial{\gamma_{j}}}
\end{equation}
All the tangent vectors to the surface $\delta{F}$ at x denoted
form a tangent vector space denoted by $T_{x}\delta{F}$, it is
parallel to the tangent plane to $\delta{F}$ at x.

The metric tensor is given in a bilinear form by the inner product
of two basis vectors, $g_{ij}=<e_{i},e_{j}>$. Then the first
fundamental form is
\begin{eqnarray}\label{1st-funda-form}
&&E={\textbf{X}}_{\gamma_{1}}\cdot{\textbf{X}}_{\gamma_{1}},\;\;\;F'={\textbf{X}}_{\gamma_{1}}\cdot{\textbf{X}}_{\gamma_{2}},
\;\;G={\textbf{X}}_{\gamma_{2}}\cdot{\textbf{X}}_{\gamma_{2}},\nonumber\\
&&g_{11}=E,\;\;\;g_{12}=g_{21}=F',\;\;\;g_{22}=G.
\end{eqnarray}
The second fundamental form is
\begin{eqnarray}\label{2nd-funda-form}
&&L={\textbf{X}}_{\gamma_{1}\gamma_{1}}\cdot{\textbf{n}}=-{\textbf{X}}_{\gamma_{1}}\cdot{\textbf{n}_{\gamma_{1}}}\nonumber\\
&&M={\textbf{X}}_{\gamma_{1}\gamma_{2}}\cdot{\textbf{n}}=-{\textbf{X}}_{\gamma_{1}}\cdot{\textbf{n}_{\gamma_{2}}}\nonumber\\
&&N={\textbf{X}}_{\gamma_{2}\gamma_{2}}\cdot{\textbf{n}}=-{\textbf{X}}_{\gamma_{2}}\cdot{\textbf{n}_{\gamma_{2}}}.
\end{eqnarray}
with $\textbf{n}$ as the normal vector of the surface,
\begin{equation}\label{n-normal}
\textbf{n}=\frac{{\textbf{X}}_{\gamma_{1}}\times{\textbf{X}}_{\gamma_{1}}}
{|{\textbf{X}}_{\gamma_{1}}\times{\textbf{X}}_{\gamma_{1}}|}.
\end{equation}
Considering the second order of phase transition, the first
fundamental form and the second fundamental form includes more
information about the variation of the free energy than
Ehrenfest's definition. It is more reasonable to choose the
Gaussian curvature,
\begin{equation}\label{1/2=curvature}
\Omega=\frac{LN-M^{2}}{EG-{F'}^{2}},
\end{equation}
It is positive for spheres, negative for one sheet hyperbolic
surface and zero for planes. The principal curvatures,
$\kappa_{1}(\gamma^{0})$ and $\kappa_{2}(\gamma^{0})$, of
${\textbf{X}}$ at ${\textbf{X}(\gamma^{0})}$ are defined as the
maximum and the minimum normal curvatures at
${\textbf{X}(\gamma^{0})}$, respectively. The directions of the
tangents of the two curves that are the result of the intersection
of the surface ${\textbf{X}(\gamma^{0})}$ and the planes
containing ${\textbf{n}(\gamma^{0})}$. Then
\begin{equation}\label{omega=k1k2}
\Omega({\gamma_{1},\gamma_{2}})=\sum_{p}\Omega({\gamma_{1},\gamma_{2}})|_{p}=\sum_{p}\kappa_{1}({\gamma_{1},\gamma_{2}})\kappa_{2}({\gamma_{1},\gamma_{2}})|_{p}.
\end{equation}
The cross section of the curvature ${\textbf{X}}$ at $\gamma^{0}$
may be expanded as parabola by ignoring the higher order,
\begin{equation}\label{x-cross}
X=\frac{\kappa_{1}}{2}r_{1}^{2}+\frac{\kappa_{2}}{2}r_{2}^{2}.
\end{equation}
where
$r_{i}=(d\gamma_{1}+\frac{1}{2}\Gamma_{ij}^{1}d\gamma_{i}d\gamma_{j})\sqrt{g_{ii}}+\cdots$.
When $\kappa_{1}>0$ and $\kappa_{2}>0$, it is a elliptic surface,
when $\kappa_{1}>0$ and $\kappa_{2}=0$, it is parabolic, for
$\kappa_{1}>0$ and $\kappa_{2}<0$, it is hyperbolic.

The covariant derivative may be defined as
\begin{equation}\label{A-gamma}
D_{\gamma_{i}}\phi^{j}=\partial_{\gamma_{i}}\phi^{a}+\Gamma_{ik}^{j}\phi^{k},
\end{equation}
here we have introduced the Christoffel-Levi-Civita connection
$\Gamma_{ij}^{k}$, which is defined as
\begin{equation}\label{riemman-connection}
\Gamma_{ij}^{k}=\frac{1}{2}g^{kl}(\frac{\partial{g_{il}}}{\partial\gamma_{j}}
+\frac{\partial{g_{jl}}}{\partial\gamma_{i}}+\frac{\partial{g_{ij}}}{\partial\gamma_{l}}).
\end{equation}

\section{Definition of game theory}

A game is an abstract formulation of an interactive decision
situation with possibly conflicting interests, it involves of a
number of agents or players, who are allowed a certain set of
moves or actions. The payoff function specifies how the players
will be rewarded after they have performed their actions. Let $n=
1,\ldots,N$ denote the players; The $i$th player's strategy,
$S_{i}$, is her procedure for deciding which action to play,
depending on her information. The strategy space of the $i$th
player $\mathcal{S}_{i}$=$\{e_{i1},e_{i2},\ldots,e_{im}\}$ is the
set of strategies available to her. A strategy profile $(
s_1,s_2,\ldots,s_N)$ is an assignment of one strategy to each
player. $u_i(s_1,\ldots,s_N)$ is player $i$'s payoff function
(utility function), i.e., the measure of her satisfaction if the
strategy profile $(s_1,\ldots,s_N)$ gets realized.

The strategy assigned to each player could be pure state or mixed
state. A mixed strategy is a probability distribution over pure
strategies. Playing a mixed strategy means, the players come up
with one of her feasible actions with a pre-assigned probability.
Each mixed strategy corresponds to a point $\textbf{P}$ of the
mixed strategy simplex.
\begin{equation}\label{mixstrategy}
M_{\textbf{p}}=\{\textbf{P}=(p_{1},\ldots,p_{m})\in{\mathbb{R}^{m}}:p_{q}\geq0,\sum_{q=1}^{m}p_{q}=1\},
\end{equation}
whose corners are the pure strategies. For the case of two
players, a strategy profile is a pair of probability vector
$(\textbf{p},\textbf{q})$ with $\textbf{p}\in{M_\textbf{p}}$ and
$\textbf{q}\in{M_\textbf{q}}$. The expected payoffs of player 1
and 2 are expressed as
\begin{eqnarray}\label{u1u2}
&&u_{1}(\textbf{p},\textbf{q})=\textbf{p}\cdot{\textbf{A}}\textbf{q}={\sum^{m}_{ij}}p_{i}A_{ij}q_{j},\\
&&u_{2}(\textbf{p},\textbf{q})=\textbf{p}\cdot{\textbf{B}}\textbf{q}={\sum^{m}_{ij}}p_{i}B_{ij}q_{j}.
\end{eqnarray}
This payoffs function relies on the entangled strategies of the
two players.

The essence of a game is the payoff function defined over
different strategy profile of the players. In the language of
quantum statistics, the strategy space of the $i$th player $S_{A}$
is the Hilbert space of the $i$ particle.

\section{Differential game}\label{dif-game}

The differential game is applied to model dynamic
conflicts\cite{lewin}, such as the labor-employers relations in
economic processes, the bull and fighter in bull-fighting. The
labor and employer are called players in differential game. We
take the two player game as an example. There are two players, one
is called Alice, the other player is Bob. Alice has a strategy
space ${S}_{A}(s_1,s_2,\ldots,s_N)$, and Bob's strategy space is
${S}_{B}$. We introduce the state vector $\vec{x}$ which
characterize the conflict to convert a real life conflict to a
differential game model. For example, Alice and Bob are concerned
with the motion of a point in a plane, the state vector denotes
the location of the point $x(t)$ at the time t. All possible state
vector is a subset of a $n$-dimensional Euclidean space.

The influence of the decisions of the evolution of the state is
described by the equations of motion,
\begin{equation}\label{qdxdt}
\dot{x}=f(\textbf{{x}},\textbf{{s}}_{A},\textbf{{s}}_{B}).
\end{equation}
When the two players realize their strategy
$(\textbf{{s}}_{A},\textbf{{s}}_{B})$ at time $t$, the outcome of
the differential game is functional on the state space,
$U[\textbf{{x}},\textbf{{s}}_{A},\textbf{{s}}_{B}]$. If the
outcome satisfies for a pair of strategy
$(\textbf{{s}}^{\ast}_{A},\textbf{{s}}^{\ast}_{B})$ satisfies,

\begin{equation}\label{pi-J-saddle}
U[\textbf{{x}},\textbf{{s}}^{\ast}_{A},\textbf{{s}}^{\ast}_{B}]
\leq{U[\textbf{{x}},\textbf{{s}}^{\ast}_{A},\textbf{{s}}^{\ast}_{B}]}
\leq{U[\textbf{{x}},\textbf{{s}}_{A},\textbf{{s}}^{\ast}_{B}]},
\end{equation}
then this strategy pair is the optimal play for the two players,
${U[\textbf{{x}},\textbf{{s}}^{\ast}_{A},\textbf{{s}}^{\ast}_{B}]}$
is called the optimal outcome at $\textbf{{x}}$. The optimal
outcome is a function in the Euclidean space, it is called the
value function,
\begin{equation}\label{value-optimal}
J(\textbf{{x}})={U[\textbf{{x}},\textbf{{s}}^{\ast}_{A},\textbf{{s}}^{\ast}_{B}]}.
\end{equation}
There is a theorem states that if the value function of a
differential game exists it is unique\cite{lewin}. The saddle
relation implies that a solution of a differential game is a Nash
equilibrium. Neither player Alice nor player Bob can improve their
guaranteed results.

\section{The continuous-time infinite dynamic game}\label{inf-game}

We present the basic conception of the continuous-time infinite
dynamic game in this section. Suppose for a given many-body
system, there are $n$ player which are denoted as
$N=\{{1},{2},...,{n}\}$, $\psi$ is the state function of this
game, the state space is an entangled space of many Hilbert space.
The evolution equation of this game is governed by the
differential equation,
\begin{equation}\label{dynamicgamepsi}
i\frac{d\psi}{dt}
=f(t,\psi,\gamma_{1}(t),\gamma_{2}(t),...,\gamma_{n}(t)).
\end{equation}
$\gamma_{1},\gamma_{2},...,\gamma_{n}$ is the strategy profile of
the $n$ players. $\gamma_{i}$ are real or complex numbers. A cost
function of the game is a map from the strategy space
$\Gamma=\Gamma_{1}\otimes\Gamma_{1}\otimes...\otimes\Gamma_{n}$ to
a real number, $E^{i}:\Gamma\rightarrow{\textbf{R}}(i\in{N})$ for
each fixed initial strategy profile
$\gamma^{0}_{1},\gamma^{0}_{2},...,\gamma^{0}_{n}$. The minimum
cost-to-go from any initial state and any initial time is
described by the so-called value function which is defined by
\begin{equation}\label{V(tx)}
V(t,\psi)=\min_{\gamma(t)}[\int_{t}^{T}g(s,\psi(t),\gamma(t))ds+q(T,\psi(T))],
\end{equation}
satisfying the boundary condition $V(T,\psi)=q(T,\psi)$, $g$ is a
map from $g:(t,\psi(t),\gamma(t))\rightarrow{\textbf{R}}$. The
application of the principal of optimality leads to the
Hamilton-Jacobi-Bellman equation,
\begin{equation}\label{HJB}
-\frac{\partial{V(t,\psi)}}{\partial{t}}=\min_{\gamma}[\frac{\partial{V(t,\psi)}}{\partial{\psi}}
f(t,\psi,\gamma)+g(t,\psi,\gamma)].
\end{equation}
A theorem states that, if a continuously differentiable function
$V(t,\psi)$ can be found that satisfies the
Hamilton-Jacobi-Bellman equation, with corresponding boundary
condition $V(T,\psi)=q(T,\psi)$, then it generates the optimal
strategy through the static minimization problem defined by the
right hand side of Eq. (\ref{HJB}). We introduce the Hamilton
function
\begin{equation}\label{Hami-HJB}
\mathcal{H}(t,\psi,\gamma)=\frac{\partial{V(t,\psi)}}{\partial{\psi}}f(t,\psi,\gamma)+g(t,\psi,\gamma),
\end{equation}
the minimizing $\gamma$ will be denoted by $\gamma^{\ast}$, then
\begin{equation}\label{Hami+V=0}
\mathcal{H}(t,\psi,\gamma^{\ast})+\frac{\partial{V(t,\psi)}}{\partial{t}}=0,
\end{equation}
The conjugate momentum of $\psi$ is
$p(T)=\frac{\partial{V(T,\psi^{\ast})}}{\partial{\psi}}$. For a
stochastic differential game, the Hamilton function is
\begin{equation}\label{sto-Hami-HJB}
\mathcal{H}_{s}(t,\psi,\gamma)={\nabla_{\psi}{V(t,\psi)}}f(t,\psi,\gamma)+g(t,\psi,\gamma),
\end{equation}
the Hamilton-Jacobi-Bellman equation reads
\begin{equation}\label{sto-HJB}
\frac{\partial{V(t,\psi)}}{\partial{t}}+\frac{1}{2}\sigma^{ij}\frac{\partial^{2}V}{\partial\psi_{i}\partial\psi_{j}}
+\min{\mathcal{H}_{s}(t,\psi,\gamma)}=0.
\end{equation}
For a given two person game, suppose that the strategy pair
$(\gamma_{1}^{\ast},\gamma_{2}^{\ast})$ provides a saddle-point
solution, and $\psi^{\ast}(t)$ denoting the corresponding state
trajectory, then the Hamilton function satisfies,
\begin{equation}\label{saddl-hami-HJB}
\mathcal{H}(t,\psi^{\ast},\gamma_{1}^{\ast},\gamma_{2})\leq
\mathcal{H}(t,\psi^{\ast},\gamma_{1}^{\ast},\gamma_{2}^{\ast})\leq
\mathcal{H}(t,\psi^{\ast},\gamma_{1},\gamma_{2}^{\ast}),
\end{equation}
$(\gamma_{1}^{\ast},\gamma_{2}^{\ast})$ is the solution of the
Issac equation $\min\max\mathcal{H}=-{\partial{V}}/{\partial{t}}$,
$V(\psi)$ is the value function. It represents the coexistence
hyper-surface extended in $\psi$ space.

\section{Brief introduction to cooperative game}\label{cooperative}

Coalition is a crucial ingredient in a $n$-person cooperative
game. When a subset $g$ of the $n$ player forms a coalition and
all its members act together, we assign a real number $E(g)$ to
each possible coalition, $E(g)$ measures the payoff when this
coalition acts together. If we can find an imputation vector
$\varepsilon=(\varepsilon_{1},\varepsilon_{2},...,\varepsilon_{n})$
with real components, which satisfies
$\varepsilon_{i}\geq{E(\{i\})},\forall{i\in{N}}$ and
$\varepsilon_{1}+\varepsilon_{2}+...+\varepsilon_{n}=E(N)$, it is
the Pareto optimality. An imputation $x$ represents a realizable
way that the $n$ player can distribute the total payoff $E(N)$.

Usually, two coalitions $S$ and $T$ are inclined to united, if
their union brings them better payoffs, this is the
superadditivity of a game, $E(S\cup{T})\geq{E(S)+E(T)}$. Suppose a
coalition $S$ has $m$ players, usually there is a difference
between the payoff of the coalition and the sum of payoff of each
individual in this coalition, the excess is defined by,
\begin{equation}\label{excesss}
e(S,\vec{\varepsilon}_{a})=E(S)-\sum_{i\in{S}}\varepsilon_{i},
\end{equation}
A $n$-person cooperative game has $2^{n}$ subsets, i.e., it has $
2^{n}$ coalitions. Let
$D_{2^{n}-2}:\mathbb{R}^{2^{n}-2}\rightarrow\mathbb{R}^{2^{n}-2}$
be a mapping which arranges elements of a ${2^{n}-2}$-dimensional
vector in order of decreasing magnitude. For a certain set
$\vec{\varepsilon}$ of payoff vector, the nucleolus over
$\vec{\varepsilon}$ is the solution minimizing the vector of the
excesses,
\begin{widetext}
\begin{equation}\label{Nuceous}
N_{c}(N,E,\vec{\varepsilon})=\{\varepsilon_{a}\in\vec{\varepsilon}|D_{2^{n}-2}(e(S_{1},\varepsilon_{a}),...,e(S_{2^{n}-2},\varepsilon_{a}))
{\leq}D_{2^{n}-2}(e(S_{1},\varepsilon_{b}),...,e(S_{2^{n}-2},\varepsilon_{b}))\},\;\forall{b\in{n}}.
\end{equation}
\end{widetext}
In a game $(N,E)$, the $\epsilon$-core $C_{\epsilon}(N,E)$ is the
set of all pre-imputations ${\varepsilon}$, satisfying that all
the excess function are not greater than $\epsilon$, i.e.,
\begin{equation}\label{core}
C_{\epsilon}(N,E)=\{\vec{\varepsilon}_{a}\in\varepsilon(N,E)|e(S,\vec{\varepsilon}_{a})\leq\epsilon,\forall{S\subset{N}}\}.
\end{equation}
where the set of pre-imputations $\varepsilon(N,E)$ is
$\varepsilon(N,E)=\{\vec{\varepsilon}_{a}\in{R^{n}}|\sum_{i}{\varepsilon}_{ai}=E(N),\varepsilon_{ai}\geq0\}.$

The most important quantity we want to find is the ground state
payoff vectors $\vec{\varepsilon}_{g}$, i.e., the least core
$C_{\epsilon_{g}}(N,E)$, which satisfies
$e(S,\vec{\varepsilon}_{g})\leq\epsilon_{g}=\min\max{e(S,\vec{\varepsilon})}$.
In fact, this is equivalent to the self-content mean field theory.
The ground state payoff vector can be obtained by solving the
following linear programming problem, which minimizes the maximal
excesses:
\begin{eqnarray}\label{algorithm}
&&minimize\;\;\;\; \epsilon_{g}\\
&&subject\;\;to\;\;\;\;E(S)-\sum_{i}{\varepsilon}_{i}\leq\epsilon_{g},\forall{S\subset{N}},\\
&&{\varepsilon}_{1}+{\varepsilon}_{2}+...+{\varepsilon}_{n}=E(N),\\
&&{\varepsilon}_{i}\geq0,i=1,2,...,n.
\end{eqnarray}
There is a lemma says the unique payoff vector
${\varepsilon}^{\ast}$ minimizing ${\varepsilon}$ can always be
determined by at most $n$ steps in algorithm
computation\cite{lewin}.

\section{Prisoner dilemma}\label{prisonerdilemma}

The Prisoner's dilemma is the most famous paradigm to study game
theory. It occurs between two prisoners: Alice and Bob, who are
accomplices to a crime which leads to their imprisonment. Each has
to choose between the strategies of confession  or accusation. If
neither confesses, moderate sentences (a years in prison) are
handed out. If Alice confesses and Bob accuses him, Bob is free (0
years in prison) and Alice is sentenced to $c>a$ years in prison.
If both confess, they will each have to serve $b$ years in prison,
where $a<b<c$.

\begin{tabular}{|c|c|c|c|}
  \hline
   &  & Bob & Bob \\
  \hline
  & & \;\;Not Confesses \;\;& \;\;Confesses\;\;   \\
   \hline
Alice  &  \;\;Not Confesses \;\;& (a,\;a) & (c,\;0)   \\
   \hline
Alice  & \;\;Confesses \;\;& (0,\;c) & (b,\;b)   \\
  \hline
\end{tabular}

If there is no cooperation and communication between the players,
each of the players choose the strategies to minimize his losses
as far as possible. We suppose the players Alice and Bob choose
their strategies using their loss function $u_{A}$ and $u_{B}$
from ${S_{A}\times{S_{B}}}$ to $\mathbb{R}$. The two players
choose the strategies to minimize the biloss mapping
\begin{equation}\label{biloss}
\textbf{u}(s_{A},{s_{B}}):=(u_{A}(s_{A},{s_{B}}),u_{B}(s_{A},{s_{B}}))\in\mathbb{R}^{2}.
\end{equation}
A consistent pair of strategies following the constrain
\begin{eqnarray}\label{barf}
\bar{f}_{A}({s_{B}}):=\{\bar{s}_{A}\in{S_{A}}|u_{A}(\bar{s}_{A},{s_{B}})=inf\;
u_{A}(s_{A},{s_{B}})\}.\nonumber\\
\bar{f}_{B}({s_{A}}):=\{\bar{s}_{B}\in{S_{B}}|u_{B}(\bar{s}_{B},{s_{A}})=inf\;
u_{B}(s_{A},{s_{B}})\}.
\end{eqnarray}
is called a non-equilibrium (Nash equilibrium). In the prisoner's
dilemma, the Nash equilibrium is that both Alice and Bob
confesses, they both spend $b$ years in prison.

Besides the Nash equilibrium, there is still a better strategies
for them. If Alice and Bob communicate and cooperate with each
other, they would not confess so that they only serve $a<b$ years
in prison.

\section{Cournot duopoly}

A market comprised of two sellers and many competitive buyers is
know as a duopoly. The buyers can not influence the price or
quantities offered, it is assumed that the collective behavior of
the buyers is fixed and known. The competitive and cooperative
behavior between the sellers determines the price.

Two players are each manufacturers of the same single commodity,
the loss functions are cost function which depends on the
production of the two players. We denote the quantities of this
commodity produced by the two players by $x\in\mathbb{R}_{+}$ and
$y\in\mathbb{R}_{+}$. The price $p(x,y)$ is an affine function of
the total production $x+y$,
\begin{equation}\label{px+y}
p(x+y):=\alpha-\beta(x+y),
\end{equation}
and the cost function $u_{A}$ and $u_{B}$ of each manufacturer are
affine functions of the production $f_{A}x=ax+b,\;f_{B}(y)=ay+b.$
Alice's net cost is equal to $f_{A}(x,y):=f_{A}-p(x+y)x,
f_{A}(x,y):=f_{B}-p(x+y)y$. Eliminating the constant terms which
does not modify the game, the loss function reduced to
\begin{eqnarray}\label{fAfB=xy}
&&u_{A}(x,y):=x(x+y-u),\\
&&u_{B}(x,y):=y(x+y-u).
\end{eqnarray}
The non-cooperative equilibrium may be attained algorithmically
following the scenario of two payer's game.

\section{The Atiyah-Singer index theorem}\label{Atiyah-Singer}

The Atiyah-Singer index theorem is concerned with the existence
and uniqueness of solutions to linear partial differential
equations of elliptic type. The Fredholm index is a topological
invariant of elliptic equations. By computing a small number of
fundamental examples and by showing that both functions have
similar algebraic properties, Atiyah and Singer proved that the
Fredholm index and the topological index are both topological
invariants of elliptic equations, they are equal.

According to the Atiyah-Singer index theorem\cite{atiyah1,atiyah2}
the analytic index of the operator $D$ is defined as\cite{wang}
\begin{equation}\label{indexD}
Index D=dim Ker D-dim Coker D,
\end{equation}
where Ker D is the kernel of the operator, which is defined to be
the space of zero-modular solutions. i. e., on the entire space
$\Gamma$
\begin{equation}\label{KerD}
Ker D=\{\xi\in\Gamma(E)|D\xi=0\}.
\end{equation}
On a real oriented compact smooth $n=2l$-manifold $M$, the
Atiyah-Singer index theorem states that the Euler's characteristic
$Ch{(M)}$ is the sum of the Betti number,
\begin{equation}\label{euler}
Ch(M)=Index(\Lambda,d)=\sum_{p}(-1)^{p}dim_{R}H_{dR}^{p}(T(M),R).
\end{equation}
It is just the Gauss-Bonnet-Chern theorem. on the four dimensional
manifold, there are more interesting case. Such as the Dirac
operator, there are k zero modes of fermions coupled to the
k-instanton gauge field in the fundamental representation.
Atiyah-Singer index theorem which states that the the index of the
Dirac operator is minus the first Chern class, where $\nu_{+}$ is
the number of positive chirality zero-modes and $\nu_{-}$ is the
number of negative chirality zero-modes. i.e.,
\begin{equation}\label{mu}
Index D=Ch=\nu_{+}-\nu_{-}=\frac{1}{2\pi}\int_{M}{\Omega}.
\end{equation}

\section{Topological quantity}\label{topologicalphase}

\emph{{\textbf{\textsl{Chern character}}}}
\\
For a complex vector bundle on the base space-time manifold $M$,
whose structure group is the general $k$ dimensional complex
linear group $GL(k,c)$, there exists the Chern character, which is
an invariant polynomial of the group $GL(k,c)$,
\begin{eqnarray}\label{Holo-Chern}
Ch(M)&=&Tr(\exp\frac{i}{\pi}\Omega)\nonumber\\
&=&k+\frac{i}{2\pi}Tr\Omega+\frac{1}{2!}\frac{i^{2}}{(2\pi)^{2}}Tr(\Omega\wedge\Omega)+\cdots.
\end{eqnarray}
While this invariant polynomial may be expressed into a more
familiar generalized Berry phase,
\begin{eqnarray}\label{Holo-B1+B2+...}
Ch_{B}(M)&=&B_{0}+B_{1}+\frac{1}{2!}B_{2}+\cdots.
\end{eqnarray}
\\
\emph{{\textbf{\textsl{My Generalization of Berry
Phase}}}}\label{Gberry}
\\
In this section, we generalize the conventional Berry phase to
Chern character. We define the 1-form of the eigenfunction as
\begin{equation}\label{hoho-1-form}
d|\psi(R(t))\rangle=\partial_{R_k}|\psi(R(t))\rangle{d}{R_k},
\end{equation}
the wedge product of two one form corresponds to the conventional
Berry phase, i.e., the first Chern character,
\begin{eqnarray}\label{hoho-1-form-wedge}
&&B_{1}=d\langle\psi(R(t))|\wedge{d|\psi(R(t))\rangle}\nonumber\\
&=&\epsilon_{ijk}\langle\partial_{R_j}\psi(R(t))|\partial_{R_k}\psi(R(t))\rangle{d^{2}}{R}.
\end{eqnarray}
Then the second order generalization is
\begin{eqnarray}\label{hoho-2-form-wedge}
&&B_{2}=d\langle\psi|\wedge{d|\psi\rangle}{\wedge}d\langle\psi|\wedge{d|\psi\rangle}\nonumber\\
&&=\epsilon_{tjkl}\langle\partial_{R_t}\psi_{i}|\partial_{R_j}\psi_{i}\rangle\langle\partial_{R_k}\psi_{i}|\partial_{R_l}\psi_{i}\rangle{d^{4}R}.\nonumber\\
\end{eqnarray}
It is natural to arrive at the $p$th order generalization,
\begin{equation}\label{hoho-p-form-wedge}
B_{p}=\underbrace{d\langle\psi|\wedge{d|\psi\rangle}\cdots{\wedge}d\langle\psi|\wedge{d|\psi\rangle}}_{2p}.
\end{equation}
In mind of our new definition of the general Berry phase, the
p-form expression of Eq. (\ref{hoho-p-form-wedge}) is
\begin{eqnarray}\label{Holo-berry-2th-del-Eg}
B_{i}^{1}=\prod_{j\neq{i}}\delta(E_{j}-E_{i}){d\delta{E_{i}}}\wedge{d\delta{E_{i}}}
\end{eqnarray}
\begin{eqnarray}\label{Holo-berry-pth-del-Eg}
B_{i}^{p}=\prod_{j\neq{i}}\delta(E_{j}-E_{i})d\delta{E_{i}}\wedge{d\delta{E_{i}}}\wedge\cdots{d\delta{E_{i}}}\wedge{d\delta{E}_{i}}
\end{eqnarray}
According to Eq. (\ref{Holo-B1+B2+...}), the complete Chern
character can be expressed into beautiful form,
\begin{eqnarray}\label{Holo-Chern-del-Eg-B1-p}
Ch(M)&=&Tr[\exp(\frac{i}{\pi}\delta(E_{j}-E_{i})d\delta{E_{i}}\wedge{d\delta{E_{i}}})]\nonumber\\
&=&k+\frac{i}{2\pi}B_{1}+\frac{1}{2!}\frac{i^{2}}{(2\pi)^{2}}(B_{2})+\cdots.
\end{eqnarray}
When the correction to the $i$th energy level $\delta{E_{i}}$ is
expanded up to the $p$th order, in mind of Eq.
(\ref{delEn-1,2,3,...}), it is easy to verify that
\begin{eqnarray}\label{delEXn=delEnX}
&&\underbrace{d\delta{E}{\wedge}d\delta{E}{\wedge}\cdots{\wedge}d\delta{E}{\wedge}d\delta{E}}_{m}\nonumber\\
&&=d\delta{E}^{(m)}{\wedge}d\delta{E}^{(m)}{\wedge}\cdots{\wedge}d\delta{E}^{(m)}{\wedge}d\delta{E}^{(m)}.
\end{eqnarray}
This equation suggests that if we want to find out the $p$th order
of phase transition, the highest order of perturbation to the
$i$th energy level must be extended to the $p$th order.
\\
\emph{ {\textbf{\textsl{Berry Phase}}}}
\\
For the time dependent Hamiltonian operator $\hat{H_{i}}(R(t))$,
the eigenvalues and eigenvectors at time t can be expressed as a
function of t, $E_{i}(R(t))$ and $\psi_{i}(R(t))$.
\begin{equation}\label{psi(t)}
\hat{H_{i}}(R(t))|\psi_{i}(R(t))\rangle={E_{i}}(R(t))|\psi_{i}(R(t))\rangle.
\end{equation}
According to Berry's definition\cite{berry}, there was a induced
gage potential
\begin{equation}\label{A-psi}
\vec{A}(\vec{R(t)})=i\langle\psi_{i}(R(t))|\nabla_{\vec{R}}|\psi_{i}(R(t))\rangle.
\end{equation}
The first Chern number of Berry phase can be rewritten by these
coordinates in R space as
\begin{eqnarray}\label{chern-berry}
Ch_{1}&=&\frac{1}{2i\pi}\int_{R}\nabla\times\vec{A}({R(t)}),\nonumber\\
&=&i\epsilon_{ijk}\langle\partial_{R_j}\psi(R(t))|\partial_{R_k}\psi(R(t))\rangle.
\end{eqnarray}
The Chern number is actually the topological quantization of the
magnetic field
\begin{eqnarray}\label{dual-berry}
B^{1}&=&\nabla\times\vec{A}=i\epsilon_{tjk}\langle\partial_{R_j}\psi_{i}|\partial_{R_k}\psi_{i}\rangle\nonumber\\
&=&i\epsilon_{tjk}\sum_{p\neq{i}}\frac{\langle\psi_{i}|\partial_{R_t}H|\psi_{p}\rangle\langle\psi_{p}|\partial_{R_j}H|\psi_{i}\rangle}{(E_{p}-E_{i})^{2}}\nonumber\\
&=&-Im\epsilon_{tjk}\sum_{p\neq{i}}\frac{\langle\psi_{i}|\partial_{R_t}H|\psi_{p}\rangle\langle\psi_{p}|\partial_{R_j}H|\psi_{i}\rangle}{(E_{p}-E_{i})^{2}}.\nonumber\\
\end{eqnarray}
There has been some evidences pointing out that the Berry
curvature is singular at the points where the energy bands
touches\cite{niuqian}.

\section{Linear response and Green function}\label{Lresopnse }

Let $H_{0}$ be the full Hamiltonian describing the system in
isolation. One way to test its properties is to couple the system
to a weak external perturbation and to determine how the ground
state and the excited states are affected by the perturbation. The
Hamiltonian H for the system weakly coupled to an external
perturbation, which we will represent by a Hamiltonian $H_{e}$, is
$H=H_{0}+H_{e}.$ Let $\hat{O}(x, t)$ be a local observable, such
as the local density, the charge current, or the local
magnetization. The expectation value of the observable in the
exact ground state $|0\rangle$, under the action of the weak
perturbation $H_{e}$, is modified as
\begin{eqnarray}\label{<0>}
\langle0|\hat{O}|0\rangle_{e}=\langle0|\hat{O}|0\rangle+i\hbar^{-1}\int_{t_{0}}^{t}dt'\langle0|[H_{e},\hat{O}]|0\rangle+\cdots.
\end{eqnarray}
If we only retain the linear term in $H_{e}$, the first-order
change in a matrix element arising from the external perturbation
is expressed in terms of Heisenberg operator of the interacting,
$\delta{\langle0|\hat{O}|0\rangle}\equiv\langle0|\hat{O}|0\rangle_{e}-\langle0|\hat{O}|0\rangle$,
i.e. \begin{eqnarray}\label{delO}
\delta{\langle0|\hat{O}|0\rangle}=i\hbar^{-1}\int_{t_{0}}^{t}dt'\langle0|[H_{e},\hat{O}]|0\rangle.
\end{eqnarray}
This change represents the linear response of the system to the
external perturbation. It is given in terms of the ground state
expectation value of the commutator of the perturbation and the
observable. If $\hat{O}$ is a local observable, $H_{e}(t)$
represents an external source which couples linearly to the
observable, $H_{e}=\int{\hat{O}f(x,t)}$. The coefficient of
proportionality between the change in the expectation value
$\langle0|O(x, t)|0\rangle$ and the force $f(x,t)$ defines a
generalized susceptibility
\begin{equation}\label{susceptibility}
\chi=-\frac{i}{\hbar}\int_{-\infty}^{0}d\tau{e^{i\omega\tau}}\langle0|[\hat{O}(x',t'),\hat{O}(x,t)]|0\rangle.
\end{equation}
$G(x',x)_{o}=[\hat{O}(x',t'),\hat{O}(x,t)]$ is the retarded Green
function. Under Fourier transforms, this Green function can be
mapped into its momentum space,
\begin{equation}\label{fourier}
G(k,\omega)=\int{d(x-x')}{\int}dte^{ik(x'-x)}e^{i\omega{t}}G(x',x)_{o}
\end{equation}

In fact, the correction to the ground state energy can also be
expressed by Green function in momentum space. Usually, it is easy
to get $\delta{E_{g}}$ from the familiar formula,
\begin{equation}\label{goldstone}
\delta{E_{g}}=\langle0|H_{e}|0\rangle+
\sum_{n\neq0}\frac{\langle0|H_{e}|n\rangle\langle{n}|H_{e}|0\rangle}{E_{0}-E_{n}}+\cdots.
\end{equation}
We can write the Hamiltonian with a variable coupling constant
$\lambda$ as $\hat{H}(\lambda)=\hat{H}_{0}+\lambda\hat{H}_{e},$so
that $\hat{H}(1)=\hat{H}$ and $\hat{H}(0)=\hat{H}_{0}$. Then the
correction to the ground state energy is,
\begin{eqnarray}\label{delE-lambda}
&&\delta{E_{g}}=\pm\frac{1}{2}i\frac{V}{(2\pi^{4})}\int_{0}^{1}\frac{d\lambda}{\lambda}\int{d^{3}k}\int_{-\infty}^{\infty}d\omega{e^{i\omega}\eta}\nonumber\\
&&(\hbar\omega-\frac{\hbar^{2}k^{2}}{2m})TrG^{\lambda}(k,\omega).
\end{eqnarray}
Therefore, the topological quantum phase transition is intimately
related to the topology of momentum space. In fact, the correction
to the ground state energy is not the only physical quantity that
can be used to study the quantum phase transition, other physical
observables also present a sudden change at the phase transition
point. The thermal dynamic potential is another good candidate,
\begin{eqnarray}\label{del-omega-green}
&&\delta{\Omega}=\Omega-\Omega_{0}=\pm\int_{0}^{1}\frac{d\lambda}{\lambda}\nonumber\\
&&\int{d^{3}x}\lim\lim\frac{1}{2}[-\hbar\frac{\partial}{\partial\tau}+\frac{\hbar^{2}k^{2}}{2m}+\mu]TrG^{\lambda}(x\tau,x'\tau').
\end{eqnarray}
There also exist many other physical observables, such as the
single-particle current operator, $\hat{J}=\int{d^{3}x}j$,
$j_{\alpha\beta}=\sum_{\alpha\beta}\psi^{\dag}_{\alpha}J_{\alpha\beta}\psi_{\beta}$,
whose expression in terms of Green function is
$<j>=\pm{i}\lim\lim{Tr}[J(x)G(xt,x't')].$ The number density
operator of particles is
$<\hat{\rho}(x)>=\pm{i}Tr{{\rho}G(xt,x't')}.$ The spin density
operator $<\hat{\sigma}(x)>=\pm{i}Tr{{\sigma}G(xt,x't')}.$

\section{Green function}\label{selfE}

We first show that the Green function follows a similar
Schr${\ddot{o}}$dinger equation. For a system described by a
Hamiltonian $H$ which can not be solved exactly, the usual
approach is to set, $H=H_{0}+V,$ where $H_{0}$ is the unperturbed
part which may be solved exactly. The term $V$ represents all the
interactions. The wave function is governed by the interaction,
\begin{equation}\label{tpsi=Vpsi}
i\partial_{t}\psi(t)=\hat{V}(t)\psi(t).
\end{equation}
The wave functions of the Schr$\ddot{o}$dinger equation,
$i\partial_{t}\psi=H\psi$, are time dependent,
$\hat{\psi}(t)=U(t)\psi(0)=e^{iH_{0}t}e^{-iHt}\psi(0). $The
operator $U(t)$ obeys a differential equation which can written in
the interaction representation,
$i\partial_{t}U(t)=\hat{V}(t)U(t).$ The operator $U(t)$ in the
interaction representation has a expansion,
$U(t)=1+\sum_{n=1}^{\infty}\frac{(-i)^{n}}{n!}\int_{0}^{t}dt_{1}\int_{0}^{t}dt_{2}{\cdots}
dt_{n}\int_{0}^{t}T[\hat{V}(t_{1})\hat{V}(t_{2})\cdots\hat{V}(t_{n})],$
$T$ is the time-ordering operator. $U(t)$ may be abbreviated by
writing it as $U(t)=T\exp[-i\int_{0}^{t}dt_{1}\hat{V}(t_{1})].$.
The $S$ matrix changes the wave function $\psi(t')$ into $\psi(t)$
may be defined from $U(t)$, $\hat{\psi}(t)=S(t,t')\hat{\psi}(t').$
The time-ordered operator also obeys $S(t,t')=U(t)U(t')$. It is
easy to verify $S(t,t')$ obeys
\begin{equation}\label{tS=VS}
i\partial_{t}S(t,t')=\hat{V}(t)S(t,t').
\end{equation}
This equation is intimately related to the computation of ground
state (or vacuum) expectation values of time ordered products of
field operators in the Heisenberg representation
\begin{equation}\label{Ngreen}
G^{N}(x_{1},x_{2},...,x_{N})=\langle0|T[\hat{\phi}(x_{1})\hat{\phi}(x_{2})...\hat{\phi}(x_{N})]|0\rangle,
\end{equation}
In particular the 2-point Green function,
$G^{2}(x_{1},x_{2})=\langle0|T[\hat{\phi}(x_{1})\hat{\phi}(x_{2})]|0\rangle,
$ is known as the Feynman Propagator. For an interacting case, the
Green function reads\cite{mahan}
\begin{equation}\label{2green-expan}
G^{12}(x_{1},x_{2})=\frac{\langle0|T[\hat{\phi}(x_{1})\hat{\phi}(x_{2})S(+\infty,-\infty)]|0\rangle}
{\langle0|S(+\infty,-\infty)|0\rangle}.
\end{equation}
The Green function of energy is defined by the usual Fourier
transformation with respect to the time invariable:
\begin{equation}\label{GpE}
G(\textbf{p},E)=\int_{-\infty}^{+\infty}dte^{iE(t-t')}G(\textbf{p},t-t').
\end{equation}
According to the Goldstone's theorem,
\begin{equation}\label{E-E0=0U0}
{\delta}E=E-E_{0}=\frac{\langle0|H_{e}S(+\infty,-\infty)|0\rangle}{\langle0|S(+\infty,-\infty)|0\rangle}.
\end{equation}
The terms in the series for
${\langle0|S(+\infty,-\infty)|0\rangle}$ are called $vacuum$
polarization terms. There was theorem states that the vacuum
polarization diagrams exactly cancel the disconnected diagrams in
the expansions for $G(p,t-t')$.

A great simplification is that we sum over only linked (connected)
distinct graphs. It turn out that this corresponds exactly to
cancelling the factor of ${\langle0|S(+\infty,-\infty)|0\rangle}$
in the denominator. From the discussions above, one can that the
unperturbed part of the Hamiltonian $H_{0}$ plays a trivial role
in the quantum phase transition, it is the interaction part
$H_{e}$ that decide the orientation of the evolution of a physical
system. If we introduce the self-energy function $\Sigma(p,E)$ to
absorb all the interaction, the exact Green function can obtained
from the Dyson's equation,
\begin{equation}\label{Dyson-feynamn}
G(p,E)^{-1}={G^{-1}_{0}(P,E)}[{1-G_{0}(p,E)\Sigma(p,E)}].
\end{equation}
we expand $G$ in the power series
$G(p,E)={G_{0}}+{G_{0}}^{2}\Sigma+{G_{0}}^{3}\Sigma^{2}+\cdots.$
The self-energy is a summation of an infinite number of distinct
diagrams in the series. However it is impossible to get
$\Sigma(p,E)$ exactly, one must be contend with an approximation
results. Usually, the higher order of phase transition means the
higher order of self-energy terms must be included. The Green
function of an interacting electron system is
$G_{r}(p,\omega)=[{\omega-E_{k}-\Sigma}]^{-1}.$ The corresponding
spectra function is $A(p,\omega)=-\frac{1}{\pi}ImG_{r}$,
\begin{equation}\label{A=limit-0}
A(p,\omega)=\lim_{Im\Sigma_{r}\rightarrow0}\frac{-1/{\pi}Im\Sigma_{r}}{(\omega-E_{q}')^{2}+Im\Sigma^{2}_{r}}=\delta(\omega-E_{q}').
\end{equation}
where $E_{q}'=E_{q}-Re\Sigma_{r}(p,\omega)$. The spectra function
represents a resonance peak with width $2\Gamma_{q}$. One can
associate each peak with a quasi-particle. The lifetime of
quasiparticle is infinite for a vanished $Im\Sigma_{r}$,
$\tau_{q}=\lim_{Im\Sigma_{r}}[{2Im\Sigma_{r}}]^{-1}\rightarrow\infty.$
So the quasiparticles on the fermi surface is a kind of
topological excitation, the external perturbation does not shorten
its life. As all know, $E_{0}={G_{0}(P,E)}^{-1}$ is the
unperturbed part of the system. Therefore it is the
self-interacting part $\Sigma(p,E)$ that determines a phase
transition, in the frame work of out theory, one can denote the
self-energy as the perturbation part to exactly solved part,
\begin{equation}\label{G-G0+=sigma}
\delta{{E'}}=G(p,E)^{-1}-{G_{0}(P,E)}^{-1}=\Sigma(p,E).
\end{equation}
Usually $\Sigma(p,E)$ is a complex matrix, it may be rewritten in
the Dirac's bra and ket representation,
$\Sigma(p,E)=|\Sigma(p,E)\rangle$ and
$\Sigma(p,E)^{\dag}=\langle\Sigma(p,E)|$.

\end{document}